\documentclass[11pt,tightenlines,eqsecnum,floats,aps,amsmath,amssymb,nofootinbib,prd,shownopacs]{revtex4}

\usepackage{amsmath,amssymb,amsfonts,amsthm}
\usepackage{graphicx}
\usepackage{enumerate} 
\usepackage{color} 
\usepackage{tikz}

\usepackage[mathscr]{eucal}
\usepackage[latin1]{inputenc}
\usepackage{amsmath}
\usepackage{amsfonts}
\usepackage{amssymb}
\usepackage{graphicx}
\usepackage{enumerate}
\def\lp{{\ell}_{\rm Pl}}

\newcommand{\p}{\partial}

\newcommand{\mpl}{M_{\rm Pl}}

\newcommand{\f}{\frac}

\newcommand{\fref}[1]{Fig.\,\ref{#1}}
\newcommand{\eref}[1]{eq.\,(\ref{#1})}

\def\f{\frac}

\def\tpl{t_{\rm Pl}}

\usepackage{enumerate}

\newcommand{\be}{\nopagebreak[3]\begin{equation}}
\newcommand{\ee}{\end{equation}}
\newcommand{\bfig}{\nopagebreak[3]\begin{figure}}
\newcommand{\efig}{\end{figure}}
\newcommand{\ba}{\nopagebreak[3]\begin{eqnarray}}
\newcommand{\ea}{\end{eqnarray}}

\newcommand{\bmult}{\nopagebreak[3]\begin{multline}}
\newcommand{\emult}{\end{multline}}

\def\lp{{\ell}_{\rm Pl}}
\def\mpl{{m}_{\rm Pl}}

\def\Heff{\mathcal{H}_{\rm eff}}
\def\Hmatt{\mathcal{H}_{\rm matt}}
\def\mua{\bar{\mu}_1}
\def\mub{\bar{\mu}_2}
\def\muc{\bar{\mu}_3}
\def\lp{l_{\rm Pl}}

\begin{document}

\title{A quantum gravitational inflationary scenario in Bianchi-I spacetime}
\author{Brajesh Gupt}
\email{bgupt1@lsu.edu}

\author{Parampreet Singh}
\email{psingh@phys.lsu.edu}
\affiliation{Department of Physics and Astronomy, Louisiana State University, Baton Rouge, 70803}


\begin{abstract}
We investigate the $\phi^2$ inflationary model in the Bianchi-I spacetime using the effective spacetime description of loop quantum cosmology to understand the issues of the resolution of initial singularity, isotropization, effect of anisotropies on the amount of inflation, and the phase space attractors in the presence of non-perturbative quantum gravitational modifications. A comparative analysis with the classical theory by including more general initial conditions than the ones previously considered in the latter is also performed. We show that, in general, the classical singularity is replaced by a bounce of the mean scale factor in loop quantum cosmology. Due to the underlying quantum geometric effects, the energy density of the inflaton and the anisotropic shear remain bounded throughout the non-singular evolution. Starting from arbitrary anisotropic initial conditions, a loop quantum universe isotropizes either before or soon after the onset of slow-roll inflation. We find a double attractor behavior in the phase space dynamics of loop quantum cosmology, similar to the one in classical theory, but with some additional subtle features.  Quantum modifications to the dynamical equations are such that, unlike the classical theory, the amount of inflation does not monotonically depend on the initial anisotropy in loop quantum cosmology. Our results suggest that a viable non-singular inflationary model can be constructed from highly anisotropic initial conditions in the Planck regime.

\end{abstract}

\maketitle


\section{Introduction}

The inflationary paradigm provides an excellent explanation of the homogeneity and the flatness problems of the standard model of cosmology and the origin of the observable large scale structure in our universe. However, various questions about the physics of the pre-inflationary stage of the universe remain unanswered. In particular, in general relativity (GR), inflationary spacetimes have been shown to be past incomplete \cite{borde}, and little is known about the initial conditions and the geometry of the pre-inflationary epoch. Both of these issues need to be carefully addressed in any complete model of inflation. Unlike the
issue of the past singularity, which can only be faithfully addressed using a quantum theory of gravity, the issue of the role of pre-inflationary geometry on the onset and predictions of inflation
has been previously studied using GR (and also in modified theories of GR). In this setting, various investigations have been carried
out to understand the onset of inflation in the presence of anisotropic shear in the pre-inflationary epoch, often by assuming a homogeneous and anisotropic patch of spacetime using Bianchi models
\cite{collins,barrowturner81,barrow1,wald,steigturner,staro,demianski,rothmad,gron,turnerwidrow, gonzalezjones,jensenstein,rothellis,sahni1,modak,burd,ford,campa,mss,barrow_hervik1,pitrou,kofman,barrow_hervik2,soda1}. Many results point towards isotropization and genericness of inflation starting from initial anisotropic conditions, however counter-examples have also been found (see for eg. \cite{barrow_hervik1,barrow_hervik2,soda1}). Since the anisotropic shear diverges  in classical gravity  close to the initial singularity, it has been argued that a consistent treatment of
anisotropies in the pre-inflationary epoch is incomplete without inputs from quantum gravity \cite{rothmad,rothellis}. Due to these reasons, it is important to study the pre-inflationary dynamics in a quantum gravitational anisotropic spacetime to understand the way quantum gravity effects
resolve the classical singularity, and affect the isotropization and the onset of inflation in presence of anisotropies.

The goal of this article is to address these issues in loop quantum cosmology (LQC) \cite{as1}, a framework to
quantize homogeneous spacetimes based on loop quantum gravity (LQG). A key prediction of LQC is that the backward evolution of a universe, starting from classical initial data, by the quantum Hamiltonian constraint leads to a   non-singular bounce \cite{aps1,aps2,aps3}. The resolution of singularity in LQC is a result of the underlying discreteness of the quantum geometry at the Planck scale, and its occurrence has been confirmed in a variety of isotropic and anisotropic models (see for eg. \cite{apsv,warsaw_closed,warsaw_flat,acs,kv,szulc_open,bp,kp1,ap,awe2,chioub1,b1madrid1,awe3,we1}), including the $\phi^2$ inflationary model in the isotropic setting \cite{aps4}. A powerful tool in understanding the detailed physics of these models is the effective spacetime description of the underlying quantum dynamics \cite{jw,vt}, which using sophisticated
numerical simulations has been tested to the  Planck curvature scale for various matter models for both isotropic \cite{aps1,aps2,aps3,apsv,ap,bp,kv} as well as anisotropic spacetimes \cite{b1madrid2,szulc2} (see Refs.\cite{ps12,khanna} for a review of numerical methods in LQC). The effective spacetime description is derived using an effective Hamiltonian obtained via  geometric methods of quantum mechanics \cite{schilling}, by assuming states which become semi-classical at late times, describing a macroscopic universe in the classical regime. It is found that GR is an excellent approximation to quantum dynamics at curvature scales small compared to the Planck value, however departures between the two theories become significant in the Planck regime which is described by the effective dynamics to an excellent accuracy. Using effective Hamiltonian method, bounds on expansion and shear scalars have been derived \cite{cs09,gs1}, and
generic results on the absence of strong singularities in isotropic \cite{ps09,sv} and Bianchi-I spacetime \cite{ps11} have been obtained.

In this manuscript, we analyze the dynamics of $\phi^2$ inflation model in the Bianchi-I spacetime in the effective spacetime description of LQC. We first show that unlike in the classical Bianchi-I model, the past evolution in LQC is non-singular. Singularity resolution is confirmed  for a large set of  initial conditions, in confirmation with the
generic result on resolution of singularities in Bianchi-I spacetime in LQC \cite{ps11}.
In contrast to the isotropic inflationary models in LQC where there is a single bounce of the isotropic scale factor \cite{svv,zhang-ling,lalak,asloan_prob1,herrera,ck1,asloan_prob2,rs} always occurring at the maximum allowed value of the energy density $\rho_{\rm max} = 0.41\,\rho_{\rm Pl}$, the resolution of singularity in Bianchi-I spacetime occurs via non-singular bounces of the three directional scale factors $(a_1, a_2, a_3)$ occurring at a range of values of
energy density and anisotropic shear scalar $\sigma^2$, bounded by $\rho_{\rm max}$ (same as in the isotropic model) and $\sigma^2_{\rm max} = 11.57/l_{\rm Pl}^2$. In the isotropic LQC, $\rho = \rho_{\rm max}$ at the bounce constrains the value of the
 inflaton velocity at the bounce for a given initial value of the inflaton field.  On the other
 hand, in Bianchi-I spacetime in LQC, there is no such restriction. The bounce of the mean scale factor in Bianchi-I spacetime (defined as $a = (a_1 a_2 a_3)^{1/3}$ can take place
 even if the energy density is not at its maximum value at the bounce. In this way,
 presence of non-vanishing shear leads to sets of initial conditions which were not
 allowed in the isotropic spacetime.

The availability of a non-singular inflationary model in the presence of anisotropies sets the stage to answer some important questions which so far could only be addressed with in the limitations of the classical theory. These include, the way slow-roll inflation in $\phi^2$ model begins, starting from highly anisotropic geometries; whether slow-roll inflation is an attractor of trajectories starting from arbitrary anisotropic conditions in the Planck regime; and the way non-zero anisotropic shear affects the number of e-foldings. Though these issues have been addressed to some extent earlier in the classical theory, for completeness and comparative analysis, we analyze both the classical and the effective dynamics of LQC for the $\phi^2$ model in Bianchi-I spacetime. In the process, apart from answering the above questions in the non-singular anisotropic inflationary model in LQC, we gain new insights on some of these issues in the classical theory. Below, we  summarize the main results on the study of pre-inflationary dynamics and the comparison between the classical theory and LQC.

It is known that due to the presence of anisotropic shear, the Hubble friction during cosmic expansion is enhanced which, in turn, leads to faster decay of the kinetic energy of the inflaton
\cite{mss}. This helps the isotropic slow-roll conditions arrive more quickly. If the inflaton is
initially taken to roll down the potential at some initial time $t=t_i$ ($\dot\phi(t_i)<0$), then due to the enhanced
Hubble friction, higher anisotropy leads to an increase in the number of e-foldings.  In this article, in the discussion of classical dynamics,  we consider a more generic initial condition on
the initial velocity of the inflaton, including the case when the inflaton is initially rolling up the potential in the pre-
inflationary era. We find that, for such an initial condition i.e.\ $\dot\phi(t_i) > 0$, the number of e-foldings  decrease with an increase in the  anisotropic shear in the classical theory in comparison to the corresponding isotropic evolution.
Further, in the regime of low anisotropy, there is a significant difference in the number of e-foldings depending on whether  initial value of $\dot\phi$ is positive or negative. In contrast, we find that the large anisotropy gives rise to the same number
of e-foldings for both the signs of initial $\dot \phi$.  The latter behavior of the amount of
inflation can be attributed to the strength of the Hubble friction which, at large anisotropy, is so
strong that the sign of $\dot \phi$ becomes insignificant.

In the classical theory, for a given type of initial condition, the number of e-foldings ($N$)
show a monotonic variation with anisotropic shear i.e.\ $N$ either decreases (for an inflaton rolling up the potential) or increases (for an inflaton rolling down the potential) with an increasing initial value of the anisotropic shear.
In LQC, however, it turns out that the amount of inflation does not monotonically vary with the
anisotropic shear. In the case when initial $\dot\phi$ is chosen to be negative\footnote{The initial conditions in most of the simulations for LQC discussed in this work are provided at the bounce of the mean scale factor.}, the number of e-foldings
increase with increasing shear if the anisotropic shear scalar $\sigma^2$ is less than a particular value $\sigma^2{}_*$, and attain a maximum value when $\sigma^2=\sigma^2{}_*$. Thus, unlike the classical theory, there exists a range of $\sigma^2$: $\sigma^2{}_* < \sigma^2 \leq \sigma^2_{\rm max}$, for which an increase in anisotropy decreases the number of e-foldings.  On the other hand, if the initial velocity of the scalar field is positive, the
behavior of $N$ is opposite. That is, $N$ decreases for $\sigma^2< \sigma^2{}_*$, increases
if $\sigma^2> \sigma^2{}_*$ and attains a minimum value at $\sigma^2 =  \sigma^2{}_*$.
Interestingly, the value of $\sigma^2{}_*$ turns out be independent of whether the initial
velocity of the scalar field is positive or negative (though it depends on the absolute values of
initial $\dot\phi$ and initial $\phi$).


 We also study the attractor behavior of the phase-space trajectories of the classical and LQC
 effective dynamics. In the classical theory, it is known that all the classical trajectories join the isotropic slow-roll inflationary
 trajectory in their future evolution provided the initial value of $\phi$ is high enough to allow inflation (see for eg. \cite{pitrou}). Also, in the limit of initial shear scalar tending to zero,
 Bianchi-I trajectories approach the isotropic Friedmann-Robertson-Walker (FRW) trajectory, irrespective of the initial energy
 density. As in GR, it turns out that in LQC, irrespective of the initial anisotropic content of the
 spacetime, all the Bianchi-I spacetime trajectories,  meet the isotropic slow-roll inflation in their
 future evolution, if the initial value of the inflaton field is high enough. In this way, slow-roll inflation
 is an attractor for all the trajectories in the effective spacetime of Bianchi-I LQC with a $\phi^2$ potential starting at the bounce.
 Unlike in the classical theory where there exists an isotropic trajectory
  for every given value of energy density in the initial data, in LQC,
 the energy density at the bounce is fixed at $\rho=\rho_{\rm max}$ in the isotropic spacetime.
 Therefore, in order to obtain the isotropic limit of Bianchi-I spacetime in LQC at the bounce, one has to
 consider $\rho\rightarrow\rho_{\rm max}$ in addition to $\sigma^2\rightarrow0$.
 If the energy density at the bounce is fixed at any value other than $\rho_{\rm max}$ (which is allowed in Bianchi-I LQC), then
 it turns out, that the shear scalar can not be decreased to zero. In that case, there is a minimum
 non-zero value of the shear scalar, depending on the energy density at the bounce.
 In this sense, the approach to the isotropic limit in the effective description of LQC is subtle  in comparison to the classical theory. That is, in the classical theory, isotropic spacetime can be
 approached by decreasing $\sigma^2$ to zero for any fixed value of  energy density in the
 initial data, whereas, in LQC, the energy density at the bounce must also tend to  $\rho_{\rm max}$ in addition to $\sigma^2\rightarrow0$.

This manuscript is organized as follows, in section-II, we present the dynamical equations of
Bianchi-I spacetime in the classical and the effective description of LQC in the Ashtekar variables. We discuss the features of anisotropic geometries, slow
roll parameters, conditions for accelerated expansion, and behavior of equation of state of the
scalar field in the classical theory.
Based on the modified dynamical equations we discuss some key properties of the mean Hubble
rate and the shear scalar in the effective description of LQC.
In section-III, we discuss the numerical studies of the model where we present the results of
various numerical simulations of classical and the LQC dynamics. In all the simulations, the mass of the inflaton is fixed to $m = 1.21\times10^{-6}\mpl$, consistent with the WMAP data \cite{wmap}. With the help of explicit
numerical simulations, we show the occurrence of non-singular bounces for various values of initial
anisotropic shear in the effective dynamics of Bianchi-I spacetime. We explore the process of isotropization, behavior of the number of e-foldings for
various initial conditions and study the phase space trajectories of the classical and LQC
Bianchi-I spacetime. To discuss the qualitative behavior of trajectories, the initial value of inflaton in various plots is taken as $\phi = 3.14 \mpl$. Since this initial value is also required in the classical isotropic inflationary model for sufficient number of e-foldings, it serves as a good representative of the inflationary
 trajectories to study the effects of presence of anisotropy on the amount of inflation as
 compared to the classical isotropic slow-roll. Summary of results from  various other initial values of the inflaton is presented  in the tabular form, in
 Tables- \ref{classefoldtab} and \ref{efolddatavaryphi}. We summarize with the discussion of results in Sec. V.


\section{Classical and Effective Dynamics: Basic equations}
In this section we present the description of Bianchi-I spacetime in terms of the loop quantum variables -- the Ashtekar-Barbero connection $A_a^i$ and the triads $E_i^a$. In the following,
 the first subsection is devoted to the classical theory. After discussing the relationship of connection and triad variables with the metric variables, we outline the derivation of classical dynamical equations from the Hamiltonian constraint, and discuss in detail the conditions for accelerated expansion in the classical Bianchi-I spacetime sourced with a $m^2 \phi^2$ inflationary potential (with $m = 1.21 \times 10^{-6}$ $m_{\rm{Pl}}$).
In the second subsection we outline the derivation of effective dynamical equations in the effective spacetime description of Bianchi-I spacetime in LQC, and note key non-trivial features in comparison to the classical theory which play an important role in  distinctions of dynamical behavior of trajectories in GR and LQC.


\subsection{Classical theory}
We consider a homogeneous Bianchi-I spacetime with the spatial manifold having the topology $\Sigma (= \mathbb R^3)$, thus endowing $\Sigma \times \mathbb R$ topology to the spacetime manifold. The metric for Bianchi-I spacetime is given as
\be
d s^2 = - N^2 \, dt ^2\,+\,a_1^2\,dx^2\,+\, a_2^2\,dy^2\,+\,a_3^2\,dz^2,
\ee
where $a_i$ denote the directional scale factors.
 In order to define a symplectic structure on the manifold we need to introduce a fiducial cell $\mathcal V$ on the spatial manifold of the spacetime, with fiducial volume  $V_o=l_1l_2l_3$. Here $l_i$ denote the co-ordinate lengths in the three spatial directions.
The edges of the fiducial cell are chosen to lie along the fiducial triads $\mathring e_a^i$. The fiducial metric compatible with the fiducial co-triads $\mathring\omega_a^i$ is denoted as $\mathring{q}_{ab}$.

Utilizing the underlying symmetries of the homogeneous spatial manifold of Bianchi-I spacetime, the connection $A_a^i$ and triads $E_i^a$ can be written in terms of the symmetry reduced connections ($c_i$) and triads ($p_i$) as
\be
A_a^i = c^i (l_i)^{-1} \mathring\omega_a^i \quad {\rm and} \quad E_i^a=p_i(l_i) V_o^{-1}\sqrt{\mathring q},
\ee
where $\mathring q ={\rm det}\,( \mathring q_{ab})$ is the determinant of the fiducial metric and the index $i$ runs from $1$ to $3$. The symmetry reduced connection $c_i$ and triad $p_j$ are conjugate to each other and they satisfy the following Poisson bracket relation
\be
\{ c^i, \, p_j\} = 8 \pi G \gamma \delta_j^i,
\ee

where $\gamma=0.2375$ is the Barbero-Immirzi parameter whose value is fixed using the black-hole entropy computation in LQG. The triads $p_i$ are kinematically related to  the usual metric variables of the spacetime in the following manner,
\be
p_1 \,=\, \varepsilon_1 \, l_2 l_3 |a_2 a_3|, \, p_2 \, =\, \varepsilon_2 \, l_1 l_3 |a_1 a_3|, \, p_3 \, = \, \varepsilon_3 \, l_1 l_2 |a_1 a_2|
\ee
where $\varepsilon_i = \pm 1$ depending on whether the triads have positive or negative orientation. In the following, without any loss of generality, we choose $\varepsilon_i = +1$, and the coordinate lengths $l_i$ to be unity.


\subsubsection{Dynamical equations}

The classical Hamiltonian constraint for Bianchi-I spacetime, with lapse $N=1$, in terms of the connections $c_i$ and triads $p_j$ can be written as follows  \cite{chioub1}
\be
\mathcal H_{\rm cl} = -\f{1}{8\pi G\gamma^2 V}\left(c_1c_2p_1p_2\,+\,{\rm cyclic\, terms}\right) + \Hmatt,
\ee
where $V=\sqrt{p_1p_2p_3}$ stands for the physical volume of the cell $\mathcal V$ and $\Hmatt$ denotes the  Hamiltonian of the matter sector. Considering the form of the self interacting potential given via $V(\phi)= \f{1}{2}m^2\phi^2$, the matter Hamiltonian of the system can be written as
\be
\label{hmatt} \Hmatt =\f{P_\phi^2}{2\sqrt{p_1p_2p_3}} + \f{1}{2}m^2\phi^2\sqrt{p_1p_2p_3},
\ee
where $P_\phi$ is the conjugate momentum of the scalar field $\phi$.
Starting from the total classical Hamiltonian constraint, the dynamical equations can be derived using Hamilton's equations of motion
\ba
\dot{p}_i &=& \{p_i,\mathcal H_{\rm cl}\}=-8\pi G\gamma\, \f{\p \mathcal H_{\rm cl}}{\p c_i}, \nonumber \\
\dot{c}_i &=& \{c_i,\mathcal H_{\rm cl}\}=8\pi G\gamma\, \f{\p \mathcal H_{\rm cl}}{\p p_i}.
\ea
Further, the directional Hubble rates are given in terms of the time derivatives of the triads $p_i$ as follows:
\be
H_1 = \f{1}{2}\left(\f{\dot p_2}{p_2}+\f{\dot p_3}{p_3}-\f{\dot p_1}{p_1}\right),
\ee
and similarly for $H_2$ and $H_3$. Using the equations of motion and the expression for the directional Hubble rates it is now straightforward to write the classical Einstein's equations for Bianchi-I spacetime as
\ba
\label{einsteinbianchi}
\nonumber H_1H_2+H_2H_3+H_3H_1 &=& 4 \pi G (\dot{\phi}^2 + m^2\phi^2), \\
\dot H_2 + \dot H_3 + H_1^2 + H_2^2 + H_3^2 &=& - 4 \pi G (\dot{\phi}^2 - m^2\phi^2),
\ea
and the cyclic permutation of these two equations. 
Equations of motion for the scalar field can also be derived in a similar fashion which yield the  Klein-Gordon equation,
\be
\label{KG} \ddot \phi + 3 H \dot\phi + V_{,\phi} = 0 ~,
\ee
where $H = (H_1+H_2+H_3)/3$ is the mean Hubble rate and $V_{,\phi}$ is the derivative of the self interacting potential with respect to the field $\phi$. The Klein-Gordon equation governing the evolution of the scalar field is equivalent to the conservation equation $\dot\rho = -3H(\rho + P)$, where $\rho=\dot\phi^2/2 + V(\phi)$ is the energy density and $P=\dot\phi^2/2 - V(\phi)$ is the pressure of the scalar field. In the inflationary scenario,
the term $3 H \dot \phi$, also known as Hubble friction, plays an important role by causing the slow-roll of the inflaton in an expanding universe.
A bigger friction term results in faster decay of the kinetic energy of the inflaton thus leading to an early start of the slow-roll conditions.
 This turns out to be the key to understanding the effect of the anisotropy on the amount of inflation obtained, as the anisotropic shear interacts with the dynamics of the scalar field through \eref{KG}.

The anisotropic shear is measured by a scalar $\sigma^2$ which in terms of the directional Hubble rates is given as
\be
\sigma^2  = \f{1}{3}\left(\left(H_1-H_2\right)^2+\left(H_2-H_3\right)^2+\left(H_3-H_1\right)^2\right).
\ee
Using the expression of shear scalar $\sigma^2$, the  vanishing of the Hamiltonian constraint ${\cal H}_{\rm{cl}} \approx 0$, results in the generalized Friedmann equation for diagonal Bianchi-I spacetime:
\be
\label{fried} H^2 = \f{8\pi G}{3} \rho + \f{1}{6} \sigma^2 ~.
\ee
Further, using Einstein's equations for diagonal Bianchi-I spacetime (given by eq.\,(\ref{einsteinbianchi})) one can derive the Raychaudhuri equation as follows
\be
\label{raychaudh} \f{\ddot a}{a} = - \f{4\pi G}{3} \left(\rho + 3P\right) - \f{1}{3} \sigma^2.
\ee
\ Eq.\,(\ref{raychaudh}) can also be obtained by computing the time derivative of the Friedmann equation given by eq.\,(\ref{fried}) and utilizing the conservation equation $\dot\rho= - 3H(\rho +P)$ along with $\dot\sigma^2= - 6 H \sigma^2$ \cite{misner1, jacobs2}. Note that in the classical theory, $\sigma^2\propto a^{-6}$.

\subsubsection{Kasner exponents and Jacobs' parameter}

To understand the structure of spacetime in the very early stage of evolution (including the pre-inflationary epoch),  it is convenient to use the Kasner exponents, $k_i$, which are related to the directional scale factors via $a_i \propto t^{k_i}$. The Kasner exponents have been extensively used to understand structure of the spatial geometry during the approach to classical singularity, which can be classified into  point, barrel, pancake and cigar types.
When all the Kasner exponents are positive ($k_1, k_2, k_3 > 0$), then all the directional scale factors approach singularity together and the spatial geometry of the spacetime tends to a point like structure. Similarly, the approach to singularity is pancake when two of the Kasner exponents are zero and one of them is positive ($k_1,k_2=0\,{\rm and}\, k_3>0$), barrel, when one of the exponents is zero while other two positive ($k_1=0$, $k_2,k_3>0$), and cigar, when one of the exponents is negative and two of them are positive ($k_1<0$ and $k_2,k_3>0$).

Using the definition of the shear scalar ($\sigma^2$) and the expansion scalar $\theta = H_1+H_2+H_3$, one can write the quantity $\sigma^2/\theta^2$ in terms of the Kasner exponents as follows
\be
\label{hierarchy}\f{\sigma^2}{\theta^2} = \f{2}{3}\left(1-3\left(k_1k_2+k_2k_3+k_3k_1\right)\right).
\ee
The ratio of the shear scalar and the expansion scalar takes the minimum value when all of the three directions of the spacetime are expanding i.e. all the three Kasner exponents are positive.
It can be easily proved that if $\sigma^2/\theta^2 < 1/6$, then the only solution possible is that of all of the Kasner exponents are positive. This corresponds to a point-like approach to the classical singularity in the backward evolution.
Similarly, pancake type structure
is characterized by $\sigma^2/\theta^2 = 2/3$, the cigar type structure
takes place when $\sigma^2/\theta^2>1/6$ and the barrel type structure
is formed when $1/6\leq\sigma^2/\theta^2<2/3$.
Thus, $\sigma^2/\theta^2$ describes a hierarchy of the spatial geometry based on the anisotropic shear present in the spacetime \cite{gs2}. Since the expression for $\sigma^2/\theta^2$ does not assume any particular form of matter, the spatial structure of the geometry given by \eref{hierarchy} holds true irrespective of the matter content. As an example, for the inflationary spacetime, if $\sigma^2/\theta^2 < 1/6$ in the future evolution, then all the Kasner exponent must take positive values and hence all the directions of the spacetime must be expanding.

For the simplicity of the following calculations, we use the Jacobs' parameter $\epsilon_{\rm J} = \sigma/(4\pi G\rho)^{1/2}$ \cite{jacobs2}.
The parameter $\epsilon_{\rm J} $ denotes the dominance of the anisotropic shear over the matter density. A large value of $\epsilon_{\rm J} $ refers to a highly anisotropic universe while $\epsilon_{\rm J}  \rightarrow 0$ corresponds to isotropization (i.e. when the contribution of matter density to the spacetime curvature is far greater than that of the anisotropic shear).
Moreover, the value of $\epsilon_{\rm J}$ can be related to the structure of the spatial geometry near the classical singularity. As an example, for $\epsilon_{\rm J}  < 2/\sqrt{3}$ (which corresponds to $\sigma^2/\theta^2 < 1/6$), all the directional Hubble rates are necessarily positive and the approach to singularity in the backward evolution is point like. For very large anisotropy, with $\epsilon_{\rm J}  \gg 2/\sqrt{3}$, the approach to singularity is cigar like.
 For a detailed discussion on the relation between anisotropy and the hierarchy of the approach to classical singularity in diagonal Bianchi-I spacetime, see Ref. \cite{gs2}.

\subsubsection{Accelerated expansion}

The condition for an accelerated expansion in terms of the equation of state and the parameter $\epsilon_{\rm J}$ can be obtained by putting the condition $\ddot a/a>0$ in the Raychaudhuri equation (\eref{raychaudh}) for Bianchi-I spacetime. It yields:
\be
\label{inflcond} w < -\f{1+\epsilon_{\rm J}^2 }{3},
\ee
where $w= P/\rho$ is the varying equation of state of the scalar field. For $\epsilon_{\rm J} =0$ the above expression reduces to the condition for inflation in isotropic spacetime $(w < -1/3)$.

\begin{figure}
\includegraphics[width=0.47\textwidth]{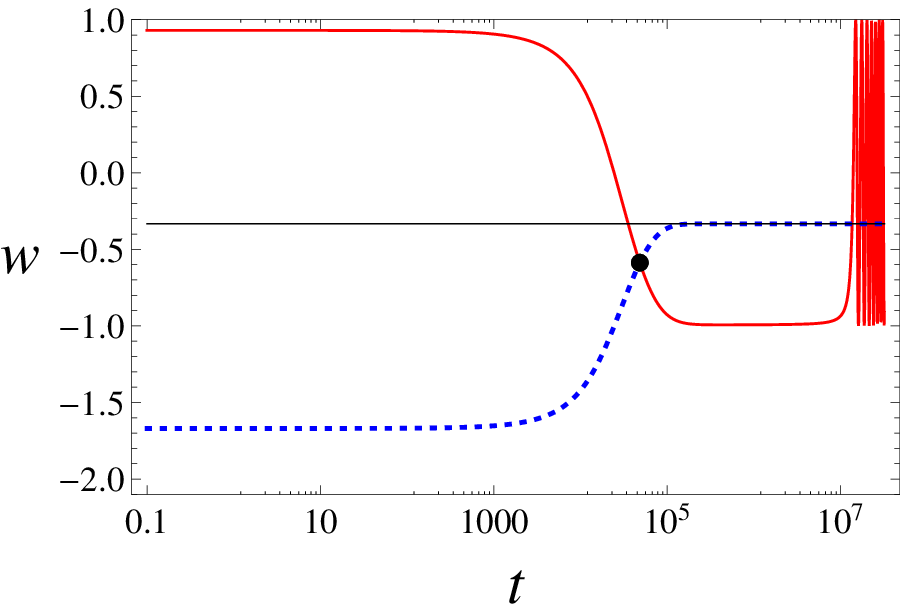}
\hskip0.5cm
\includegraphics[width=0.47\textwidth]{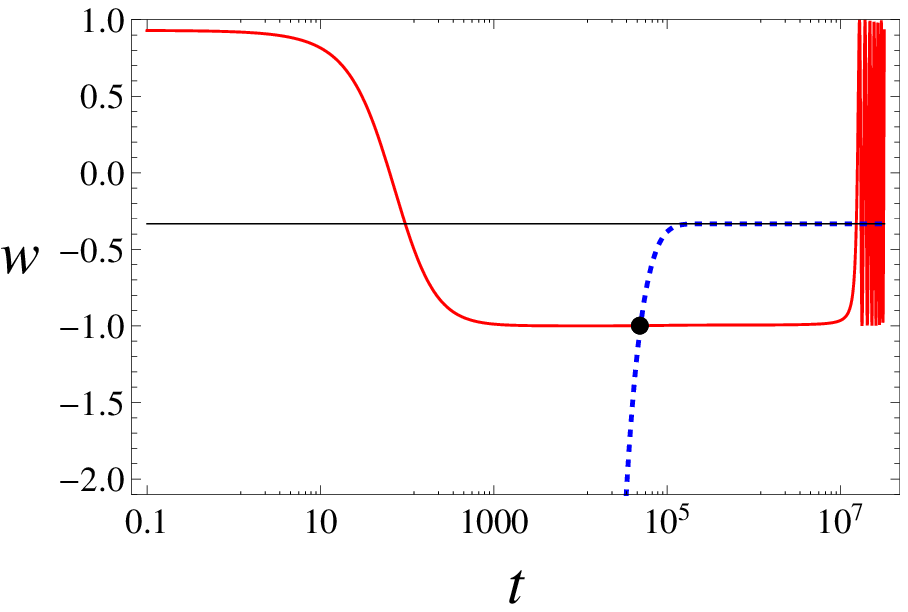}
\caption{In this figure the (red) solid curve shows the equation of state, the (blue) dashed curve denotes $-(1+\epsilon_{\rm J}^2)/3$, the thin black horizontal line shows the $w=-1/3$ line and the black dot marks the onset of accelerated expansion (inflation). The accelerated expansion starts when equation of state satisfies the condition given by \eref{inflcond}. The initial data for these plots are $\dot\phi(0) = -2\times10^{-5}\,\mpl^2$ and $\phi(0) = 3.14\, m_{\rm Pl}$, with $\epsilon_{\rm J}^2(0) = 4.01$ for the left, and $\epsilon_{\rm J}^2(0) = 1.17\times10^{5}$  for the right plot. The mass of the inflation is taken to be $m=1.21\times10^{-6}\,\mpl$.}
\label{inflcondfig}
\end{figure}

According to the inequality (\ref{inflcond}), whether or not a Bianchi-I universe undergoes accelerated expansion depends on the position of the field in the inflationary potential and also on the anisotropic shear in the universe.  This is in contrast to the isotropic model, where the
position of the inflaton in the inflationary potential is sufficient to determine whether the expansion of the universe accelerates or not. We demonstrate this in \fref{inflcondfig}, which shows two plots  depicting the validity of the inequality given in \eref{inflcond}.
In this figure, the initial values specified at $t=0$ are : $\dot\phi(0)=-2\times10^{-5}\,\mpl^2$ and
$\phi(0)=3.14\,\mpl$. (The qualitative behavior of the plot remains the same for other values
of initial $\dot\phi$ and $\phi$). We choose $\phi(0)=3.14\,\mpl$, just as a representative value to discuss
the behavior of inflationary trajectories, as this value is close to the value of the inflaton field near the
onset of slow-roll in the isotropic spacetime, to generate approximately $60$ e-foldings.
In the rest of the article, unless a different initial condition is specified, we will take $\phi(0)=3.14\,\mpl$, wherever required to
discuss qualitative behavior of amount of inflation and dynamical trajectories, shown in the plots in the following sections.
The left plot in \fref{inflcondfig} is for a small value of initial anisotropic shear and the right one corresponds to a
higher anisotropic shear. In these plots, the initial conditions are taken such that the kinetic
energy is greater than potential energy and the equation of state is initially\footnote{\hskip-0.1cm By starting with kinetic dominated initial conditions we make sure that we start with
conditions not favorable for inflation to occur immediately. The initial trajectory, although away
from the inflationary trajectory, joins the inflationary trajectory in the forward evolution (as discussed later in this article).} $w \approx 1$.
 As the further evolution takes place, the kinetic energy decays and  the equation of state
 decreases. After some time, the potential energy dominates over the kinetic energy and the
 equation of state becomes $w \approx -1$.
 In the right plot of \fref{inflcondfig}, we see that though the equation of state $w$ attains the value $-1$ quite
 early, around $t \approx1,000\,\tpl$, but, the accelerated expansion does not start until
 \eref{inflcond} is satisfied, which occurs around $t\approx 50,000\,\tpl$. In contrast, in the left
 plot since the initial anisotropic shear is weak, the accelerated expansion starts close to $w < -1/3$.


Using \eref{inflcond}, one may be tempted to conclude that a non-vanishing value of the parameter $\epsilon_{\rm J}$
is detrimental to the amount of inflation in comparison to the isotropic model, since more negative value of the equation of state is required for an accelerated expansion of the universe. However, as earlier pointed out in Ref. \cite{mss}, this expectation turns out to be incorrect. It turns out that the presence of anisotropy actually helps inflation by increasing the Hubble rate, which in turn enhances the friction term ($3H\dot\phi$) in the Klein-Gordon equation.  Due to this enhanced Hubble friction, the kinetic energy of the scalar field decays at a faster rate. This brings the condition for accelerated expansion to be achieved earlier, in presence of anisotropy, causing a longer phase of accelerated expansion.

\begin{figure}[tbh!]
\includegraphics[width=0.5\textwidth]{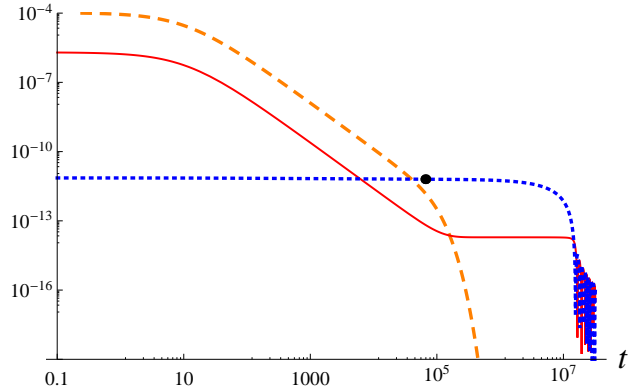}
\caption{This figure shows the evolution of kinetic $(\dot \phi^2/2)$, potential $(m^2 \phi^2/2)$ and anisotropic shear $(\sigma^2/16 \pi G)$ terms with time.   The (red) solid curve shows the kinetic energy contribution, the (blue) short dashed curve corresponds to potential energy contribution, and the (orange) dashed curve shows the shear term. The black dot marks the onset of inflation. Inflation takes place in the regime when potential energy dominates over the kinetic energy as well as the shear term.}
\label{kepe}
\end{figure}

The condition for accelerated expansion can also be written in terms of the contributions from kinetic and  potential energies, and the shear scalar, by using \eref{raychaudh}. Substituting the expressions for energy density and pressure in terms of the kinetic ($\dot\phi^2/2$) and potential energy $V(\phi)$, \eref{raychaudh} yields,
\be
 \f{\ddot a}{a} = -\f{4\pi G}{3} \left(2\dot\phi^2 - 2V(\phi)+\f{\sigma^2}{4\pi G}\right).
\ee
The condition for accelerated expansion can then be written as:
$$
V(\phi) > 2\left(\f{\dot\phi^2}{2} + \f{\sigma^2}{16 \pi G}\right) .
$$
%
This relation 
  tells us that for inflation (accelerated expansion) to take place, the potential energy must dominate both the kinetic energy and the anisotropic shear term in the above equation.\ In \fref{kepe}, we show the evolution of the kinetic and potential energy, and
$\sigma^2/16 \pi G$ with respect to time. 
It is evident that inflation does not start (denoted by a black dot in \fref{kepe}), until the sum of shear term $\sigma^2/16 \pi G$ and the kinetic energy term becomes less than half of the potential term. We also see that
following the onset of inflation, there is a short period where the shear term dominates over the kinetic energy, before it decays quickly during accelerated expansion.\
Thus, there may be a short duration of anisotropy during the classical inflationary phase of Bianchi-I spacetime. However, this is not a generic feature of evolution in Bianchi-I spacetime.\ If the initial conditions are such that the shear energy is less than the kinetic energy in the beginning, then there will not be any duration of such dominance of shear over kinetic energy.
\vskip0.5cm

 For the subsequent discussion of the numerical results, it is useful to define the slow-roll parameters. In the classical isotropic inflationary spacetime, the slow-roll parameters $\epsilon\,{\rm and}\, \eta$ are defined as
\be
\epsilon = 3\f{\dot\phi^2}{\dot\phi^2+2 V(\phi)}\,\quad {\rm and}\, \quad \eta = -\f{\ddot \phi}{H\dot\phi}~.
\label{slowroll}
\ee
The slow-roll inflation is defined as the phase of evolution where  $\epsilon, |\eta| \ll 1$. Smallness of these parameters implies that the Hubble rate ($H$) varies very slowly. In the discussion of the numerical results, we consider above definition of slow-roll parameters in the inflationary Bianchi-I spacetimes.

\bfig
\includegraphics[width=0.43\textwidth]{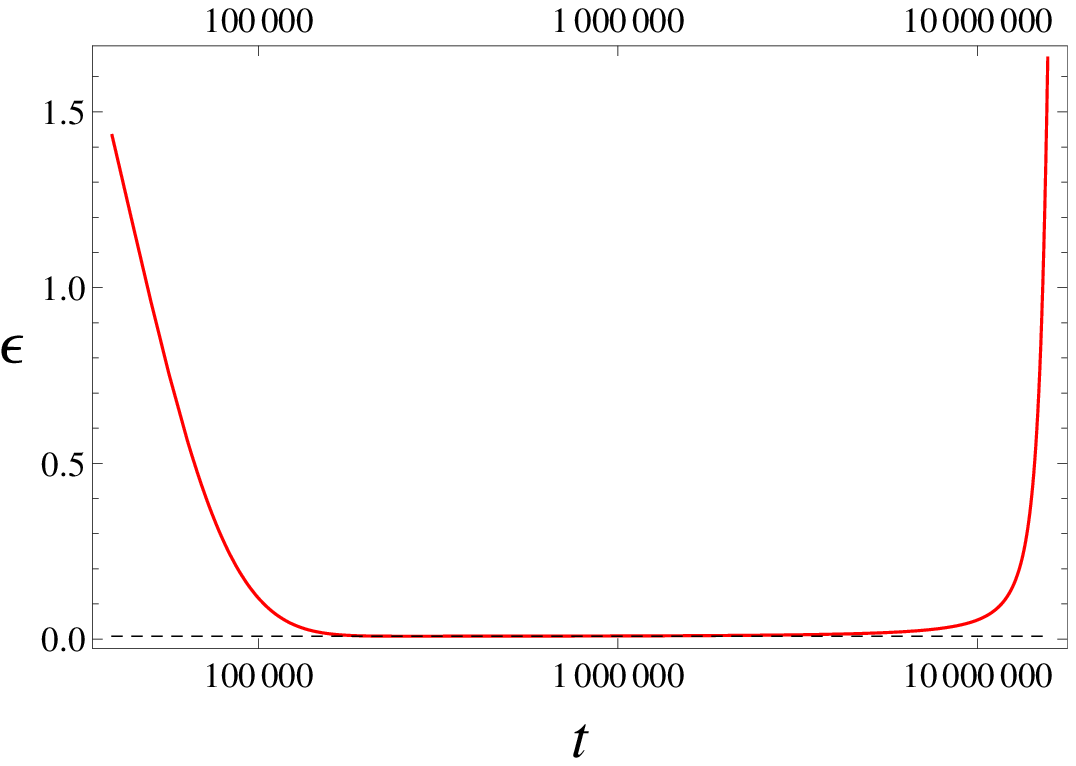}
\hskip0.5cm
\includegraphics[width=0.48\textwidth]{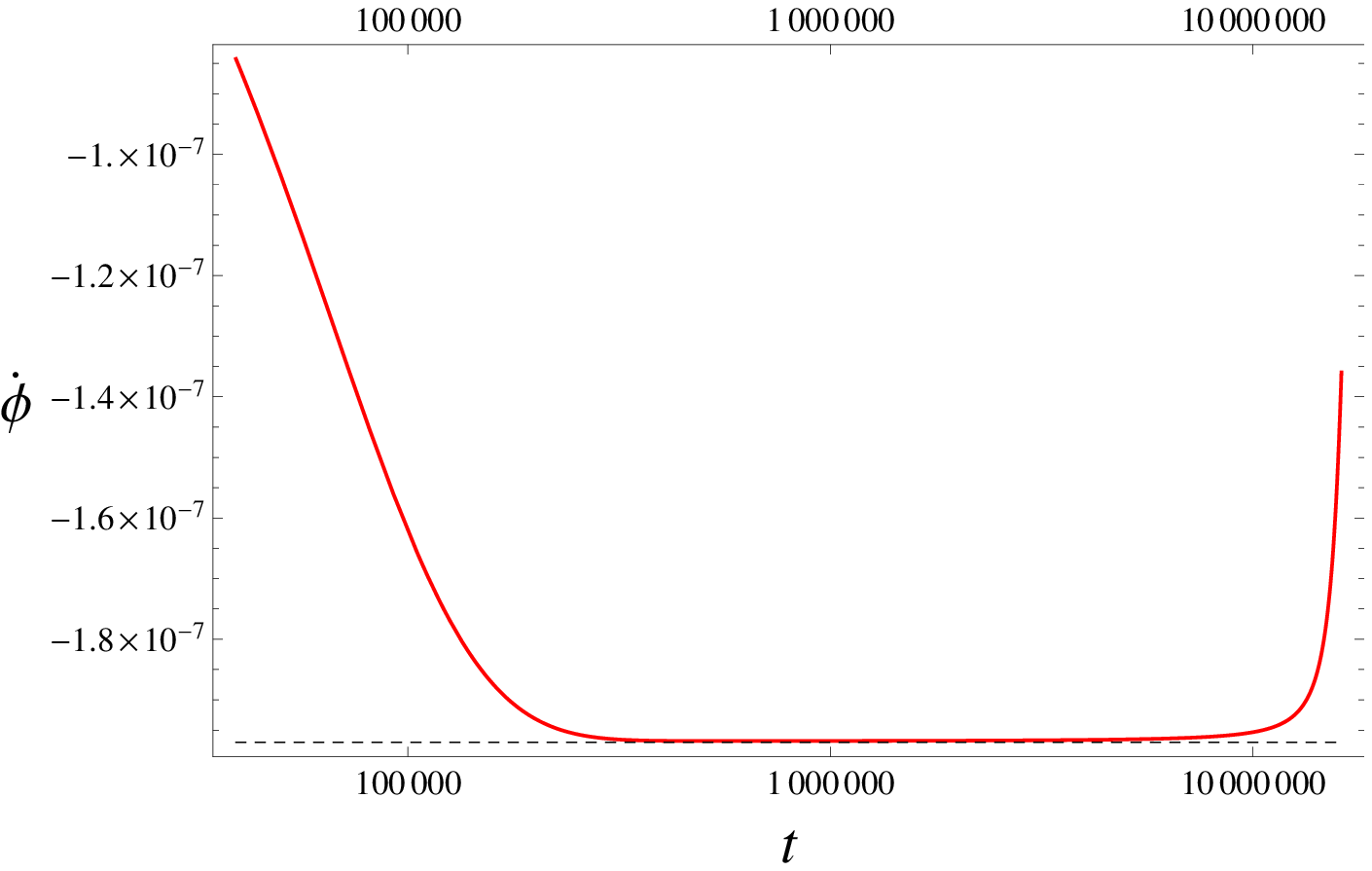}
\caption{This figure shows an example of the evolution of the slow-roll parameter $\epsilon$
and corresponding field velocity $\dot\phi$. The dashed black line in the left plot marks
$\epsilon \approx 0.008$ and in the right plot it shows the corresponding value of the field velocity
$\dot\phi\approx 1.97\times10^{-7}\mpl^2$. The initial conditions are taken as $\phi(0)=3.14\,\mpl,\, \dot\phi(0)=0.00002\,\mpl^2\, {\rm and}\, \epsilon_{\rm J}^2(0)=1.58\times10^{14}$.
It is clearly seen that even for initial anisotropy as high as $\epsilon_{\rm J}^2(0)=1.58\times10^{14}$, the
slow-roll conditions are met in the future evolution.}
\label{slowrollfig}
\efig

It is evident from the \eref{slowroll} that for small slow-roll parameters, $\epsilon\,,|\eta|\ll1$, the value of both $\dot\phi$ and $\ddot\phi$ will be small. This implies that during the slow-roll phase, the inflaton rolls down the potential with a small and almost constant velocity. An example of the numerical evolution of $\epsilon$ and $\dot\phi$, in Bianchi-I spacetime, starting from $\phi(0)=3.14\,\mpl,\, \dot\phi(0)=0.00002\,\mpl^2\,{\rm and}\, \epsilon_{\rm J}^2(0)=1.58\times10^{14}$, is shown in \fref{slowrollfig}. We see that, once $\epsilon \ll 1$, the slow-roll inflation sets in, and the field velocity varies very slowly. For these initial conditions, during this phase, the value of velocity of the inflaton is $\dot\phi\approx1.97\times10^{-7}\mpl^2$ and the slow-roll
parameter is $\epsilon\approx0.008$.

\subsection{LQC Effective equations }

Canonical quantization of cosmological models in LQC is based on techniques in LQG. The classical gravitational Hamiltonian constraint of a given model is first expressed in terms of the holonomies of connections $A_i^a$ and fluxes of the triads $E_a^i$, which are then expressed as appropriate quantum operators. The quantum Hamiltonian constraint hence obtained gives rise to a non-singular quantum difference equation \cite{as1}.
Based on the geometric formulation of quantum mechanics, an approximate continuous description of the underlying quantum geometry, known as the effective description of LQC, can be obtained through a faithful embedding of the classical phase space into the quantum phase space of LQC \cite{schilling}.
Through an appropriate choice of  semiclassical states, those which correspond to a macroscopic universe at late times, an effective Hamiltonian constraint can be derived by calculating the leading order corrections introduced by the underlying discrete geometry of the spacetime \cite{jw,vt}.
The effective Hamiltonian constraint hence obtained gives rise to the modified equations of motion, which introduce corrections to the classical Friedmann and Raychaudhuri equations.
Extensive numerical simulations for
various matter models, including the Bianchi-I anisotropic spacetimes, by considering the LQC evolution of a
semi-classical states show that the effective theory agrees very well
with the full quantum dynamics at almost all scales, including the deep
Planck regime and the classical low curvature regime \cite{aps2,aps3,apsv,acs,kv,bp,ap,b1madrid2,szulc2,ps12,khanna}.

The effective Hamiltonian constraint for the Bianchi-I spacetime, with lapse $N=1$, is given via \cite{awe2,cv,csv,corichi_aniso} \footnote{The effective Hamiltonian considered here does not incorporate ``inverse triad corrections'' which are only meaningful in the models with a compact topology. As mentioned earlier, the topology for the Bianchi-I spacetime, assumed to be $\Sigma\times\mathcal{R}$, is non-compact in the present context. Even if one considers a compact topology, since the physical volume of the fiducial cells in the numerical simulations remain large compared to the Planck volume, the corrections due to the inverse triad terms would be negligible.}
\be
\label{heff}{\mathcal H}_{\rm eff} = -\f{1}{8\pi\gamma^2V}\left(\f{\sin(\mua c_1)}{\mua}\f{\sin(\mub c_2)}{\mub}p_1p_2 + {\rm cyclic\,terms}\right) + \Hmatt,
\ee
where $V=\sqrt{p_1p_2p_3}$ is the physical volume, $\Hmatt$ is the matter Hamiltonian taken to be same as given in \eref{hmatt} and $\bar\mu_i$ are lengths of the plaquette around which the holonomies are calculated \cite{awe2}:\footnote{Sometimes in literature, one finds a different functional dependence of $\bar \mu_i$ on triads. It turns out that the form $\mu_i(p_1,p_2,p_3)$ given by the equation above is the only known choice where resulting physics is independent of the rescaling freedoms of the fiducial cell when the manifold is non-compact (see ref. \cite{cs09} for details).}
\be
\mua = \lambda \sqrt{\f{p_2p_3}{p_1}}\,, \quad \, \mub = \lambda \sqrt{\f{p_1p_3}{p_2}}\, \quad \,{\rm and} \quad \muc = \lambda \sqrt{\f{p_2p_1}{p_3}} ~.
\ee
Here $\lambda^2 = 4\sqrt{3}\pi\gamma\lp^2$ denotes the minimum area gap which results from the quantum discreteness of spatial geometry in LQG.\

The effective Hamiltonian constraint given via \eref{heff} generates the dynamics of the spacetime for a given matter part of the Hamiltonian constraint, $\Hmatt$. The equations of motion of the triads and the connections are given via the Hamilton's equation of motion for $\Heff$ (see for eg. \cite{ps11}):
\be
\label{p1dot} {\dot p}_1 = \f{p_1}{\gamma\lambda}\left(\sin(\muc c_3)+\sin(\mub c_2)\right)\cos(\mua c_1)
\ee
and
\ba
{\dot c}_1 &=& \f{1}{\gamma\lambda p_1}\big[\cos(\mub c_2)\left(\sin(\mua c_1)+\sin(\muc c_3)\right)c_2p_2+\cos(\muc c_3)\left(\sin(\mua c_1)+\sin(\mub c_2)\right)c_3p_3 \nonumber \\ && -\cos(\mua c_1)\left(\sin(\mub c_2)+\sin(\muc c_3)\right)c_1p_1 \nonumber - \f{2\mua p_2p_3}{\lambda}\\ &&\left(\sin(\mub c_2)\sin(\muc c_3) +\,{\rm cyclic\, terms}\right)\big] + 8\pi G\gamma \f{\partial \Hmatt}{\partial p_1} ~.
\ea
The equations of motion of other directional triads and connections can also be computed in a similar fashion. The equations of motion hence obtained, along with the initial data which satisfies the effective Hamiltonian constraint given by \eref{heff}, lead to a well posed initial value problem.\ 
The dynamical equations obtained from the effective Hamiltonian constraint modify the classical Einstein's field equations in a
significant way. 
It is evident from \eref{p1dot} that $\dot p_i/p_i$ never diverges during the evolution as the value of $\sin(\bar\mu_i c_i)\, {\rm and } \cos(\bar\mu_i c_i)$ are bounded above by unity.
This further leads to non-diverging curvature scalars of Bianchi-I spacetimes such as energy density $(\rho)$, shear $(\sigma^2)$ and expansion $(\theta)$ scalars for the comoving observers which are bounded above by their respective maxima \cite{gs1}. These values are given by:
$$ \rho_{\rm{max}} \approx 0.41 \rho_{\rm{Pl}}, ~~~ \theta_{\rm{\max}} \approx \f{2.78}{l_{\rm{pl}}}, ~~ \mathrm{and} ~~ \sigma^2_{\rm{max}} \approx \f{11.57}{l^2_{\rm{Pl}}} ~.$$
Moreover, it has been shown that all the curvature invariants of Bianchi-I spacetime are
bounded and all the strong singularities of Bianchi-I spacetime, for the matter satisfying null
energy condition, are resolved \cite{ps11}. The resolution of singularity results in a
non-singular bounce in the evolution of Bianchi-I spacetime in the effective description
of LQC. Far way from Planckian regime, where the spacetime curvature is small, the
effective dynamical trajectory agrees well with the classical theory. In the backward evolution
as the spacetime curvature grows, the effective trajectory differs from the classical
trajectory. During the subsequent backward evolution the classical trajectory goes into
singularity, whereas the effective dynamical trajectory undergoes a non-singular bounce.
In the effective dynamical evolution, the curvature scalars are always finite and bounded  by their respective maxima given in the above equation.

The finite and non-diverging values of the curvature scalars introduce interesting modifications in terms of the choice of initial conditions and the dynamics of the evolution of universe in the pre-inflationary regime in the Bianchi-I spacetime in LQC. There are two important modifications relevant to our discussion.
\vskip0.5cm
\begin{itemize}

\item Unlike in the classical theory, it turns out that in LQC, the mean Hubble rate does not monotonically vary with the shear scalar. This behavior has been noted for the first time in the numerical simulations discussed in this paper\footnote{Due to unavailability of the modified generalized Friedmann equation for the Bianchi-I spacetime in LQC, it is difficult to obtain this result analytically. However, in the regime where anisotropic shear is weak, the modified Friedmann equation for Bianchi-I spacetime in LQC has been obtained and it contains $\sigma^4$ corrections \cite{cv}. These corrections can cause the Hubble rate to decrease if the anisotropic shear increases beyond a certain value.}.
Due to such a non-monotonic behavior of the Hubble rate as a function of the shear scalar, the Hubble friction in the Klein-Gordon equation is affected in a similar fashion.
In other words, unlike in the classical theory, {\it Hubble friction is not always enhanced by the anisotropic shear in LQC}.

\item The shear scalar in LQC during the pre-inflationary regime behaves very differently than that in the classical theory. Unlike in the classical theory, where $\sigma^2 \propto a^{-6}$, shear scalar does not have a monotonic dependence on the scale factor, near  the bounce.
 There may be instances when, depending on the conditions at the bounce, the shear scalar starts out with a very small value, and then increases to reach a local maxima. On the subsequent evolution, the shear scalar would decrease and approach the classical value in the regime where quantum gravity corrections are weak.
As we will see in next section, this has important implications in regards to the isotropic limit and the attractor behavior of Bianchi-I spacetime in the vicinity of the bounce.


\end{itemize}

\section{Classical and Effective Dynamics: Numerical Results}

In this section we present numerical studies of the Bianchi-I spacetime with an inflaton as a matter field with a self-interacting potential $V(\phi)=m^2\phi^2/2$. In the first subsection, we discuss the results of the classical theory and the following subsection is devoted to the numerical simulations of the effective spacetime description in LQC.
We discuss important features of both the classical and the LQC descriptions of the Bianchi-I inflationary spacetime, such as the effect of anisotropic shear on the amount of inflation, the process of isotropization, the isotropic limit of Bianchi-I spacetime, and the attractor behavior.
 Over a hundred simulations were performed for both the classical theory and LQC by varying different parameters including the initial shear scalar, the initial value of the inflaton and the initial velocity of the inflaton. In all of the simulations, the mass of the inflaton is taken as  $m = 1.21 \times 10^{-6}\,\mpl$ and the numerical accuracy of the vanishing of the Hamiltonian constraint is of the order of $10^{-10}$. Note that there is a symmetry associated to the signs of the value of inflaton
field and its time derivative. That is, the results remain invariant if the
sign of both $\phi$ and $\dot\phi$ are changed simultaneously. In the numerical
simulations performed in this article, we take the initial value of the
inflaton field to be positive. For an inflaton which is initially
rolling down, we take $\dot\phi(0)<0$ and for an inflaton which is initially rolling up, we
take $\dot\phi(0)>0$.

\subsection{Classical theory}

As discussed in the previous section, the dynamics of the classical Bianchi-I spacetime is governed by the generalized Friedmann and Raychaudhuri equations (\eref{fried} and \eref{raychaudh}) from where it follows that specifying the energy density, pressure and the shear scalar at a given time provides a complete initial data.
 For a scalar field the pressure and energy density can be expressed in terms of the value of the scalar field $\phi$ and its time derivative $\dot\phi$.
 Thus, giving the initial value of the scalar field, its velocity and the shear scalar forms a complete set of information required to compute the state of the universe at any later time.
 We evolve the initial data provided in a pre-inflationary phase and study various properties of the dynamical trajectories including isotropization, amount of inflation and the phase-space portraits.\
As discussed in the previous section, inflation takes place when the potential energy dominates over the sum of kinetic energy term $(\dot \phi^2/2)$ and the shear contribution $(\sigma^2/16 \pi G)$. 
 In the numerical simulations presented in this section we will, however, start with a kinetic and shear dominating initial conditions. This gives insights on the way conditions in favor of inflation are achieved, starting from  comparatively unfavorable initial conditions.
 Moreover, it brings out important features about the behavior of the pre-inflationary dynamics and its dependence on various parameters.

\subsubsection{Isotropization: behavior of the directional scale factors}
We first discuss the process of isotropization of Bianchi-I spacetime, that is the way a Bianchi-I universe starting with highly anisotropic initial conditions at the bounce with at least  one scale factor contracting turns into one with  all-expanding-directions. A Bianchi-I universe can evolve from highly anisotropic initial conditions which, depending on the strength of anisotropy, entail to various possible geometrical structures of the spatial geometry such as a cigar, pancake etc, close to the classical singularity. The key question is the way an inflationary spacetime with all expanding directions emerges from such a state.

\begin{figure}[tbh!]
\includegraphics[width=0.47\textwidth]{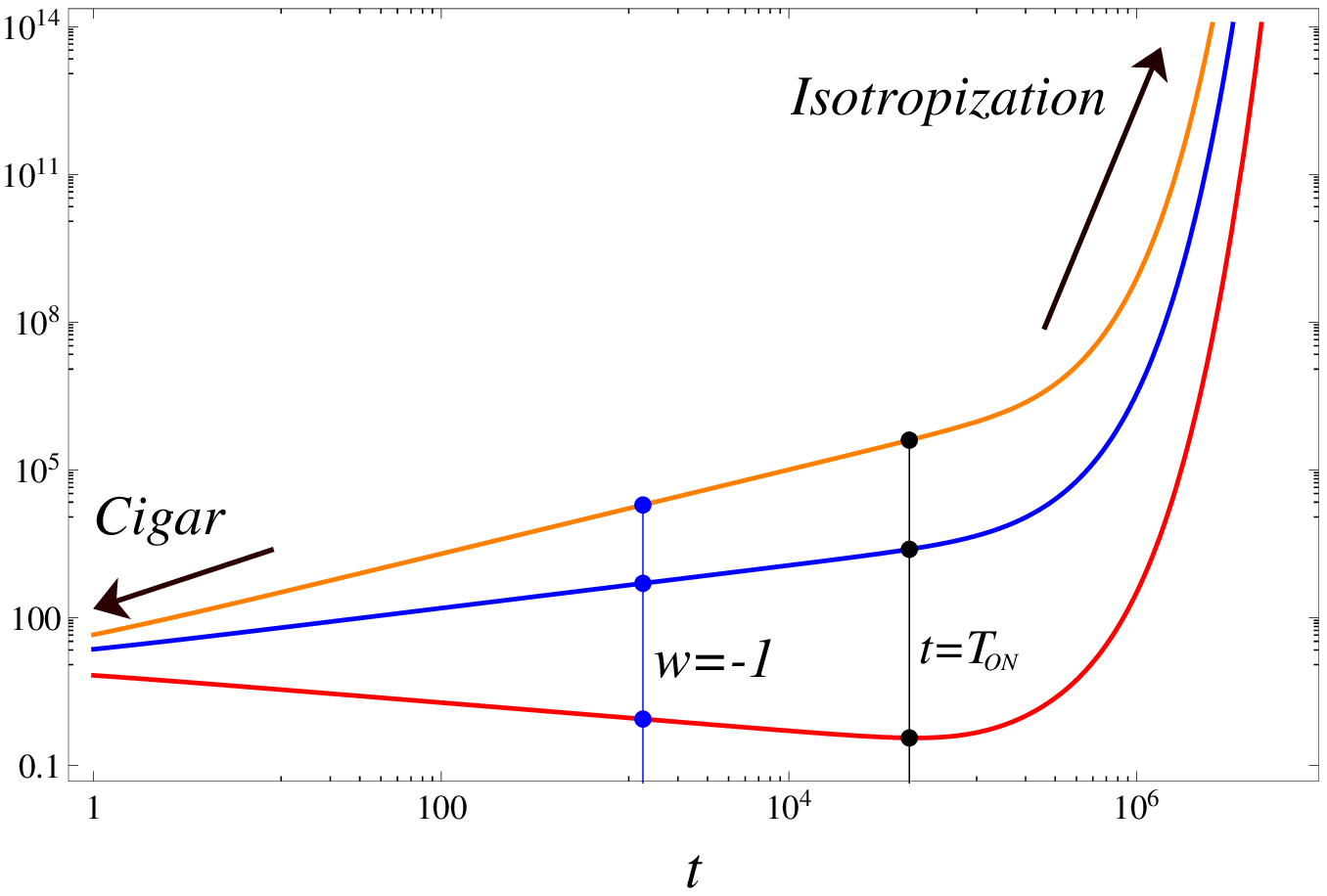}
\hskip0.5cm
\includegraphics[width=0.47\textwidth]{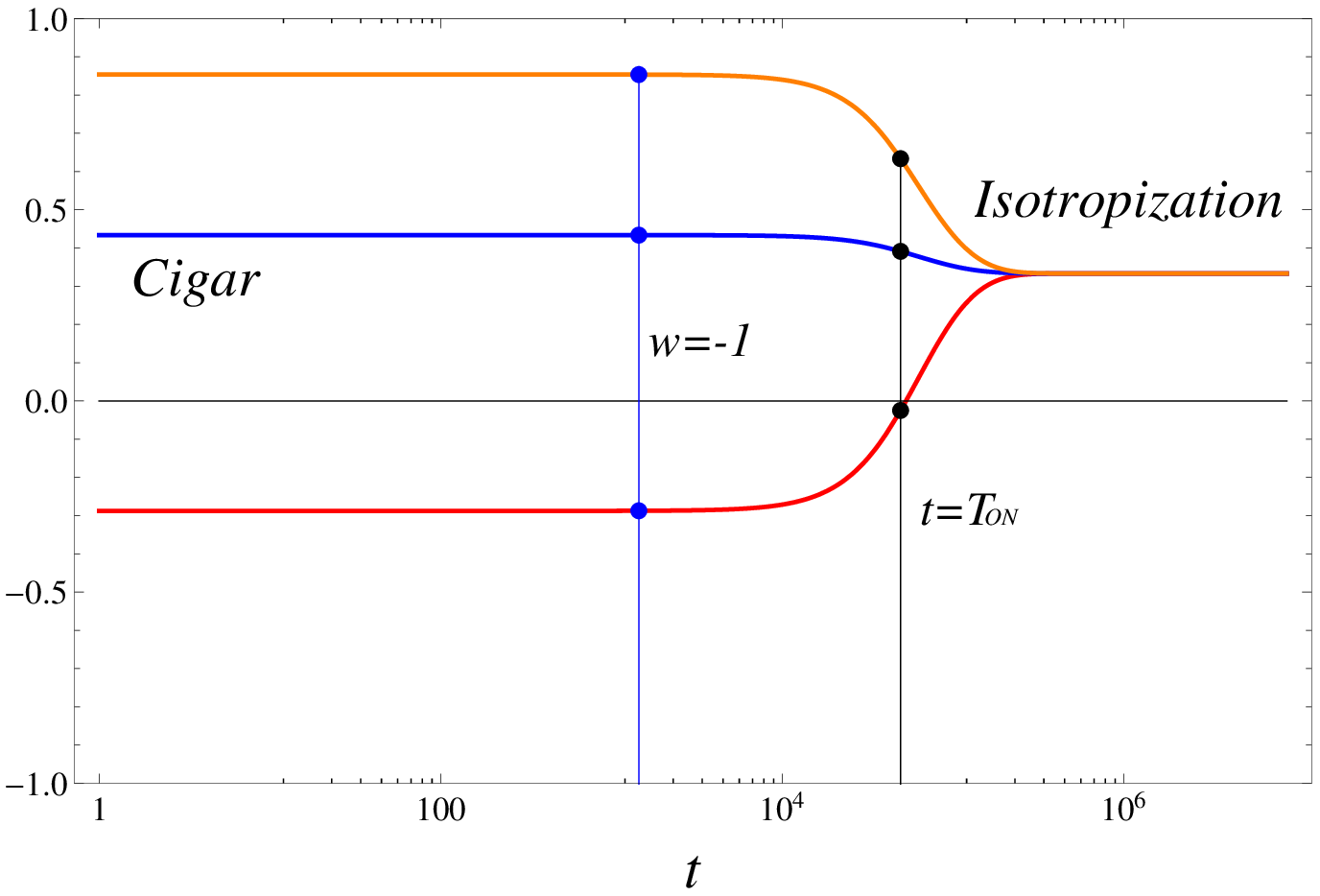}
\caption{The left plot shows the evolution of directional scale factors in classical Bianchi-I spacetime in the inflationary scenario. The initial data is such that backward evolution gives a cigar like singularity. It is evident that, in the forward evolution, the contracting direction undergoes a turn around near the onset of inflation and during most of the inflation all the scale factors expand. The adjoining plot shows the evolution of the corresponding Kasner exponents. Note that although $w=-1$ is attained earlier, inflation does not begin until $t=T_{\tiny \rm ON}$ due to the presence of anisotropy. ($T_{\tiny \rm ON}$ corresponds to the time at the onset of inflation).}
\label{directional}
\end{figure}

Recall that the structure of the spatial geometry of the spacetime can be determined via the value of the Jacobs' parameter $\epsilon_{\rm J}$. For example, $\epsilon_{\rm J}<2/\sqrt{3}$ is a sufficient condition for all the directions to expand,
whereas, for $\epsilon_{\rm J} \ge 2/\sqrt{3}$, not all of the directions will be in expanding state (as shown in the left plot of \fref{directional} where one of the directions is contracting while the other two expand in the forward evolution).
Thus, at the onset of inflation, depending on the value of $\epsilon_{\rm J}$, all the directions may either be expanding or not. From the equations of motion of Bianchi-I spacetime, it is straightforward to see that the anisotropic shear ($\sigma^2$) in an expanding classical Bianchi-I spacetime always decreases, as $\sigma^2\propto a^{-6}$.  During inflation, the mean scale factor grows exponentially which leads to a rapid decrease in $\sigma^2$, and hence in the value of $\epsilon_{\rm J}$. Due to this reason, $\epsilon_{\rm J}$ quickly becomes less than $2/\sqrt{3}$ even if it starts from a larger value.
Thus, even if one starts with a large $\epsilon_{\rm J}$ in the Bianchi-I spacetime, $\epsilon_{\rm J}$ becomes less than $2/\sqrt{3}$  either before, or shortly after, the start of inflation, and all the directional scale factors turn out to be in the expanding state.

\fref{directional} shows the evolution of the three directional scale factors starting from a cigar like structure, such that two of the scale factors are increasing while the remaining one decreases, as the universe expands. It is evident from the left plot of \fref{directional} that in the forward evolution, the decreasing scale factor undergoes a turn around shortly after the time of onset of inflation, $t=T_{\tiny \rm ON}$ (defined as the time where $\ddot a/a$ becomes positive).
After the turn around, all the scale factors undergo expansion. The adjoining right plot shows the evolution of individual Kasner exponents. In the beginning, one of the Kasner exponents, corresponding to the contracting directional scale factor, is negative which becomes positive soon after the onset of inflation.

In a similar way, one can study the isotropization of the Bianchi-I spacetime from the barrel and pancake like structures. In these cases too, irrespective of the their initial behavior, all the directional scale factors enter expansion phase either before or soon after the onset of inflation. In this way,  an inflationary Bianchi-I spacetime isotropizes, no matter what the pre-inflationary conditions are.
 Such a turn around of the scale factor is a feature of a generic initial data, except for those with point like structures for which all the three directions are already expanding. Similar turn around of directional scale factors and the evolution of Kasner exponents was also noted in Ref. \cite{pitrou}.

\subsubsection{Number of e-foldings and initial conditions}
\begin{figure}[tbh!]
\includegraphics[width=0.47\textwidth]{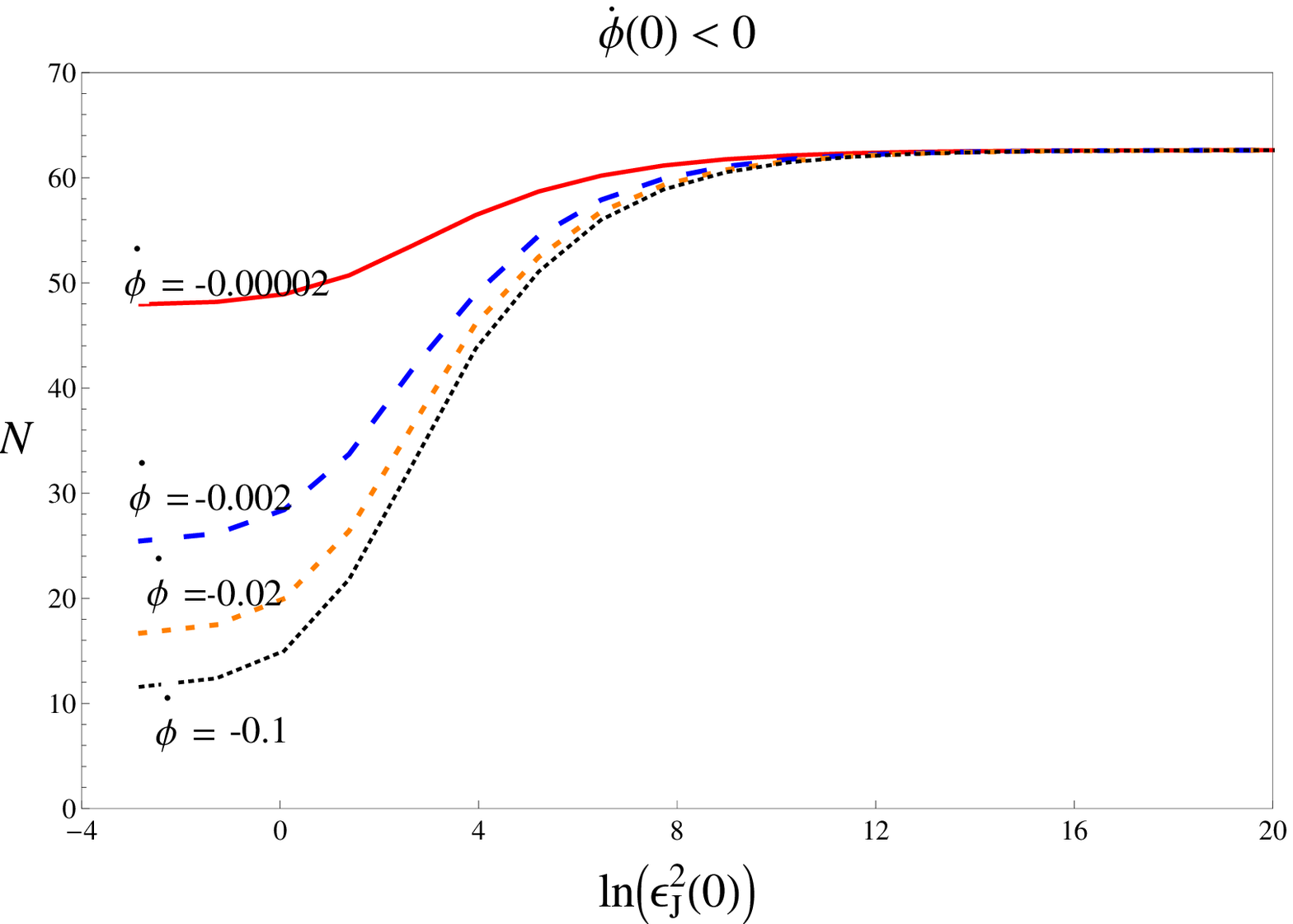}
\hskip0.5cm
\includegraphics[width=0.47\textwidth]{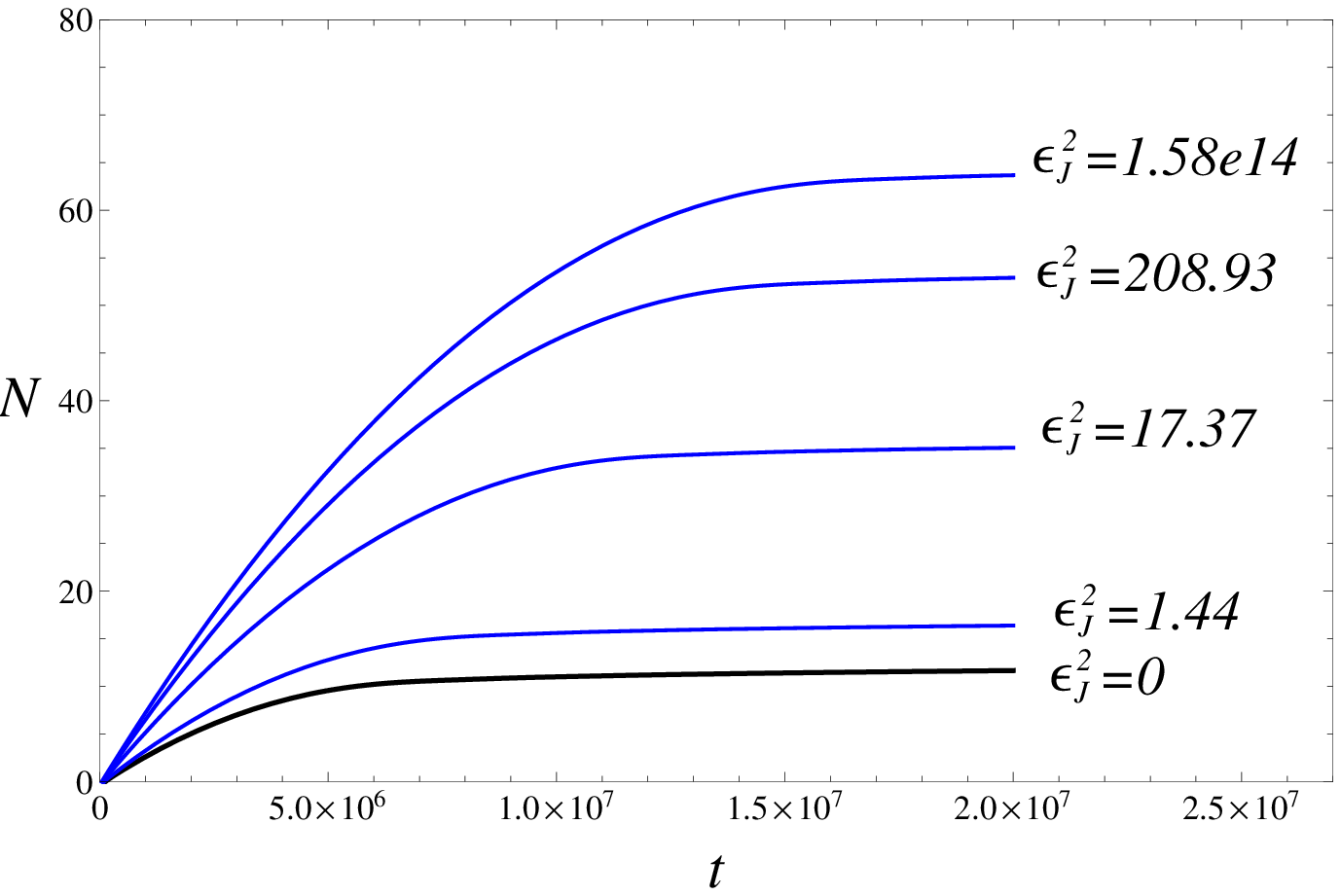}
\caption{ The left plot in this figure shows the number of e-foldings, $N$ plotted against the initial value of $\ln (\epsilon_{\rm J}^2)$ for various values of the initial $\dot\phi$. For each initial value of $\dot \phi$, the energy density is kept fixed while the anisotropic shear is varied. From top to bottom, curves correspond to $\dot\phi(0)=(-2\times10^{-5},\,-2\times10^{-3},\,-2\times10^{-2},-0.1)\,\mpl^2$. For large values of initial $\epsilon_{\rm J}$, all the trajectories tend to produce the same number of e-foldings. The right plot shows the variation of the number of e-foldings with time for various initial anisotropy parameters $\epsilon_{\rm J}$, but starting with the same initial value of $\dot \phi$: $\dot\phi(0)=-0.1\,\mpl$. In  both the plots, the initial value of the inflaton field is taken as $\phi(0)=3.14\,m_{\rm Pl}$.}
\label{efoldvaryphidot}
\end{figure}

The number of e-foldings measure the amount of inflation during the inflationary era of the universe.
This number is defined as $N = \ln(a_f/a_i)$, where $a_f$ corresponds to the value of the
mean scale factor where accelerated expansion ends, and $a_i$ denotes the value of the mean scale factor where the accelerated expansion begins.
We performed many simulations to compute the number of e-foldings for various initial values
of the anisotropic shear.  \fref{efoldvaryphidot} shows the variation of the number of e-foldings
as a function of the initial anisotropy (measured by $\epsilon_{\rm J}$), for a scalar field
having a negative initial velocity ($\dot\phi(0) < 0$, corresponding to a rolling down inflaton).
To discuss the qualitative behavior of the amount of inflation, in the plots shown in this
article, we choose the initial value of the inflaton to be $\phi(0)=3.14\,\mpl$. As mentioned earlier, this is just a
choice representing inflationary trajectories and the way anisotropic shear affects the
sufficient number of e-foldings, as this value is close to the value of inflaton at the onset of
isotropic slow-roll to generate approximately $60$ e-foldings in the isotropic
spacetime\footnote{\label{60efoldnote} In order to obtain approximately 60 e-foldings in the
classical isotropic spacetime, the slow-roll has to start at $\phi\approx3.14\,\mpl$ with the
required value of the velocity of the inflaton being
$\dot\phi\approx1.97\times10^{-7}\,\mpl^2$}.
The qualitative behavior discussed in this section also holds true for other values of
$\phi(0)$ for which inflation takes place. We will consider other values of $\phi(0)$ while discussing the quantitative details of the amount of e-foldings.
 In confirmation with the known results in classical theory in this scenario, we find that if the inflaton is initially rolling down then the number of e-foldings  increases with an increase in anisotropy. Further, different initial $\dot\phi$'s produce the similar number of e-foldings for large initial anisotropy.
That is, the spacetime tends to achieve the same number of e-foldings for large value of initial anisotropic shear irrespective of the initial kinetic energy of the scalar field. Note that for the higher values of
initial anisotropy, an inflaton which rolls down the potential without initially satisfying slow roll conditions, quickly enters into slow roll. The universe starts inflating almost exponentially, which shortens the stage of anisotropic evolution. Numerical simulations show that for high values of initial anisotropic shear, the phase of anisotropic inflation is so short that any change in anisotropy produces little effect on the number of e-foldings.

The right plot in \fref{efoldvaryphidot} shows the evolution of the number of e-foldings during the forward evolution of inflationary Bianchi-I spacetime starting with the same $\dot\phi(0)=-0.1\,\mpl^2$ but for the various values of the initial anisotropy parameters $\epsilon_{\rm J}$. For these initial conditions, the number of e-foldings in the isotropic case (the case $\epsilon_{\rm J} = 0$) is less than 10. Whereas, in the Bianchi-I spacetime, the same initial conditions on the scalar field produce significantly more number of e-foldings for higher $\epsilon_{\rm J}$.
Thus for $\dot\phi(0)<0$, anisotropic shear not only increases the number of e-foldings, but a large anisotropy also makes slow-roll conditions be achieved irrespective of the initial velocity of the inflaton due to
an increase in the Hubble friction in the Klein-Gordon equation.\footnote{In GR, irrespective of the initial inflaton velocity, one can always find a suitable value of initial anisotropic shear such that slow-roll is obtained.}

\begin{figure}[tbh!]
\includegraphics[width=0.47\textwidth]{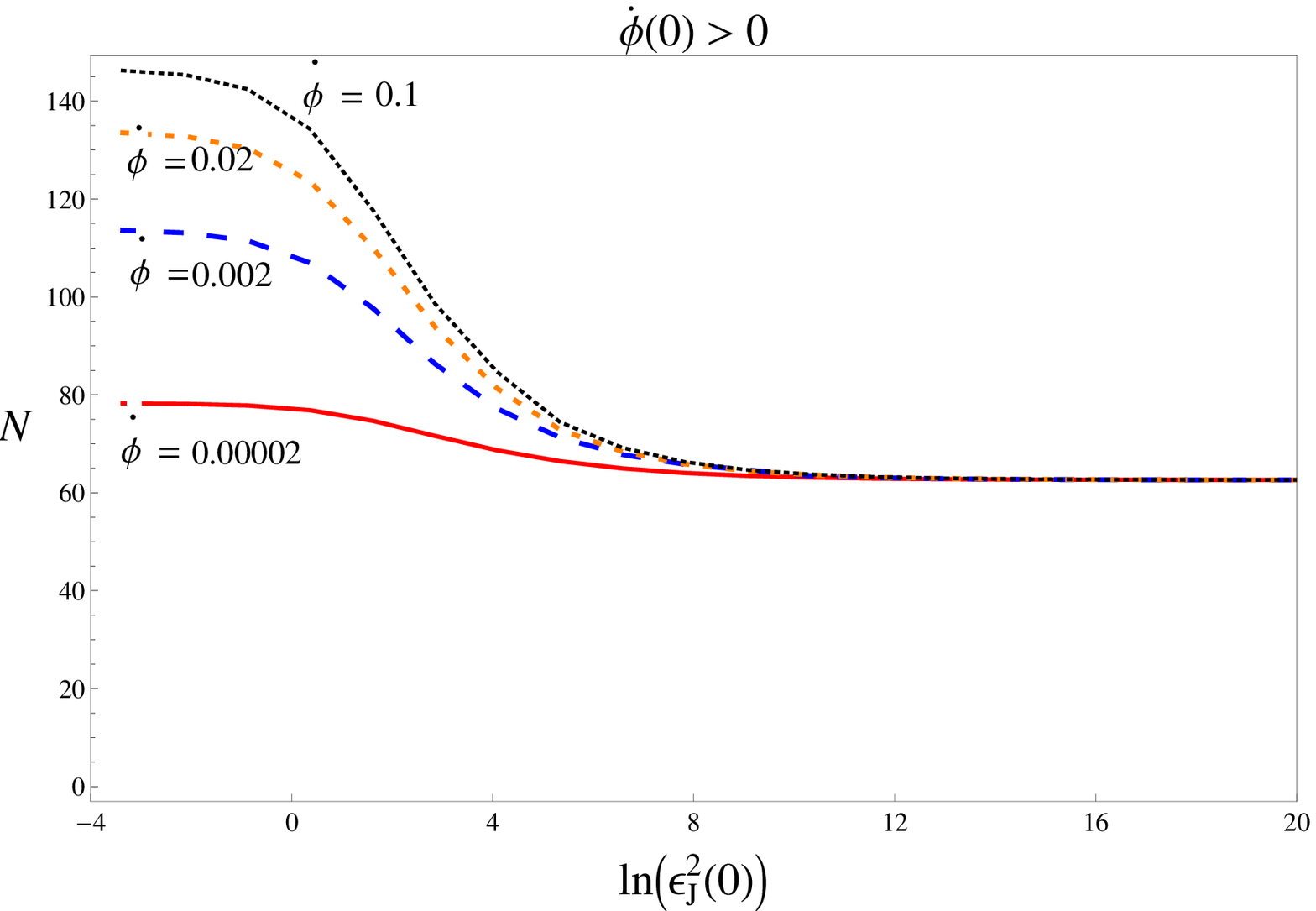}
\hskip0.5cm
\includegraphics[width=0.47\textwidth]{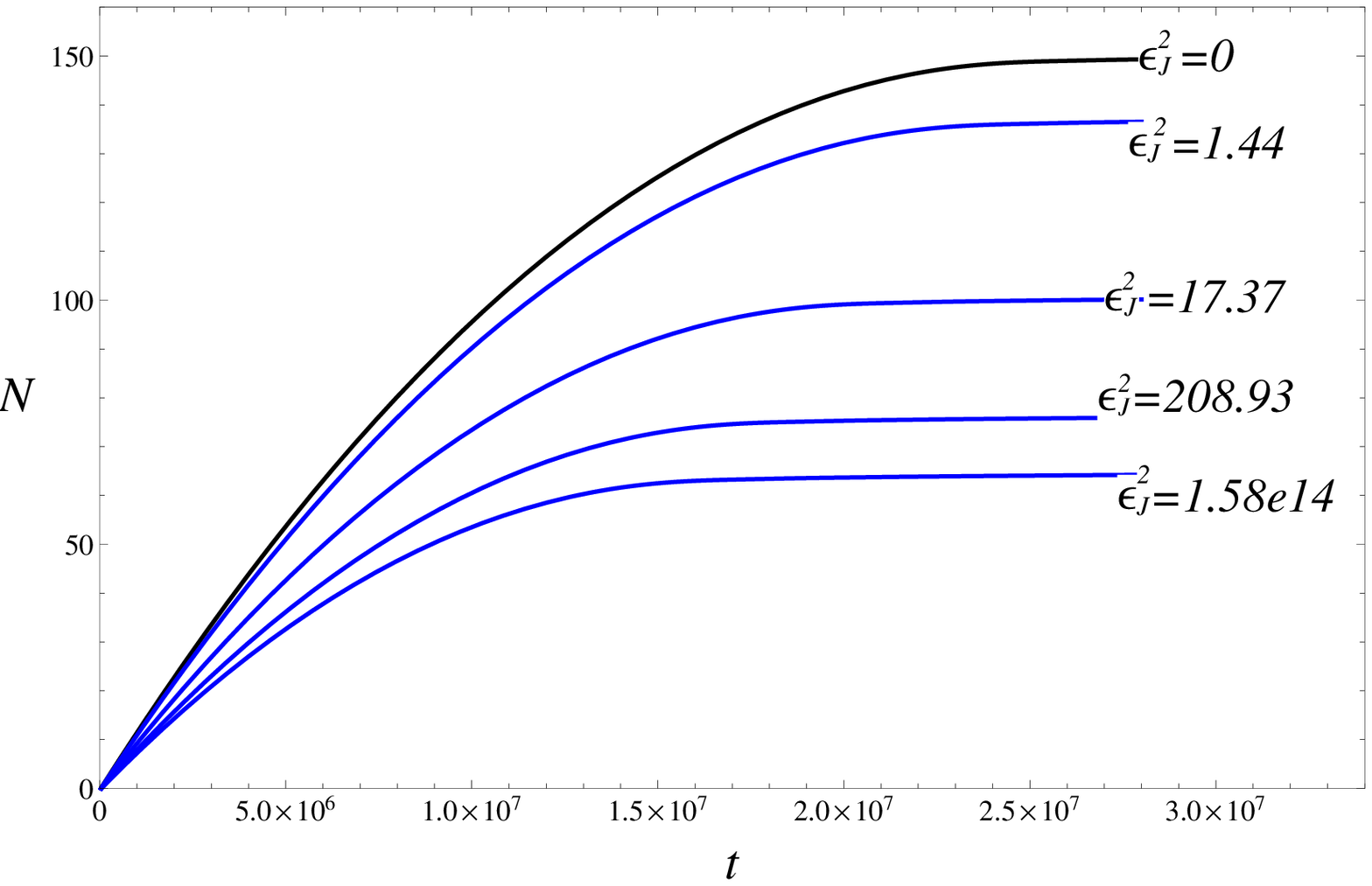}
\caption{ This figure shows the number of e-foldings, $N$ plotted against $\ln (\epsilon_{\rm J}^2)$ for various values of the initial $\dot\phi$. In the left plot, from bottom to top, the curves correspond to $\dot\phi(0)=(2\times10^{-5},\,2\times10^{-3},\,2\times10^{-2},0.1)\, \mpl^2$ and initial value of field is taken $\phi(0)=3.14\, m_{\rm Pl}$ for all the simulations. The right plot shows the time evolution of the number of efoldings for various initial anisotropy starting with the same initial value of the inflaton velocity $\dot\phi(0)=0.1\,m_{\rm Pl}$.}
\label{efoldphidnegpos}
\end{figure}

We now consider the case of inflaton initially rolling up the potential, i.e. $\dot \phi(0) > 0$. To our knowledge, this case has not been discussed elsewhere in the literature.
The left plot in the \fref{efoldphidnegpos} shows the variation of number of e-foldings with the increasing anisotropy in the initial data for various values of $\dot\phi(0) = (2\times10^{-5},\,2\times10^{-3},\,2\times10^{-2},0.1)\,\mpl^2$.
In all these simulations, the value of the inflaton field in the initial data has been taken as $\phi(0)=3.14\,m_{\rm Pl}$, on the same footing as in \fref{efoldvaryphidot}.
In contrast to the case of inflaton rolling down the potential, it turns out that, in this case, the number of e-foldings decrease as the anisotropy in the initial data increases.  For large initial anisotropic shear, all the trajectories attain the same number of e-foldings irrespective of the initial $\dot\phi$.\

To understand the decrease in number of e-foldings with an increase in anisotropy, let us for a moment consider the isotropic model. In the isotropic case, an initial $\dot\phi(0)>0$ causes the field to roll up the potential before the onset of inflation. For example, if the initial $\dot\phi$ is say $\dot \phi(0) = 0.1\,\mpl^2$ (with initial $\phi$ as $3.14 m_{\rm{Pl}}$), then the inflaton will roll up the potential and inflation starts to take place at $\phi \approx 4.8\, m_{\rm Pl}$. As a result, the amount of inflation is greater as compared to the case of initial $\dot\phi<0$.
On the other hand, in Bianchi-I spacetime due the increase in Hubble friction, there is a faster decay of the kinetic energy of the field, and the field stops its upward journey on the potential before reaching $\phi \approx 4.8\, m_{\rm Pl}$. Thus, inflation starts taking place at a comparatively smaller value of the field.
Hence, for an inflaton which is initially rolling up the potential, an increase in anisotropy reduces the
the number of e-foldings in comparison to the corresponding isotropic evolution.
At very large anisotropy, the Hubble friction is so strong that the kinetic energy of the field decays quickly, and inflation onsets very close to the initial value of the inflaton.

 It is important to note that for a large initial anisotropy, both the positive and negative initial velocity of the field give the same amount of inflation. This is evident from the comparison of
 \fref{efoldvaryphidot} with \fref{efoldphidnegpos}.
It is also clear from these plots, that although large anisotropic shear may diminish the number of e-foldings, yet, it never forbids the occurrence of inflation.
The right plot in \fref{efoldphidnegpos}, shows the time evolutions of the number of e-foldings, for $\dot\phi(0)>0$, starting with different anisotropy but with the same initial $\dot\phi(0)=0.1\,{\mpl^2}$. It is evident that the maximum number of e-foldings decreases with an increasing initial anisotropy.
 In Table-\ref{classefoldtab} we provide a representative summary for various simulations to determine the number of e-foldings obtained during the inflationary phase for various initial values of the $\dot\phi(0) = (-0.002,\,0.002,\,-0.02,\,0.02)\,\mpl^2$, and different values of initial anisotropic shear for $\phi(0)=(1.0,\,2.0,\,3.0,\,3.14,\,4.0,\,5.0)\,\mpl$.
 It is clearly seen that at large anisotropic shear, the number of e-foldings for a given
 $\phi(0)$ is almost the same for all $\dot\phi(0)$.

\begin{table}
 \begin{tabular}{|c|c|c|c|c|c|c|c|}
 \hline

 $\dot\phi(0)\, (\mpl^2)$& $\epsilon_{\rm J}^2$ & \multicolumn{6}{|c|}{$\phi(0)\, (\mpl)$} \\
\cline{3-8}
   & &1.0 & 2.0 & 3.0 & 3.14 & 4.0 & 5.0  \\
 \hline
 &0 & 1.551 & 2.755 & 20.600 & 24.195 & 52.051 & 96.749 \\
 &0.060 & 1.503 & 2.849 & 20.790 & 24.398 & 52.326 & 97.101 \\
 &1.084 & 0.907 & 4.947 & 23.587 & 27.369 & 57.364 & 102.195 \\
 -0.002&$9.665\times10^4$ & 6.388 & 25.322 & 56.727 & 62.122 & 100.662 & 157.145 \\
 &$5.994\times 10^8$ & 6.563 & 25.646 & 57.195 & 62.611 & 101.271 & 157.892 \\
 &$3.881\times 10^{19}$ & 6.566 & 25.651 & 57.203 & 62.619 & 101.282 & 157.905 \\
 &$1.565\times 10^{41}$ & 6.566 & 25.651 & 57.203 & 62.619 & 101.282 & 157.905 \\
 \hline
 &0 & 31.055 & 62.534 & 106.206 & 113.788 & 161.763 & 229.467 \\
 &0.060 & 30.853 & 62.149 & 105.86 & 113.429 & 161.349 & 228.988 \\
 &1.084 & 28.018 & 57.415 & 100.94 & 108.343 & 155.451 & 222.147 \\
0.002 &$9.665\times10^4$ & 6.744 & 25.970 & 57.6814 & 63.118 & 101.903 & 158.666 \\
 &$5.994\times 10^8$ & 6.569 & 25.657 & 57.211 & 62.628 & 101.292 & 157.918 \\
 &$3.881\times 10^{19}$ & 6.566 & 25.651 & 57.203 & 62.619 & 101.282 & 157.905 \\
 & $1.565\times 10^{41}$ & 6.566 & 25.651 & 57.203 & 62.619 & 101.282 & 157.905\\
 \hline
 &0 & 3.9404 & 0.199 & 12.284 & 15.163 & 38.690 & 78.448 \\
 &0.060 & 3.8420 & 0.194 & 12.495 & 15.393 & 39.027 & 78.901 \\
 &1.084 & 2.5564 & 1.692 & 15.653 & 20.1637 & 43.962 & 85.487 \\
 -0.02& $9.665\times10^4$ & 6.358 & 25.261 & 56.636 & 62.028 & 100.541 & 156.994 \\
 &$5.994\times 10^8$ & 6.562 & 25.645 & 57.194 & 62.609 & 101.269 & 157.89 \\
 &$3.881\times 10^{19}$ & 6.566 & 25.651 & 57.203 & 62.619 & 101.282 & 157.905 \\
 &$1.565\times 10^{41}$ & 6.566 & 25.651 & 57.203 & 62.619 & 101.281 & 157.905 \\
 \hline
 &0 & 41.544 & 77.366 & 125.623 & 133.832 & 185.725 & 258.003 \\
 &0.060 & 41.230 & 76.781 & 125.101 & 133.296 & 185.102 & 257.282 \\
 &1.084 & 36.855 & 69.644 & 117.725 & 125.679 & 176.281 & 247.045 \\
0.02 & $9.665\times10^4$ & 6.776 & 26.029 & 57.773 & 63.214 & 102.025 & 158.818 \\
 &$5.994\times 10^8$ & 6.569 & 25.657 & 57.2126 & 62.6293 & 101.294 & 157.92 \\
 &$3.881\times 10^{19}$ & 6.566 & 25.651 & 57.203 & 62.619 & 101.282 & 157.905 \\
 &$1.565\times 10^{41}$ & 6.566 & 25.651 & 57.203 & 62.619 & 101.282 & 157.905 \\
 \hline
\end{tabular}
 \caption{This table summarizes the number of e-folding in the classical Bianchi-I spacetime for various initial values of $\dot\phi(0)$ and $\phi(0)$, depending on value of anisotropic shear present in the initial data. Note that for $\epsilon_{\rm J}^2=0$, we have not shown the entries corresponding to $\phi=3.14\,\mpl$ leading to 60 e-foldings. For the conditions required to obtain 60 e-foldings in the isotropic case, see footnote-\ref{60efoldnote} in Sec.-IIIA.2}
 \label{classefoldtab}
\end{table}

\begin{figure}[tbh!]
\includegraphics[width=0.8\textwidth]{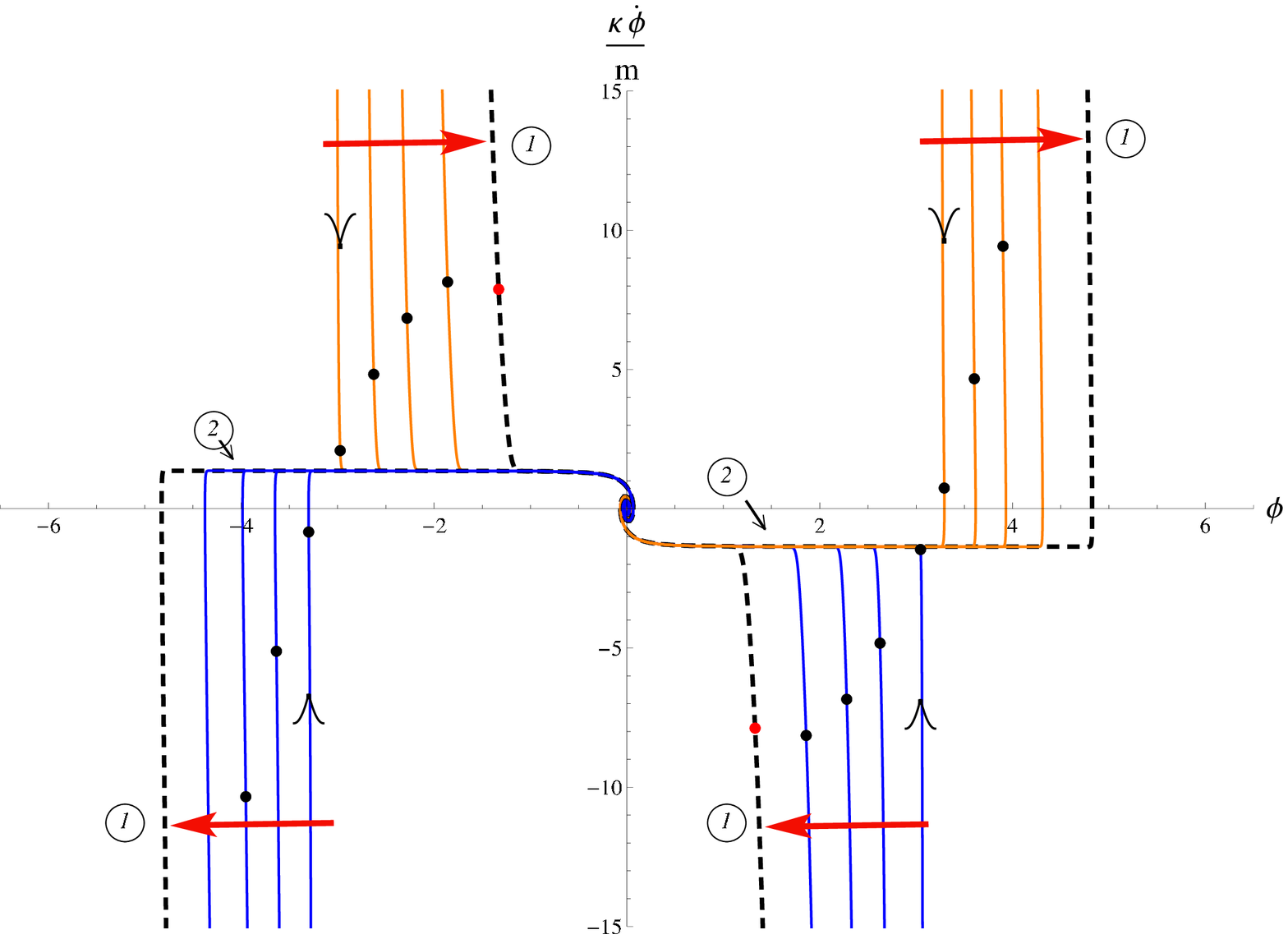}
\caption{This figure shows the two attractor behavior of the inflationary Bianchi-I spacetime. The dashed trajectories (denoted by \textcircled{1}) show the isotropic non-slow roll FRW attractor while the horizontal lines to which all the trajectories meet depict the isotropic slow-roll attractor (denoted by \textcircled{2}). The directions of (red) horizontal arrows show decreasing anisotropy (i.e. increasing parameter $\epsilon_{\rm J}$). The black dots denote the onset of accelerated expansion. $\kappa$ is $\kappa=8\pi G/3$.}
\label{2dphase}
\end{figure}

\bfig[tbh!]
\includegraphics[width=0.6\textwidth]{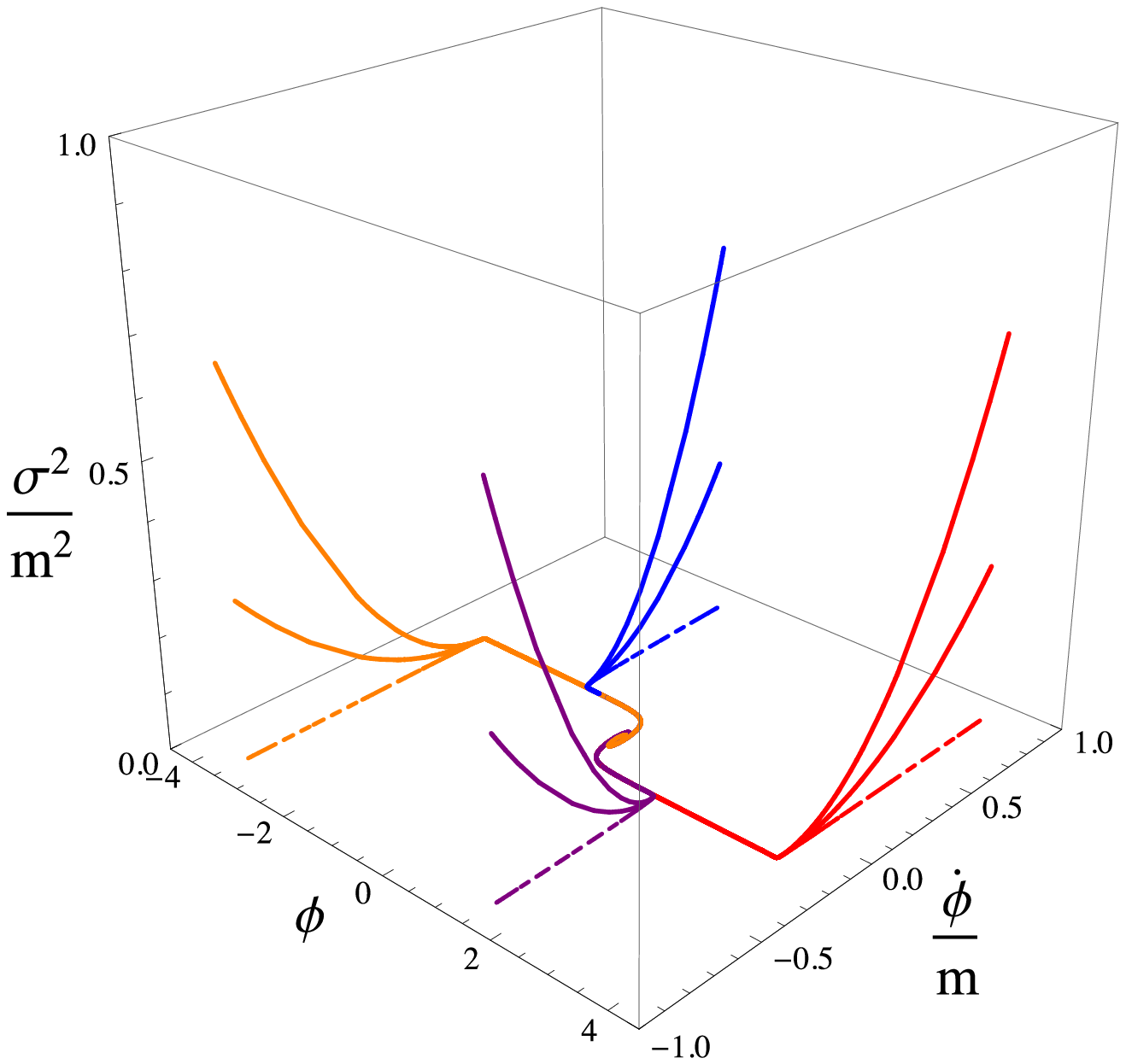}
\caption{This figure shows the 3D plot of the phase-space trajectories of the scalar field. The trajectories starting in the same quadrant of $\phi-\dot\phi$ plane correspond to the same initial values of $\phi$ and $\dot\phi$ but different initial shear scalar. The dashed trajectories in the $\sigma^2=0$ plane depicts the corresponding isotropic trajectories. It is clear that as the shear scalar is decreased the Bianchi-I trajectories tend to the corresponding isotropic trajectory.}
\label{3dphasecl}
\efig

\subsubsection{Phase portrait and attractor behavior}
We now discuss the phase-space trajectories of an inflationary Bianchi-I spacetime in the classical theory.\
\fref{2dphase} shows the evolution of various classical trajectories in the $\phi-\dot\phi$  space for Bianchi-I spacetime.
Trajectories in the same quadrant correspond to different initial anisotropies but with the same initial values of
$\phi$ and $\dot \phi$, that is, the initial energy density is fixed for all the trajectories and initial shear scalar is varied.\ The horizontal (red) arrows show the order of decreasing anisotropy. The trajectories with labels $\textcircled{1}$ and $\textcircled{2}$ correspond to the
isotropic spacetime (i.e. $\epsilon_{\rm J} = 0$). Each isotropic curve is composed of two distinct phases. The vertical part (labeled by  $\textcircled{1}$) corresponds to the phase in which there is no slow-roll. The horizontal part (labeled by $\textcircled{2}$) corresponds to the slow-roll phase in the isotropic case. The black dots on the trajectories denote the point when the accelerated expansion begins.  The trajectories in the upper half of the phase space correspond to $\dot\phi(0) > 0$ and the curves in the lower half show the trajectories for $\dot\phi(0) < 0$.
For all the curves, shown in the \fref{2dphase}, initial conditions on the inflaton are kept fixed at $|\phi|=3.14\, m_{\rm Pl}$ and $|\dot\phi(0)|=0.1\,\mpl^2$, and the anisotropic shear is varied.
It is clearly seen from \fref{2dphase}, that all the trajectories corresponding to various initial values of anisotropic shear meet the slow-roll curve in their evolution. In this sense, the isotropic slow-roll is an attractor for all such solutions.

For $\dot\phi(0),\,\phi(0) > 0$ (also for $\dot\phi(0),\phi(0) < 0$), i.e.\ the initial conditions corresponding to the inflaton rolling up the potential,  trajectories in Bianchi-I model remain close to isotropic slow roll  curve for a longer time for smaller initial value of the anisotropic shear. Furthermore, with increasing anisotropic shear, the duration of the slow-roll decreases.
This behavior is compatible with the variation in the amount of inflation with the varying initial anisotropy discussed earlier.
 On the other hand, trajectories corresponding to  $\phi(0)>0$ and $\dot\phi(0) < 0$ (or with $\phi(0) < 0$ and $\dot \phi(0) > 0$) tend to have longer duration of slow-roll for larger value of the anisotropy, which is again compatible with the observation that an increase in anisotropy increases the e-foldings for an inflaton initially rolling down the potential.

\fref{3dphasecl} shows the 3D plot of the phase space trajectories of the field in the classical Bianchi-I spacetime. It turns out that as the shear scalar tends to decrease, the phase-space trajectories in classical Bianchi-I spacetime approach that of the corresponding isotropic classical FRW spacetime (with $\sigma^2=0$). This behavior is also evident from the \fref{2dphase} where the thick (red) arrows show the order of decreasing anisotropy. In this sense, in addition to the attraction behavior towards  the slow roll trajectory, the non slow roll isotropic curve also acts as an attractor for trajectories in Bianchi-I spacetime in the limit $\sigma^2\rightarrow0$.\footnote{For an earlier discussion of the second attractor, see Ref. \cite{pitrou}.}
In other words, the classical FRW spacetime is an isotropic attractor of the classical  Bianchi-I spacetime.
That is, as the initial anisotropy is decreased, the trajectories tend closer to the classical FRW isotropic trajectories (shown by the dashed-thick lines in \fref{2dphase} and \fref{3dphasecl}).
To differentiate this attractor from the slow-roll, we call it the `non slow-roll FRW attractor'.
In this precise sense, there is a double attractor behavior in the classical Bianchi-I spacetime.
 Before we go into the discussion of numerical results for LQC, a remark about the two attractors is in order.
\vskip0.3cm
{\bf Remark:} The two attractors discussed above, namely, the non slow-roll FRW attractor and the isotropic slow-roll attractor, are two distinct properties of the Bianchi-I spacetime. The approach to the isotropic slow-roll is a property of the Bianchi-I spacetime with an inflationary potential. It is achieved in the future evolution when the slow-roll conditions are met, whereas, the non slow-roll FRW attractor can be approached in the pre slow-roll phase of the evolution when the value of the initial anisotropic shear is decreased. In contrast, the slow-roll attractor
is approached only after the onset of inflation.

\subsection{Effective loop quantum dynamics}
In this subsection we discuss the results from numerical simulations of the effective
dynamical equations of Bianchi-I spacetime in LQC with a $m^2 \phi^2$ potential.
The mass of the inflaton is taken to be same as in previous sections:
$m=1.21\times10^{-6}\, m_{\rm Pl}$. Unlike the classical theory, LQC is devoid of initial
singularity. The mean scale factor instead of vanishing in the backward evolution,
undergoes a non-singular bounce. Hence, LQC provides a way to investigate the
inflationary scenario in Bianchi-I spacetime in a non-singular setting. In this article we
present the first such study of inflationary Bianchi-I spacetime in a non-singular scenario.
The non-singular evolution, in the effective
description of LQC allows us to extend the study of dynamics of Bianchi-I spacetime
prior to the onset of inflation, in the deep Planck regime.

\begin{figure}[tbh!]
\includegraphics[width=0.6\textwidth]{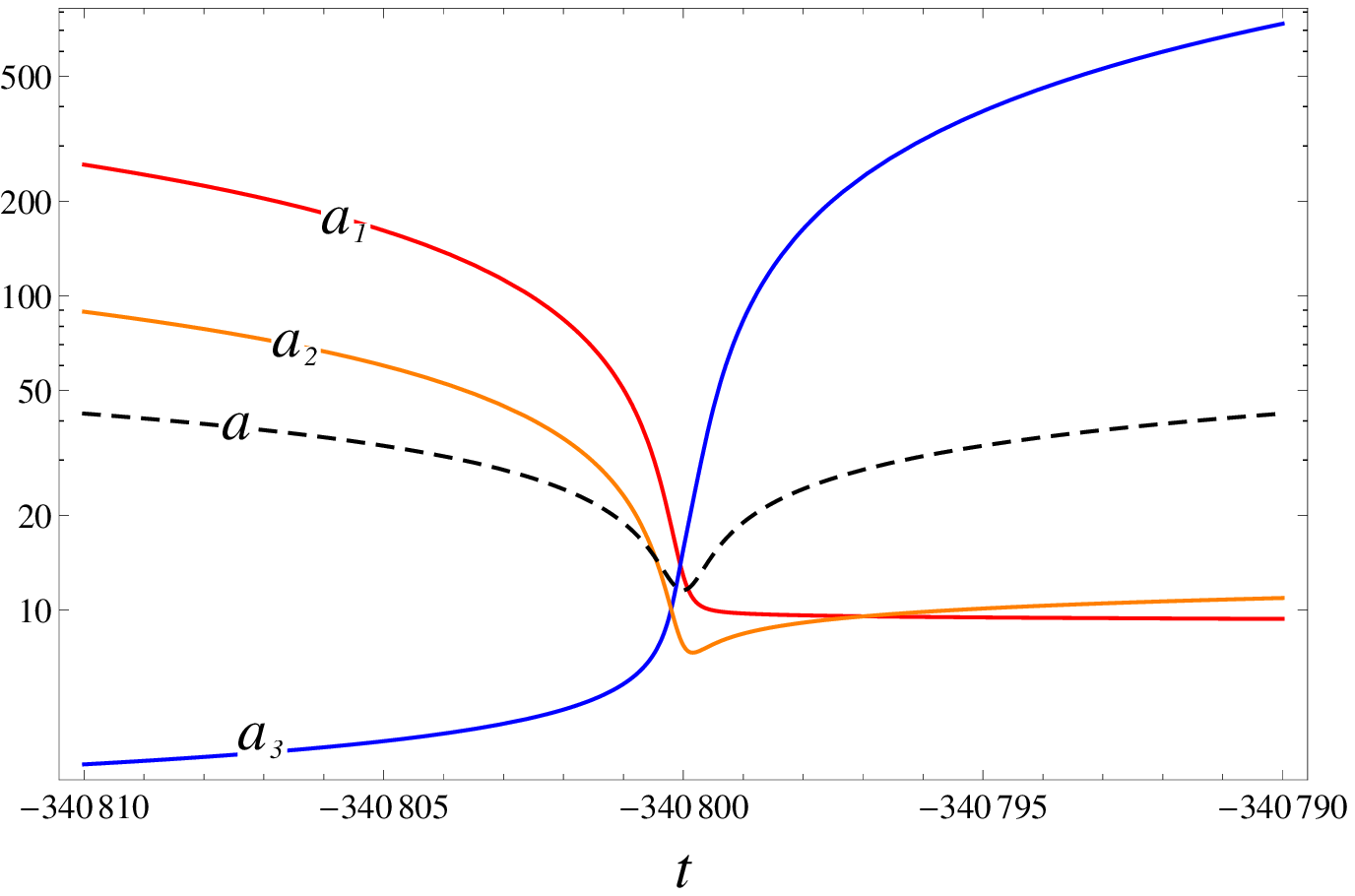}(a)
\vskip0.5cm
\includegraphics[width=0.45\textwidth]{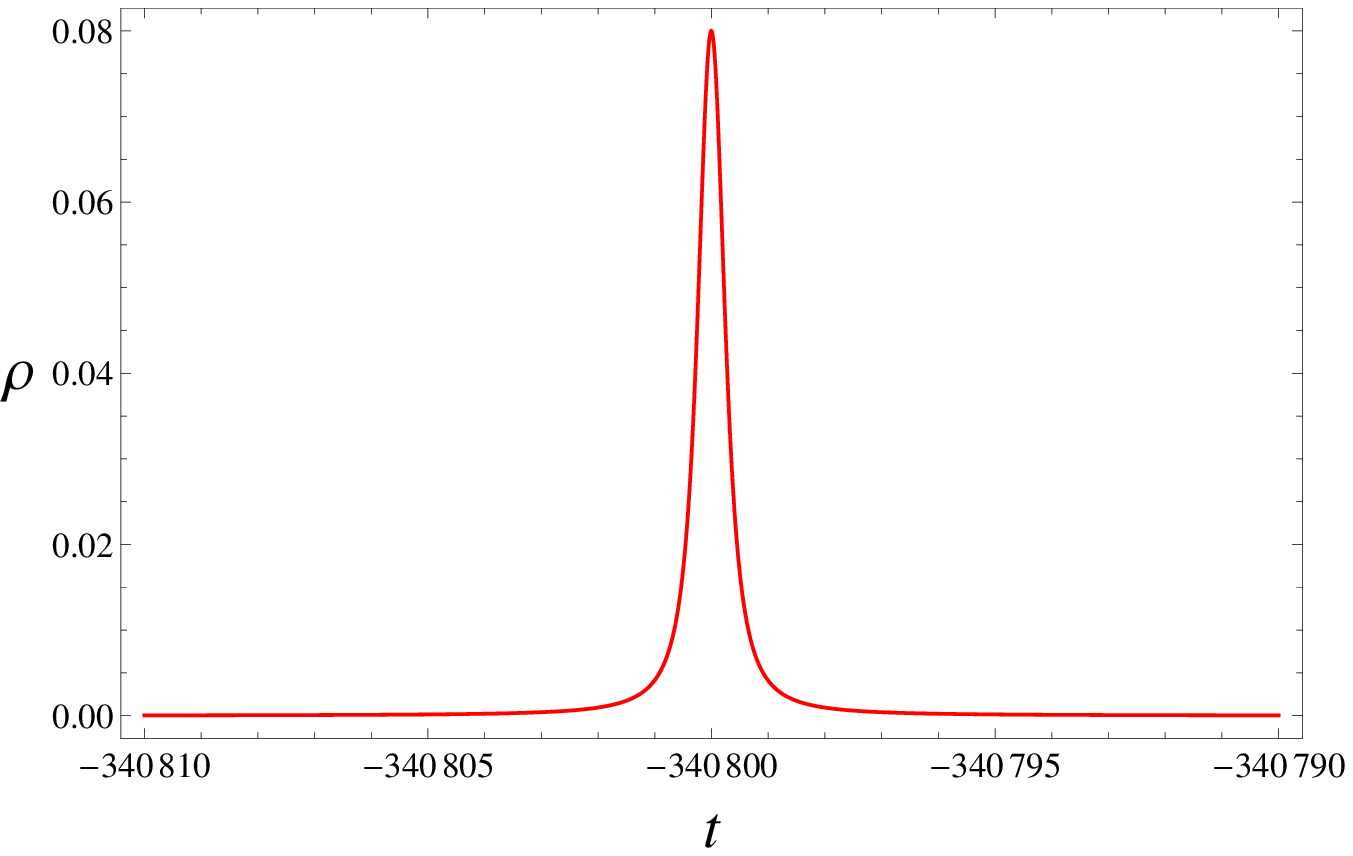}(b)
\hskip0.5cm
\includegraphics[width=0.45\textwidth]{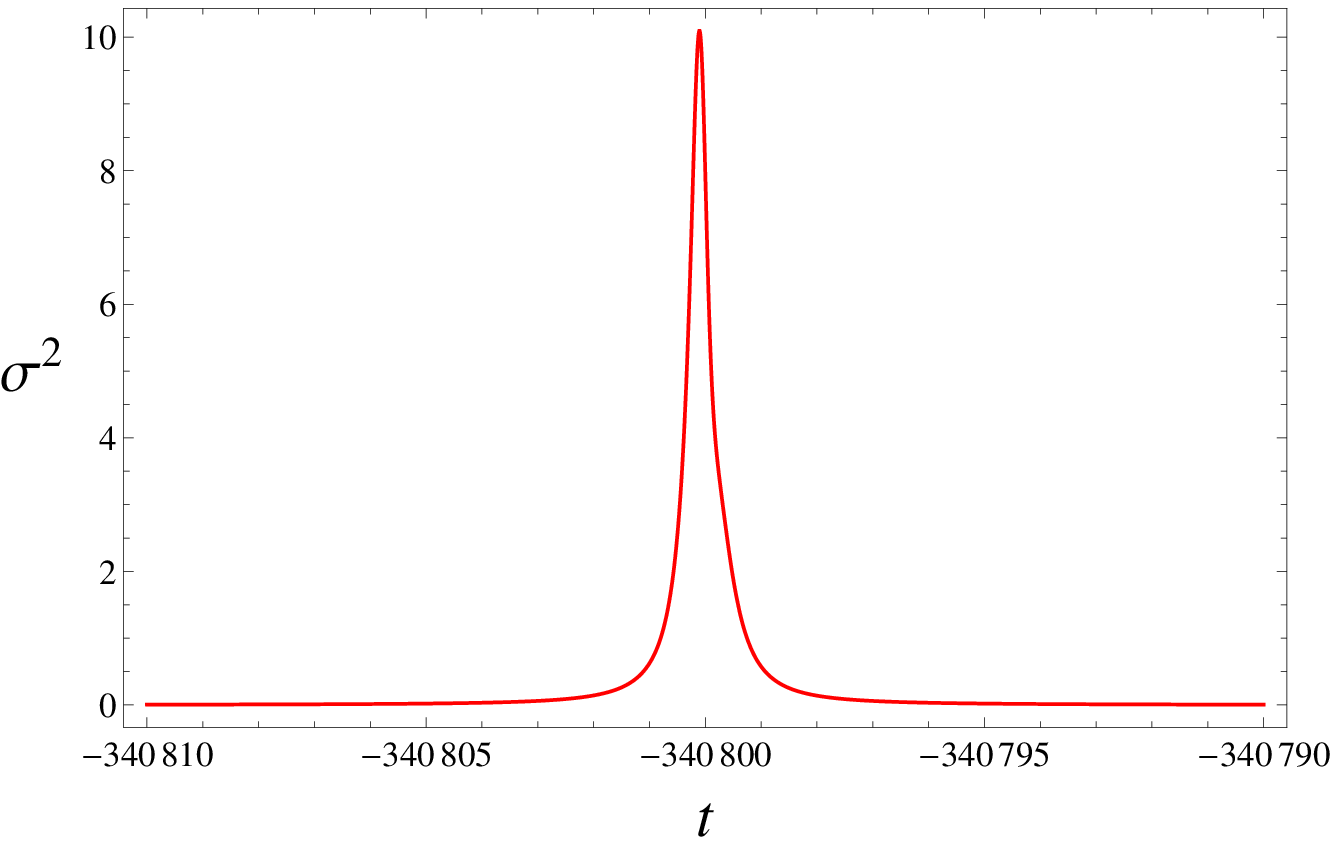}(c)
\caption{The figure (a) shows the evolution of the mean and directional scale factors, close to
the bounce, when initial conditions are given far from the bounce near the onset of inflation.
Figures (b) and (c) show the evolution of energy density and the shear scalar respectively.
The values of $\rho$ and $\sigma^2$ are in Planck units. The initial conditions for this figure
are $\phi(0)\approx3.14\,\mpl$, $\dot\phi(0)\approx-1.97\times10^{-7}\,\mpl^2$ and
$\sigma^2(0)\approx1.09\times10^{-16}\,\lp^{-2}$. It is clear from the plots that the mean
scale factor bounces while the directional scale factors undergo smooth evolution.}
\label{bounce1}
\end{figure}

\begin{figure}[tbh!]
\includegraphics[width=0.6\textwidth]{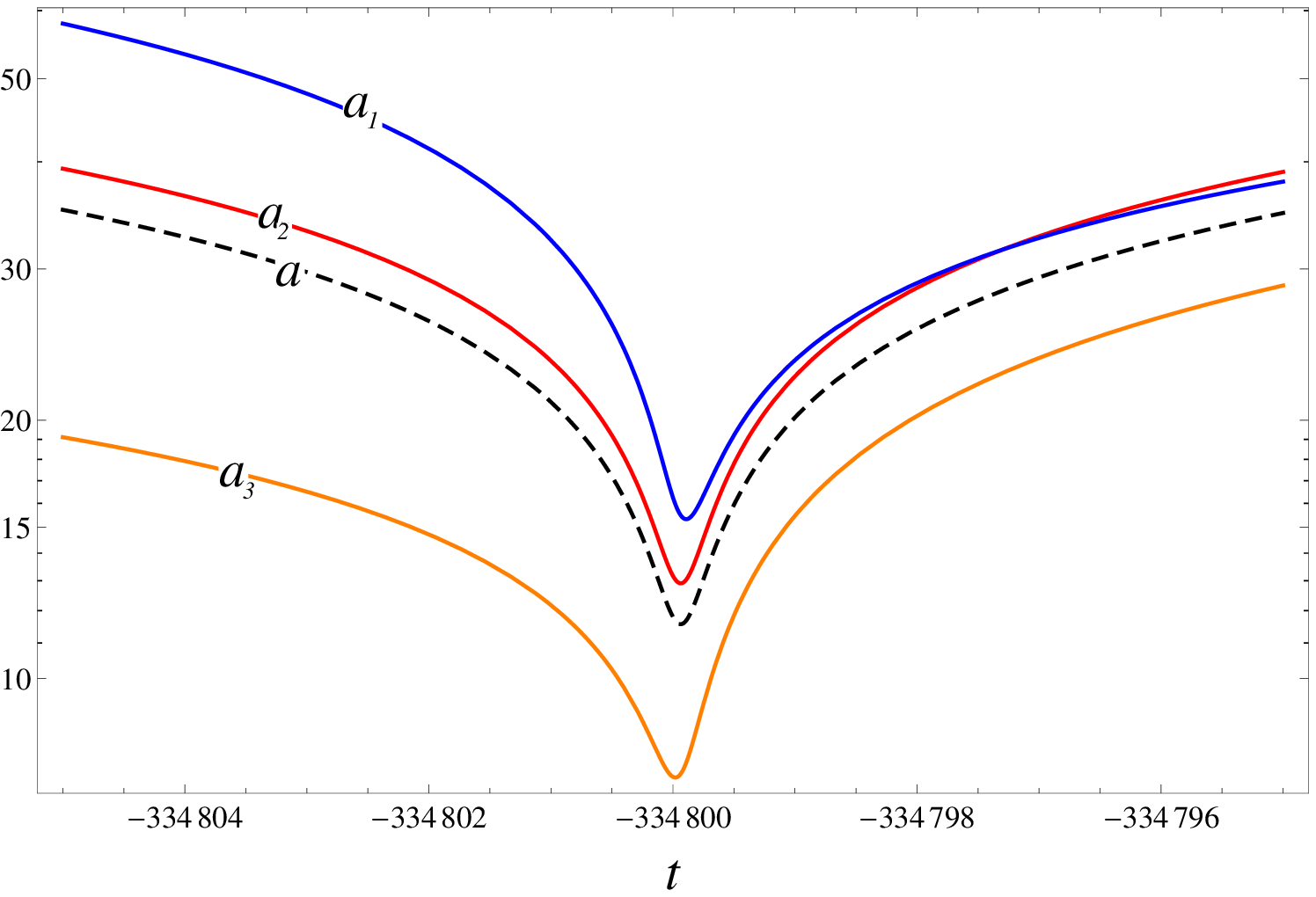}(a)
\vskip0.5cm
\includegraphics[width=0.45\textwidth]{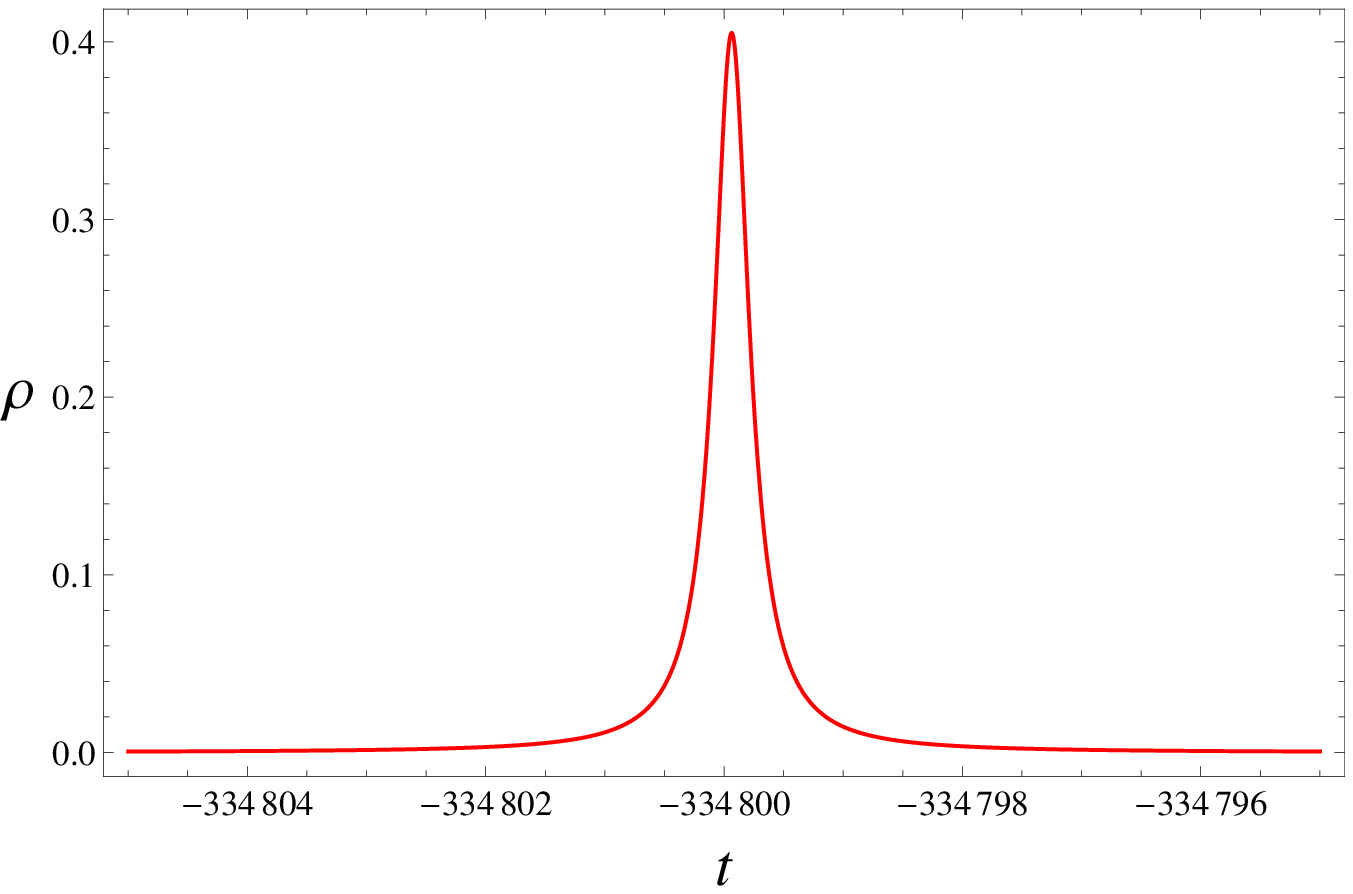}(b)
\hskip0.5cm
\includegraphics[width=0.45\textwidth]{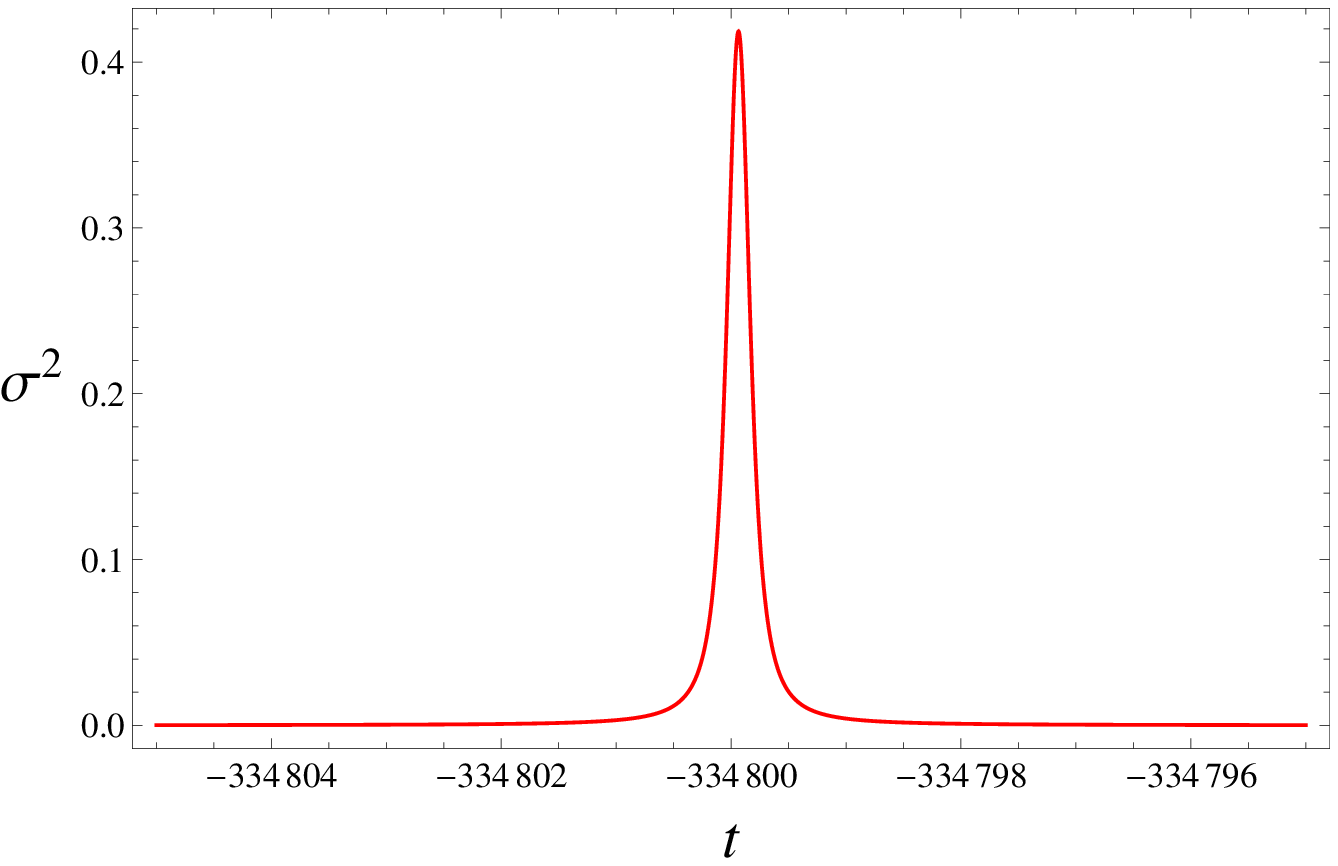}(c)
\caption{The figure (a) shows the evolution of the mean and directional scale factors, close to
the bounce, when initial conditions are given far from the bounce near the onset of inflation.
Figures (b) and (c) show the evolution of energy density and the shear scalar respectively.
The values of $\rho$ and $\sigma^2$ are in Planck units. The initial conditions for this figure
are $\phi(0)\approx3.14\,\mpl$, $\dot\phi(0)\approx-1.99\times10^{-7}\,\mpl^2$ and
$\sigma^2(0)=9.32\times10^{-19}\,\lp^{-2}$. It is clear from the plots that the mean scale
factor bounces while the directional scale factors undergo smooth evolution.}
\label{bounce2}
\end{figure}

As discussed in the previous section,
irrespective of the initial conditions, all the curvature invariants of Bianchi-I spacetime are
bounded and never diverge during the evolution given by the effective dynamical equations.
This implies that the occurrence of bounce is a generic feature of the effective description of
Bianchi-I spacetime in LQC, and bounce is obtained without violating any energy condition. If
one starts with a classical expanding universe, so that the spacetime curvature is too small
for the quantum geometric effects to be important, then during the backward evolution, initially
the classical trajectory agrees well with the effective trajectory. In the further backward
evolution, as the curvature of spacetime grows, the quantum geometric effects start becoming
more prominent. Due to this, the effective trajectory starts deviating from the classical
trajectory. In the subsequent backward evolution, the classical trajectory meets the initial
singularity, when the mean scale factor goes to zero. On the other hand, the effective
trajectory undergoes a smooth non-singular evolution, with the mean scale factor bouncing
from a finite non-vanishing value. The individual scale factors, also, undergo smooth evolution
from one side of the bounce of the mean scale factor to the other side.
Figs.\,(\ref{bounce1}(a) and \ref{bounce2}(a)) show two examples of evolution of the mean
and the directional scale factors of Bianchi-I spacetime in LQC, with initial conditions being
provided far from the bounce, close to the onset of inflation. The corresponding evolutions of
the energy density and shear scalar are shown in Figs.\,(\ref{bounce1}(b, c) and
\ref{bounce2}(b, c)). The initial conditions for the evolutions shown in
\fref{bounce1} are $\phi(0)= 3.14\,\mpl$, $\dot\phi(0)=-1.97\times10^{-7}\,\mpl^2$,
$\sigma^2(0)= 1.09\times10^{-16}\,\lp^{-2}$, and the initial
mean scale factor is $a(0)=8.02\times10^3$ while the initial energy density being
$\rho(0)=7.23\times10^{-12}\,\mpl^4$. Similarly the initial conditions for the evolution
shown in \fref{bounce2} are $\phi(0)=3.14\,\mpl$, $\dot\phi(0)=-1.99\times10^{-7}\,\mpl^2$,
$\sigma^2(0)=9.32\times 10^{19}\,\lp^{-2}$, $a(0)=8.05\times10^3$ and
$\rho(0)=7.24\times10^{-12}\,\mpl^4$. It evident from Fig. (\ref{bounce1} and \ref{bounce2})
that the energy density and the shear scalars do not diverge, and are below their maximum values $\rho_{\rm max}$ and $\sigma^2_{\rm max}$ throughout the non-singular evolution. The values of the energy density and shear scalar at the bounce can be quite different, as can be seen by comparing Fig.\ref{bounce1} and Fig.\ref{bounce2}.
It is worth emphasizing that the bounce occurs without any fine tuning of the initial conditions
or parameters.

\begin{figure}[tbh!]
\includegraphics[width=0.47\textwidth]{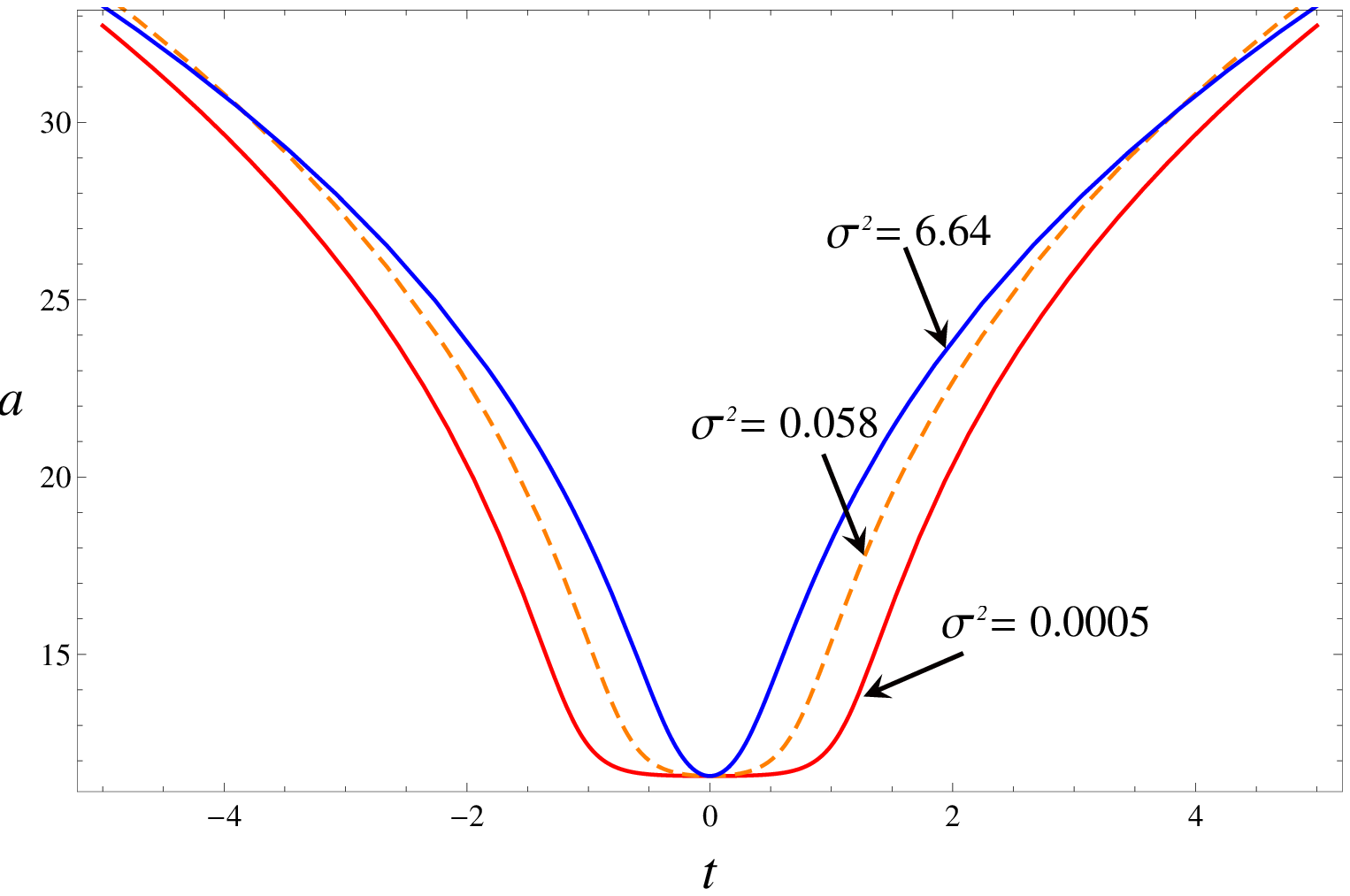}
\caption{This figure shows the evolution of the mean scale factor for different initial shear scalars (but with the same initial $\phi=3.14\,\mpl$ and $\dot\phi=-0.002\,\mpl^2$) given at the bounce. The values of the shear scalar shown in the figure correspond to the values at the bounce (in units of $\lp^{-2}$).}
\label{lqcmeanscale}
\end{figure}

Since the bounce is a generic property of all the solutions in the effective dynamics of LQC, it provides a natural point in the evolution to
specify initial data.  In all the simulations which we discuss in the following, the initial
conditions are provided at the bounce, characterized by $t=0$. The bounce is characterized by
the turn around of the mean scale factor which entails  vanishing mean Hubble rate,
$H(0)=0$ at the bounce. Therefore, the initial conditions along with satisfying the Hamiltonian
constraint, must be chosen such that the mean Hubble rate equals zero. To ensure this, we
provide $p_1(0),\, p_2(0),\, p_3(0),\, c_1(0),\, \phi(0),\, {\rm and}\, \dot\phi(0)$ at the bounce,
and compute the value of $c_2(0)\,{\rm and}\,c_3(0)$ by requiring that both the Hamiltonian
constraint and the mean Hubble rate vanish. Keeping $p_i(0),\,\phi(0),\, \dot\phi(0)$ fixed,
and by varying $c_1(0)$ at the bounce,
we can vary the initial anisotropic shear scalar at the bounce for a fixed value of the energy
density. This gives rise to a continuous set of solutions with a fixed energy density and varying
initial shear scalar, all starting at the bounce. These solutions turn out to be the key
to understand the effect of varying anisotropic shear scalar on the number of e-foldings.
To illustrate, \fref{lqcmeanscale} shows the bounce of mean scale factors for varying
$\sigma^2(0)$ for the same $\phi(0)=3.14\,\mpl$ and
$\dot\phi(0)=-0.002\,\mpl^2$ at the bounce. As in the previous sections, we choose the
initial value of the inflaton field same as above to discuss the qualitative behaviors shown in the
plots in this subsection. Results for different initial values of inflaton field will be given in the
tabular form (see Table-\ref{efolddatavaryphi}). Another way of generating initial
conditions with varying initial shear scalar is to vary the initial energy density by varying the
initial velocity of the inflaton. This method of choosing initial data is essential in order to
understand the isotropic attractor behavior of Bianchi-I spacetime in LQC, as we will see later
in this subsection.

In the following, we first study the evolution of the individual directional scale factors and
the isotropization of Bianchi-I spacetime as the future evolution takes place. Then, we turn
to compute the amount of inflation in terms of the number of e-foldings generated during the
inflationary era, for a variety of initial conditions including different initial $\dot\phi$, $\phi$ and
$\sigma^2$ at the bounce. We also study the evolution of dynamical trajectories in
the $\phi-\dot\phi$ phase space, starting from different initial conditions. The study of
the phase trajectories brings out important features of the attractor behavior of Bianchi-I
spacetime in effective description of LQC.

\subsubsection{Directional scale factors and isotropization}

We study the evolution of the spacetime starting from initial conditions characterized by the presence of a significant anisotropic shear.
Due to the non-vanishing anisotropic shear, as in the classical theory, depending on the strength of the anisotropy, the expansion of the spacetime can be highly anisotropic. That is, while the mean scale factor expands,  one or two of the directional scale factors decrease and the remaining one(s) increase in the forward evolution. However, in contrast to the classical theory, the backward evolution in LQC is non-singular.
Here, we are interested in investigating the way an anisotropically expanding universe at the bounce turns into isotropically inflating spacetime in LQC.

\begin{figure}[tbh!]
\includegraphics[width=0.47\textwidth]{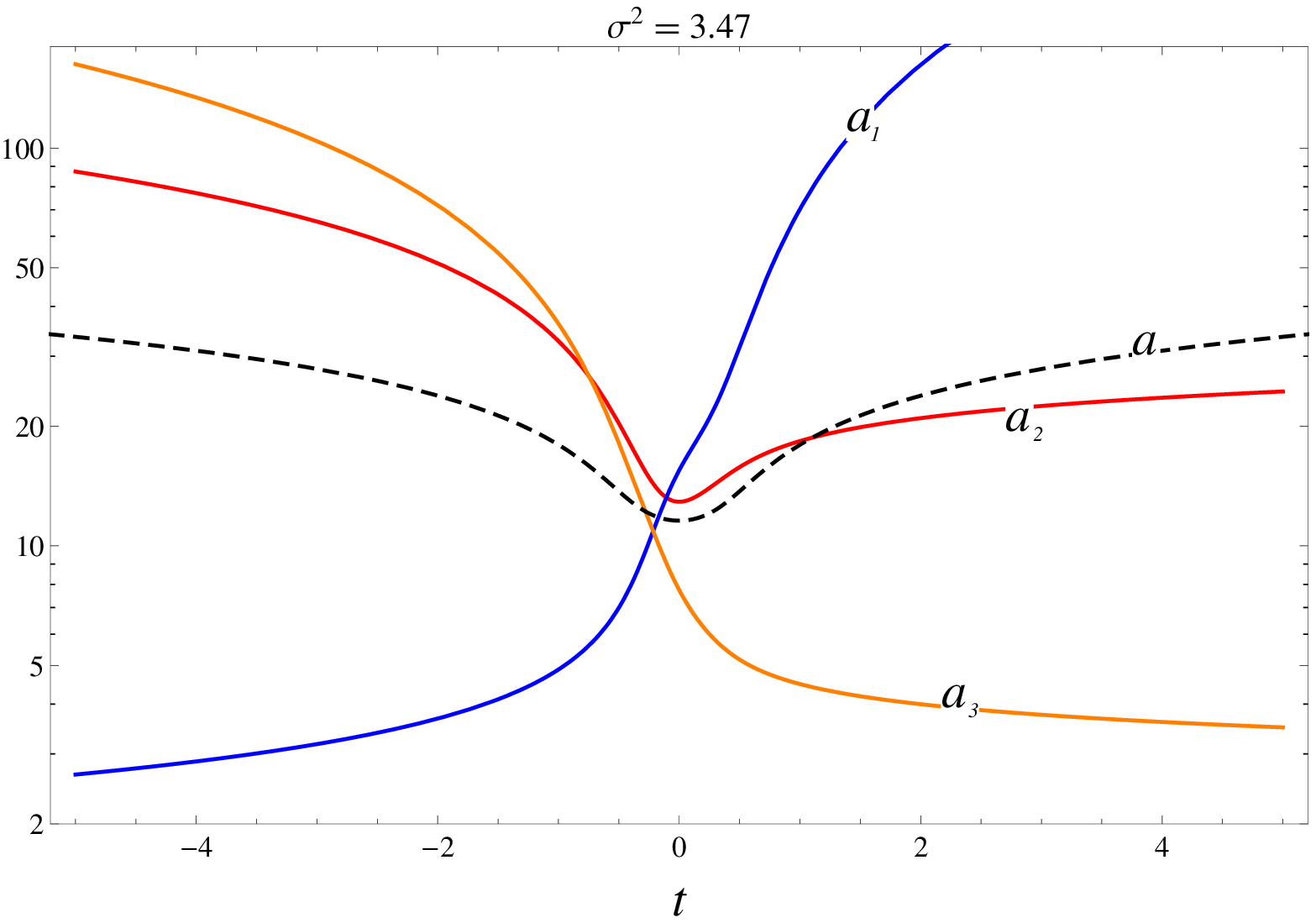}
\hskip0.5cm
\includegraphics[width=0.47\textwidth]{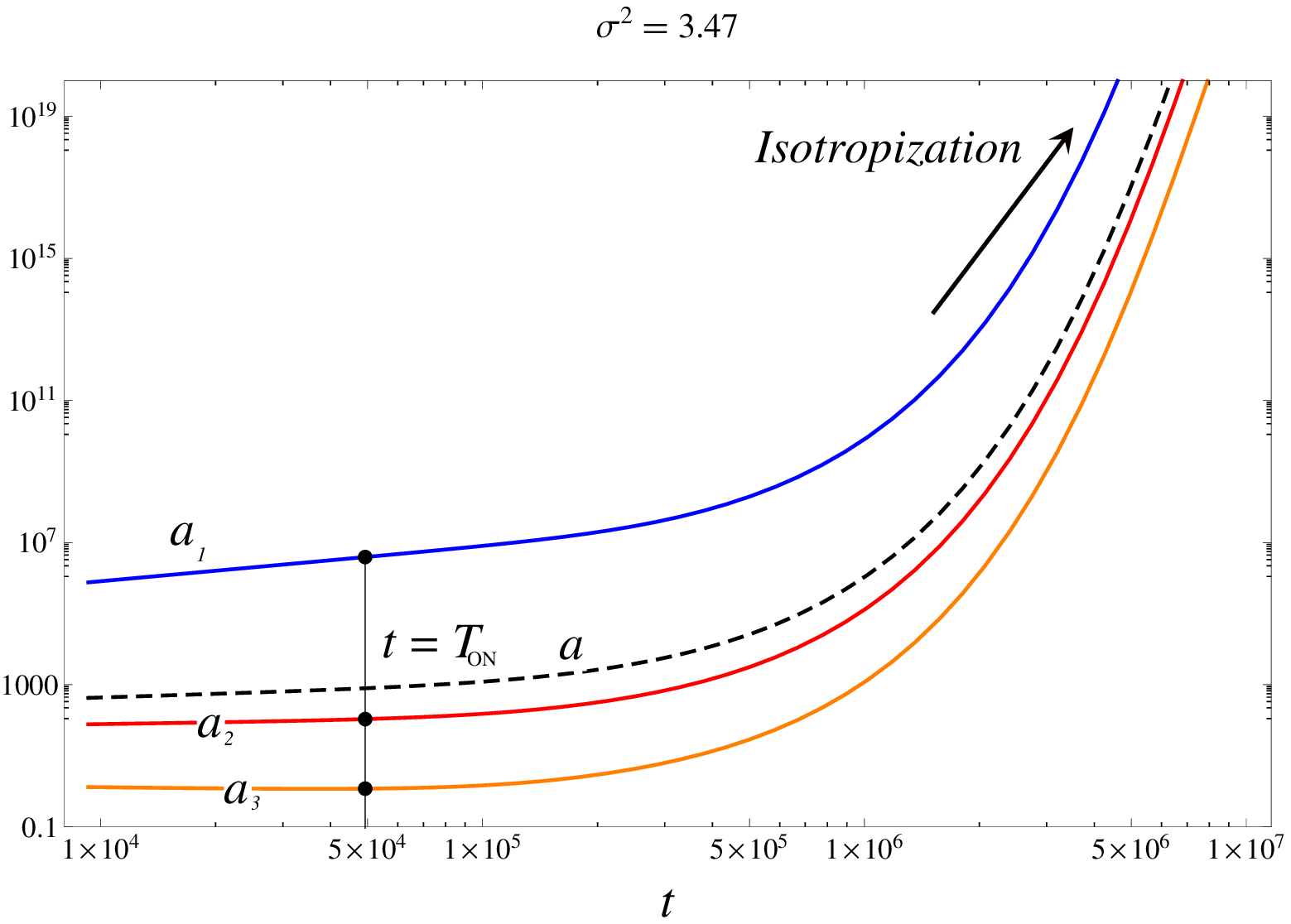}
\caption{This figure shows the evolution of directional scale factors for $\sigma^2(0)=3.47\,\lp^{-2}$ at the bounce with initial $\phi=3.14\,\mpl$ and $\dot\phi=-0.002\,\mpl^2$. The right plot shows the isotropization of the directional scale factors. At the bounce one of the scale factors start out in decreasing state but as the inflationary era progresses, all the three directions turn to expand.}
\label{lqca123}
\end{figure}

In \fref{lqca123}, we show the evolution of scale factors in a representative simulation in the Bianchi-I spacetime in LQC. In the expanding branch of the mean scale factor $a$, two scale factors are expanding and one is contracting near the bounce. This corresponds to a cigar like structure of the spatial geometry close to the bounce, in the backward evolution of the expanding branch.
It turns out that long before the onset of accelerated expansion, LQC trajectory enters into the classical domain, and the shear scalar monotonically falls as $\sim a^{-6}$.
Due to the ever decreasing shear scalar, in this domain, the parameter $\epsilon_{\rm J}$ keeps decreasing and, just like in the classical theory, reaches $\epsilon_{\rm J}<2/\sqrt{3}$ before or right after the onset of inflation. 
  As the accelerated expansion starts to take place, like in the classical theory, the parameter $\epsilon_{\rm J}$ falls even more quickly and the contracting directional scale factor turns around, causing isotropization of the Bianchi-I universe in LQC. 

\begin{figure}[tbh!]
\includegraphics[width=0.47\textwidth]{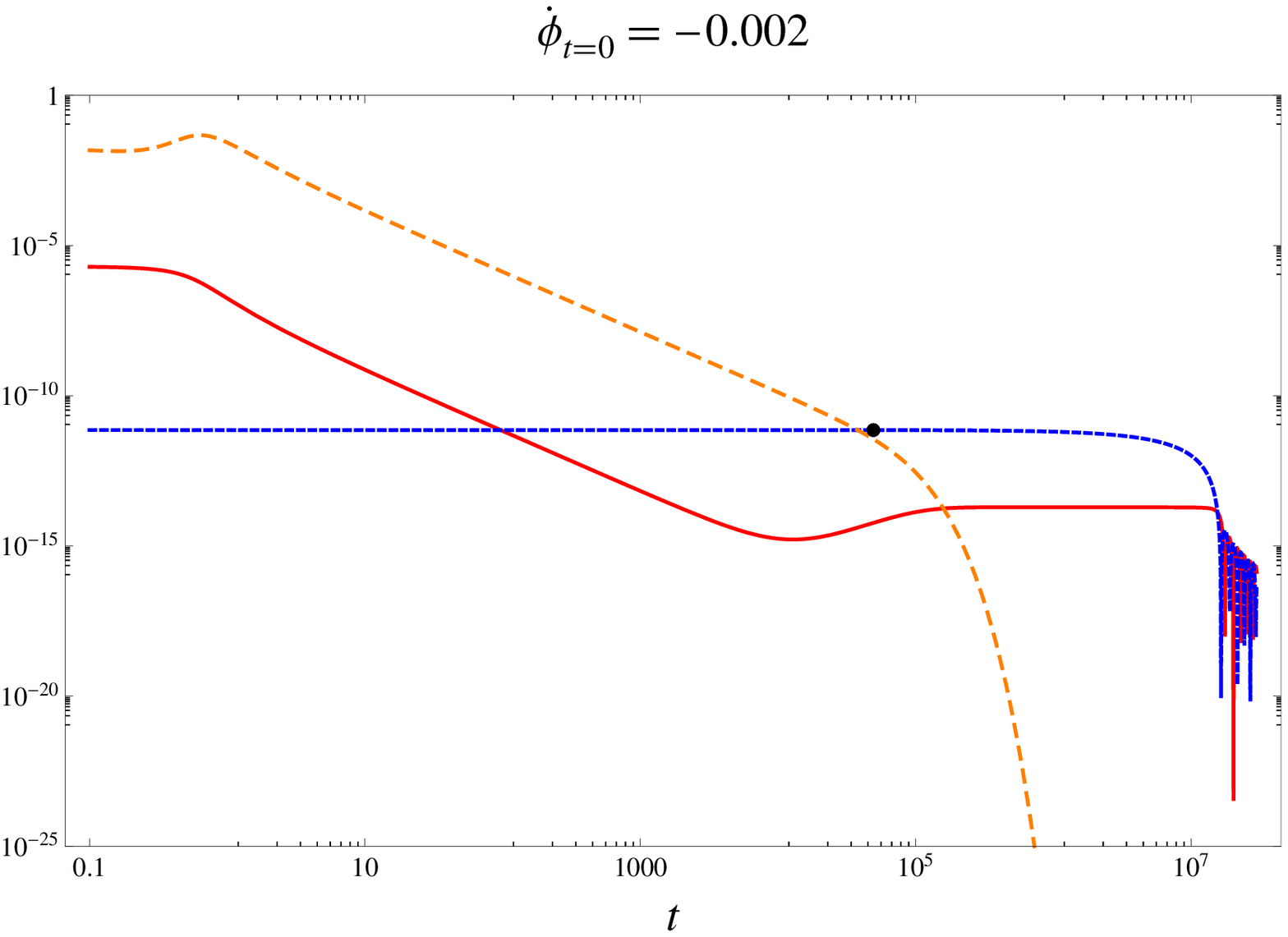}
\caption{Evolution of the kinetic $(\dot \phi^2/2)$, potential $(m^2 \phi^2/2)$ and anisotropic shear contribution $(\sigma^2/16 \pi G)$ in LQC effective dynamics. The short dashed  (blue)  curve represents potential energy, the dashed (orange) line corresponds to the shear contribution and the solid (red) line shows the evolution of kinetic energy of the inflaton.}
\label{kepelqc}
\end{figure}

In Sec. IIA3, we discussed the evolution of kinetic and potential energies and compared them with the anisotropic shear
contribution $(\sigma^2/16 \pi G)$ in the classical theory. We now study their behavior in LQC.
 \fref{kepelqc} shows an example of the evolution of all of the three in an expanding branch of inflationary Bianchi-I spacetime.\
Starting with a shear dominant initial conditions such that kinetic energy dominates over the potential energy,  the kinetic energy monotonically decays, whereas, the shear energy shows a non-monotonic behavior in the vicinity of the bounce i.e.\ it attains a local maximum and then continues to decrease. In the forward evolution, as the spacetime curvature approaches the classical conditions, the anisotropic shear decreases monotonically.
This behavior of the anisotropic shear stands in contrast with the classical evolution where the shear term $\sigma^2$ monotonically decreases in an expanding universe for generic initial conditions. \
This brings out a key difference between the process of isotropization in classical and LQC evolutions.

It is also clear from the plot that the onset of accelerated expansion (marked by the black dot) takes place when the potential energy wins over the shear and kinetic energy, just like in the classical theory.
This is not surprising because, in the above numerical simulation, at the onset of inflation the spacetime is  well approximated with a classical solution. Following the start of inflation there may be a period where the contribution of anisotropic shear is still greater than kinetic energy (while contribution of potential energy being larger than the both). However, the shear scalar quickly fades away giving rise to isotropization during the inflationary phase. In this way, similar to the classical theory, there may be a short period in the inflationary era when the anisotropic shear dominates the kinetic energy, while the potential energy remains greater than both.


\subsubsection{Amount of inflation}

\begin{figure}[tbh!]
\includegraphics[width=0.47\textwidth]{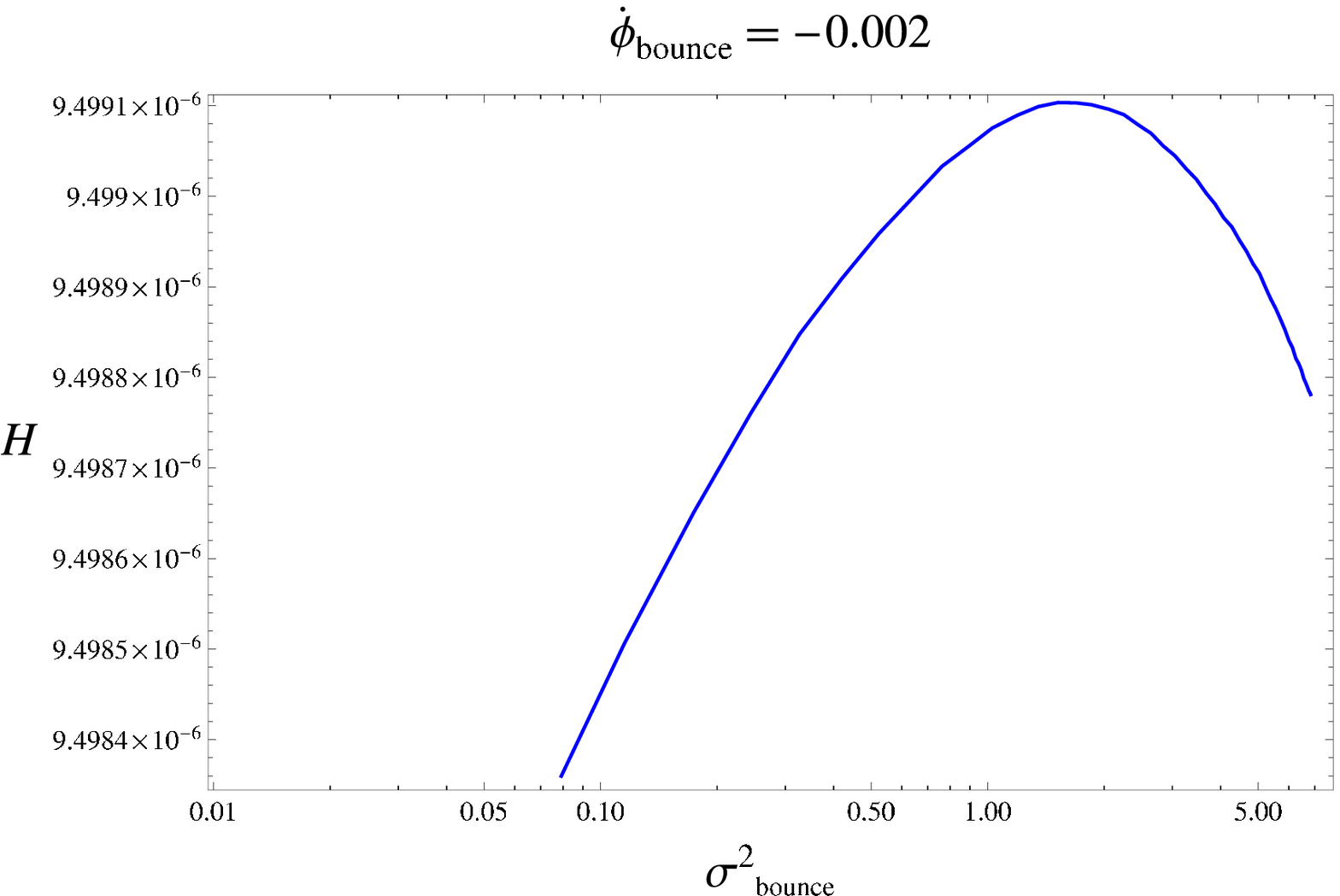}
\hskip0.5cm
\includegraphics[width=0.47\textwidth]{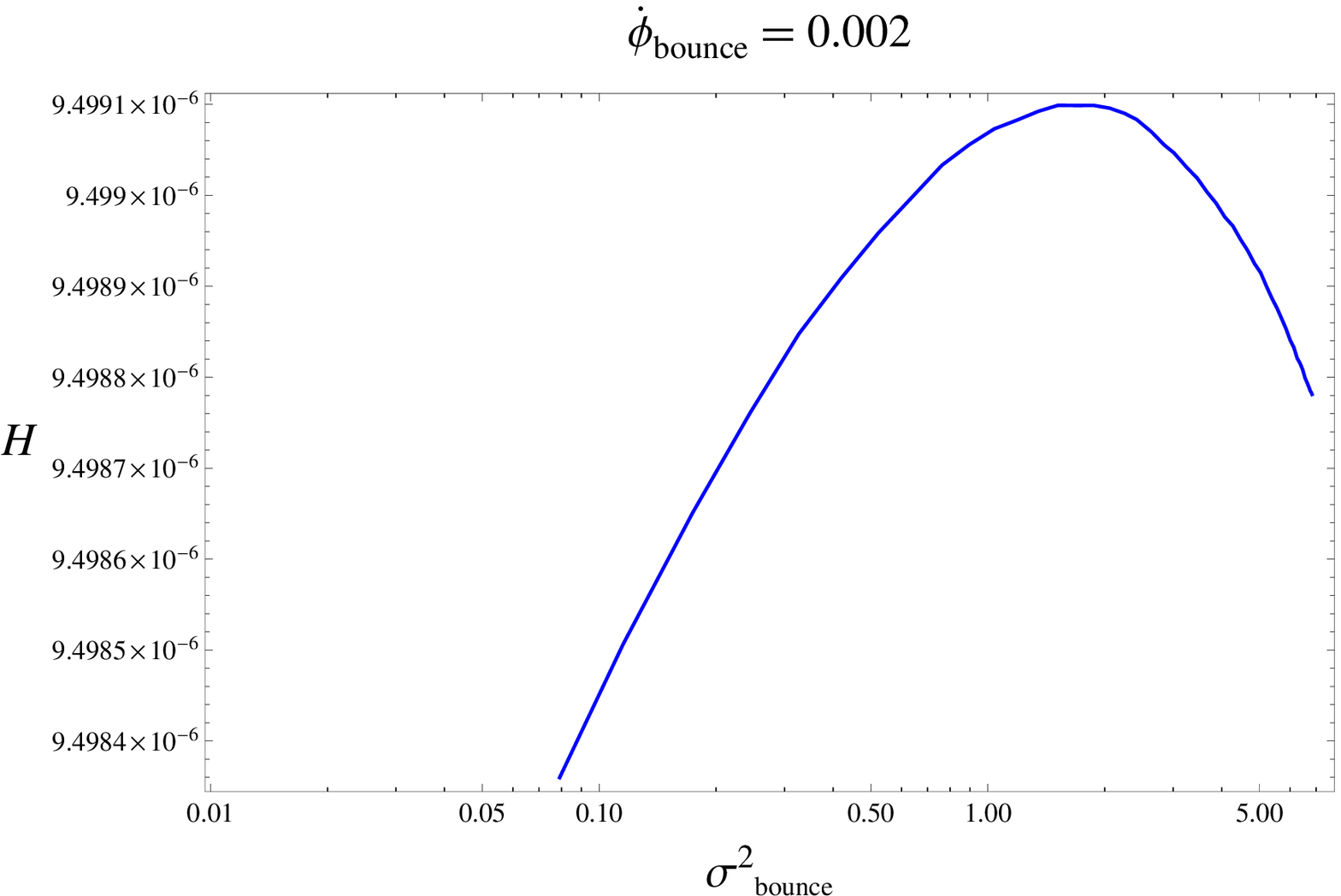}
\caption{This figure shows the non-monotonic behavior of the mean Hubble rate at the onset of inflation with varying initial value of the shear scalar at the bounce. The left plot corresponds to a negative initial $\dot\phi$ and the right plot corresponds to a positive initial $\dot\phi$ (in Planck units).}
\label{hublqc}
\end{figure}

\begin{figure}[tbh!]
\includegraphics[width=0.47\textwidth]{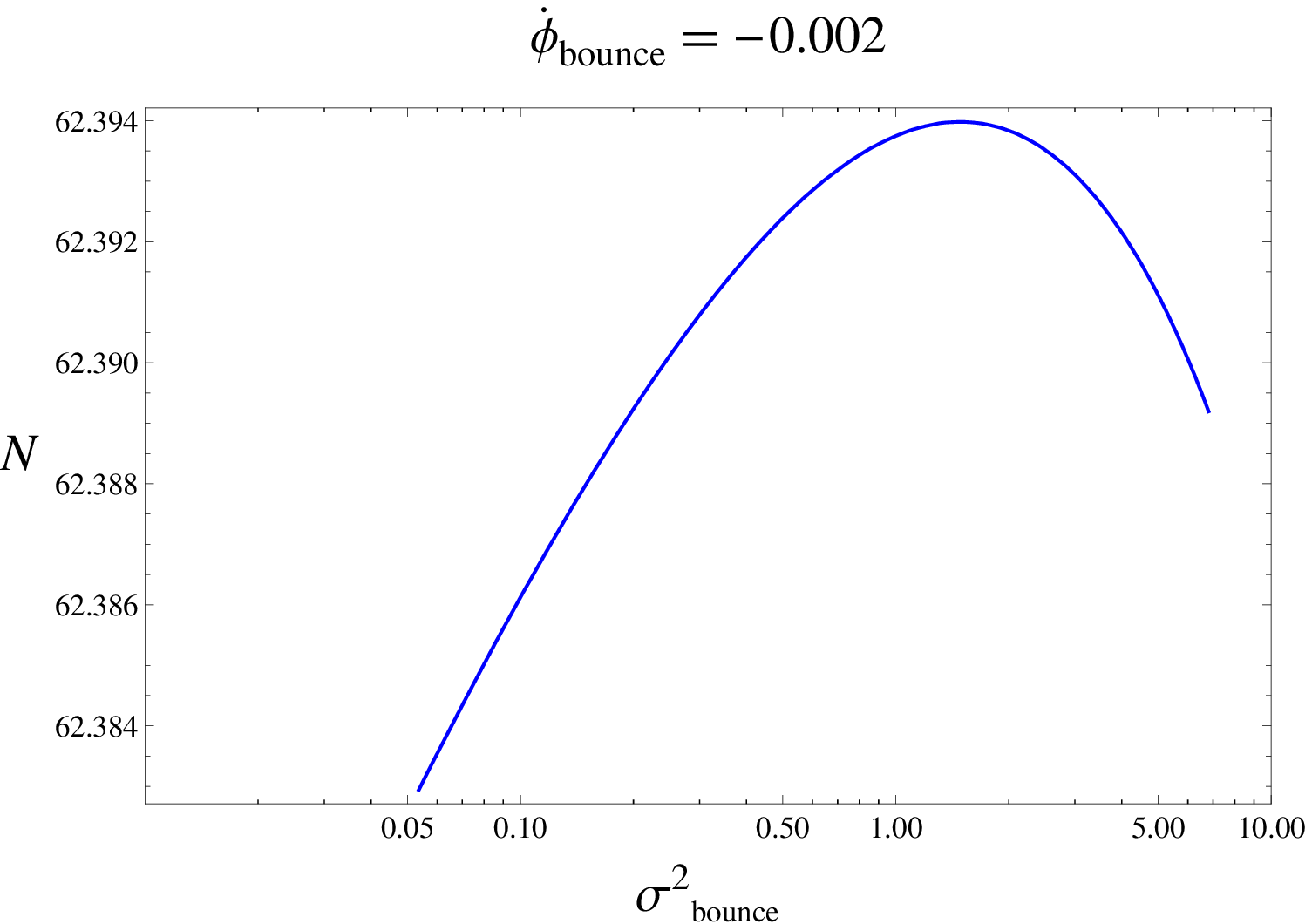}
\hskip0.5cm
\includegraphics[width=0.47\textwidth]{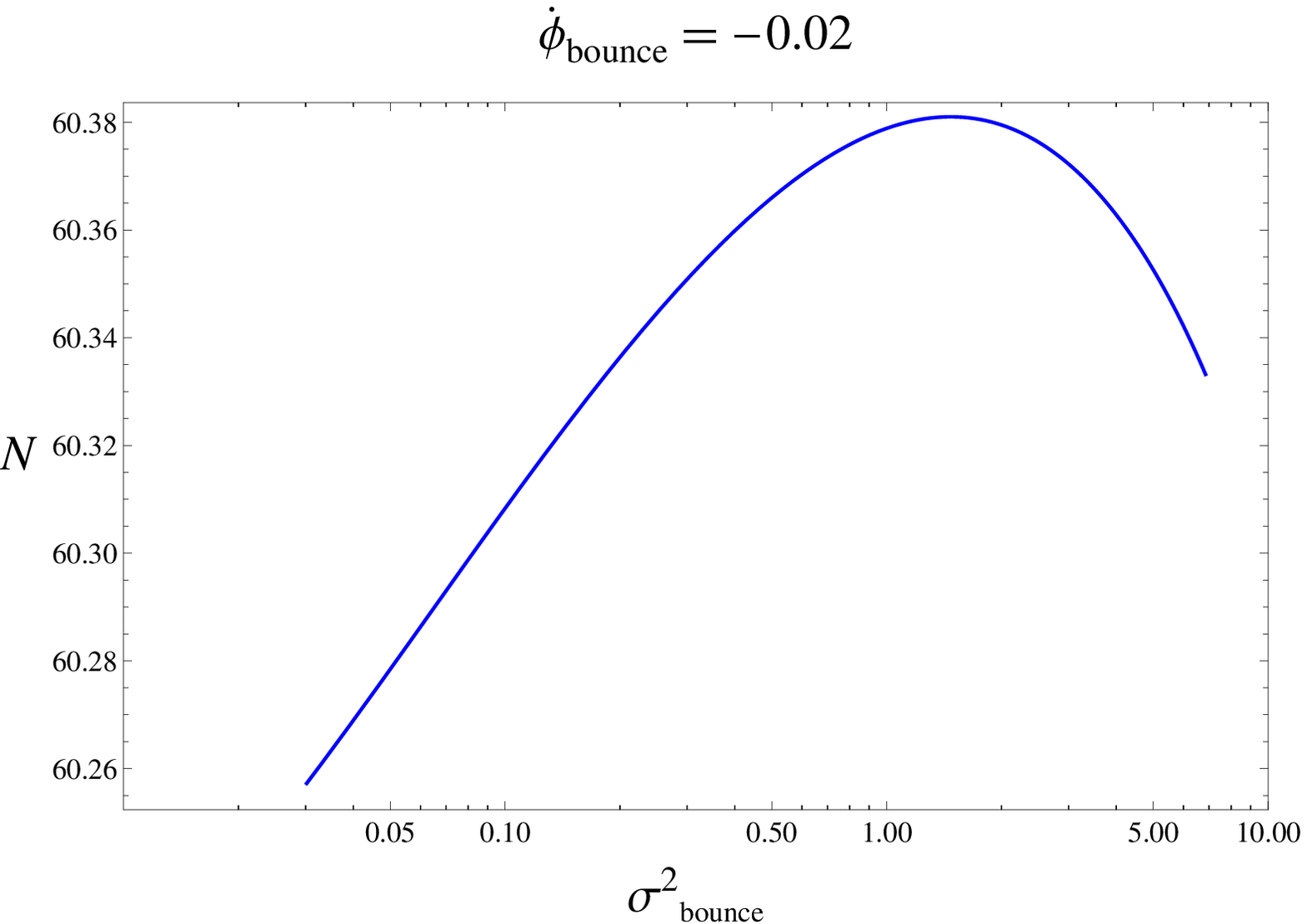}
\caption{This figure shows the variation of the number of e-foldings with the changing shear scalar at the bounce for $\dot\phi < 0$ at the bounce. The values of $\dot\phi$ is given in the units of $\mpl^2$.}
\label{efoldlqc1}
\end{figure}

We discussed earlier that the amount of inflation in the classical theory increases with increasing initial value of the shear scalar, for an inflaton which is rolling down. Since the shear scalar increases the mean Hubble rate it enhances the damping term in the Klein-Gordon equation. This in turn causes the slow roll conditions to be achieved earlier.
Unlike the classical theory, in LQC the mean Hubble rate is not necessarily a monotonic function of the shear scalar.
Due to the higher order corrections introduced by the underlying quantum geometry, the modified form of generalized Friedmann equation for Bianchi-I spacetime may contain nonlinear terms\footnote{For example, an approximate expression in the small shear limit, discussed in the Ref. \cite{cv},
contains $\sigma^4$ terms with opposite sign to $\sigma^2$ term, suggesting
non-monotonic behavior of the mean Hubble rate with shear scalar.} in $\sigma^2$.
\fref{hublqc} shows the variation of the mean Hubble rate with an increasing initial value of shear scalar. It is evident that the mean Hubble rate shows a non-monotonic behavior with increasing shear scalar, and attains a maximum at some value of $\sigma^2$.
A result of this is that, in LQC, in the presence of anisotropic shear the number of e-foldings can show a non-monotonic behavior with an increasing shear scalar (while keeping the $\dot\phi(0)$ fixed). \fref{efoldlqc1} shows the variation of the number of e-foldings for various different values of the initial energy density and $\dot\phi(0)<0$ at the bounce.
It turns out that for small shear scalar, the number of e-foldings ($N$) increases with the values of shear scalar, then $N$ attains a maximum value at some $\sigma^2=\sigma^2_*$.
For any higher value of anisotropic shear than $\sigma^2_*$, the amount of e-foldings decreases.
Note that, for the values of the parameters considered here, the changes in the number of
e-foldings is very small,
as can be seen in \fref{efoldlqc1}.
In the numerical simulations, $\sigma^2_*$ turns out to be
$\sigma^2_*\approx 1.47\, \lp^{-2}$ for $|\dot\phi(0)|< 0.02\,\mpl^2$,
 and for $|\dot\phi(0)| > 0.02\,\mpl^2$ it weakly depends on the initial data.
At this point the occurrence of $\sigma^2_*$ is only a numerical result.
Due to the unavailability of the generalized Friedmann equation for Bianchi-I spacetime
in LQC, we do not yet have analytical argument to explain its existence.

\begin{figure}[tbh!]
\includegraphics[width=0.47\textwidth]{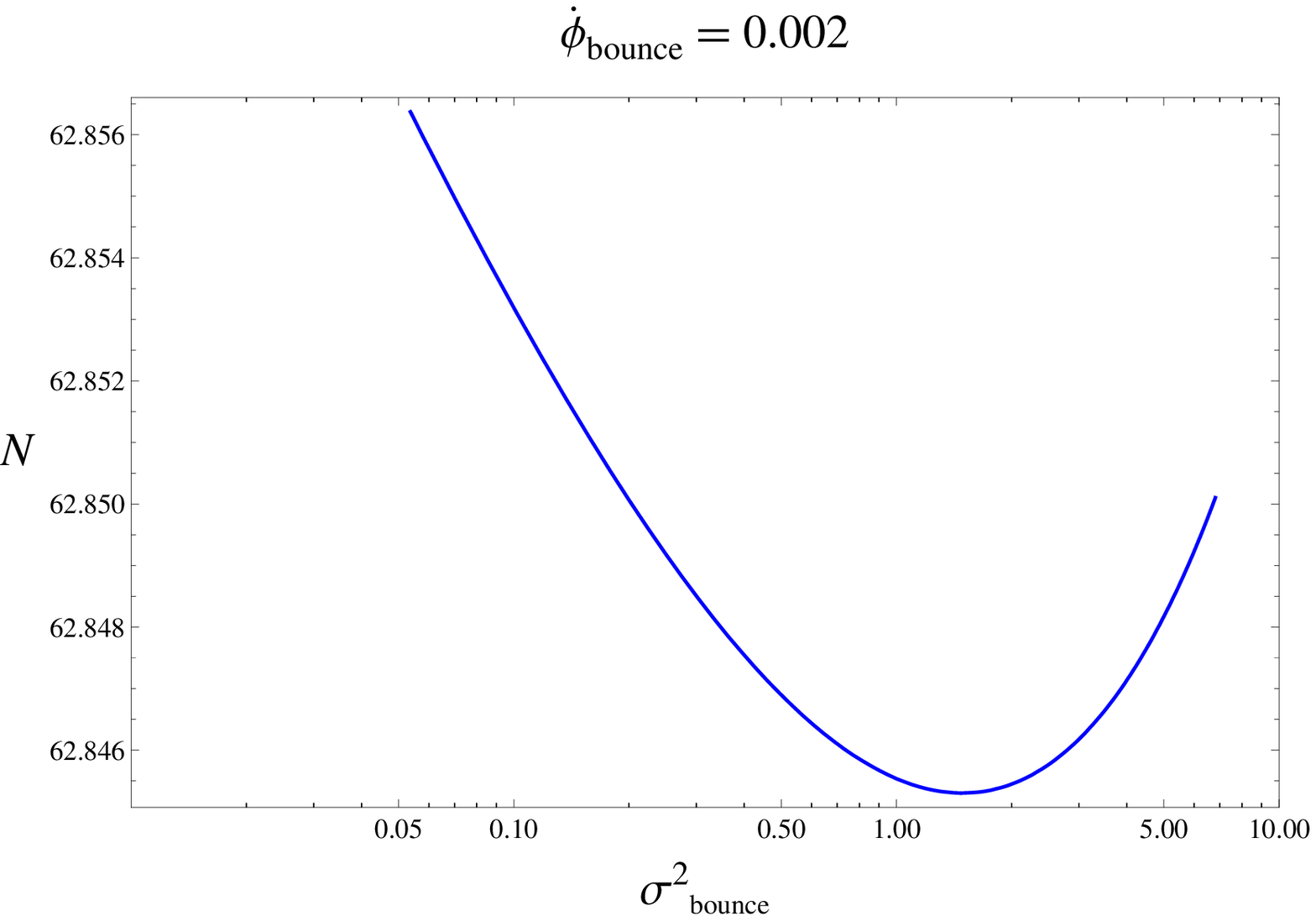}
\hskip0.5cm
\includegraphics[width=0.47\textwidth]{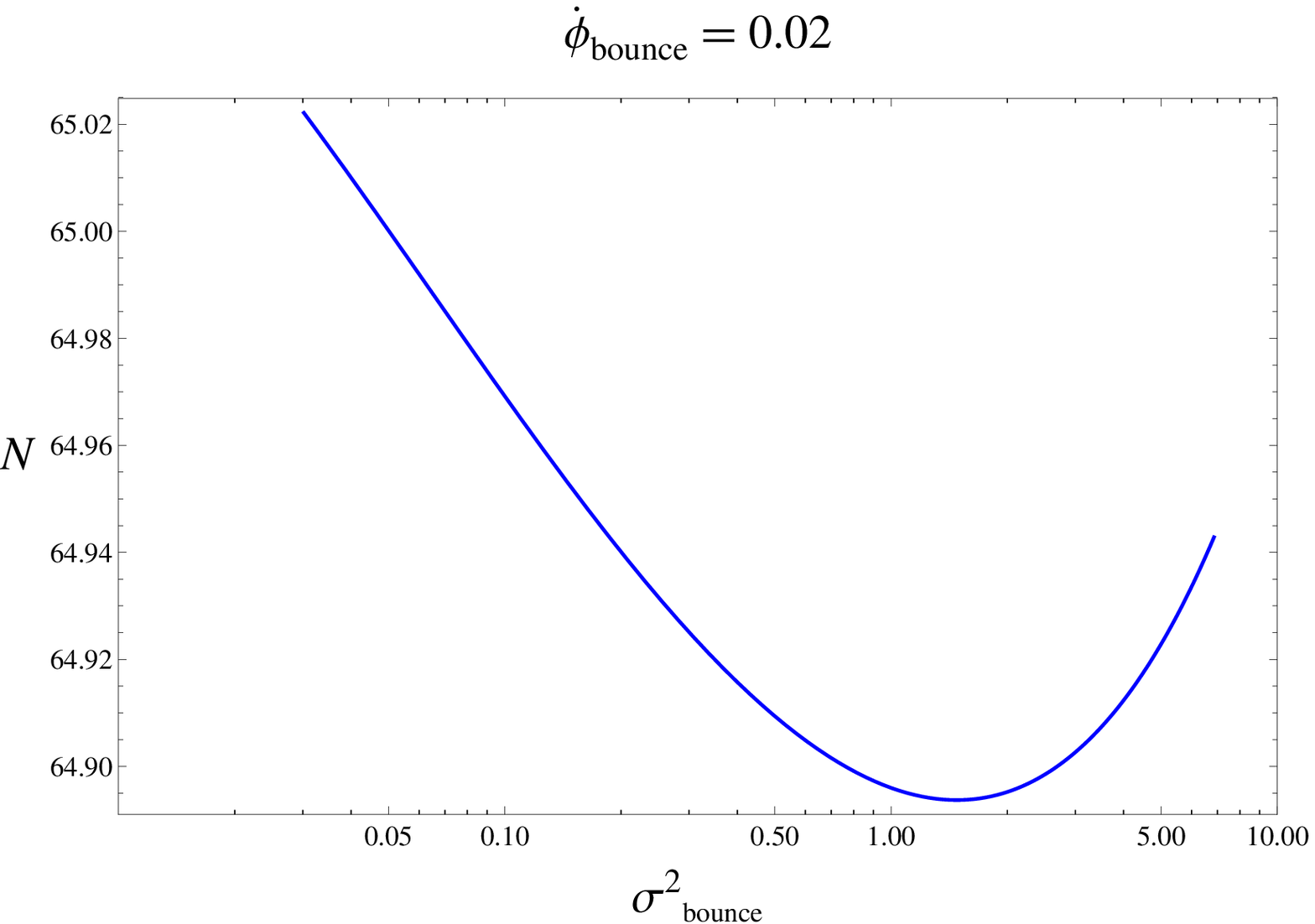}
\caption{Variation of the number of e-foldings with the changing shear scalar at the bounce for $\dot\phi > 0$ at the bounce. The values of $\dot\phi$ and $\sigma^2$ are in the units of $\mpl^2$ and $\lp^{-2}$ respectively.}
\label{efoldlqc2}
\end{figure}

As in the classical theory, for an opposite sign of $\dot\phi$ in the initial data (in this case the
initial data given at the bounce), the amount of inflation shows opposite behavior with the initial
anisotropic shear. The numerical simulations performed for $\dot\phi(0)>0$ are shown in
\fref{efoldlqc2}. It is evident that for $\dot\phi(0)>0$, in contrast to $\dot\phi(0)<0$, $N$ decreases
with increasing $\sigma^2$ in the small shear regime, while it increases for
$\sigma^2 > \sigma^2_*$. Interestingly, the value of $\sigma^2_*$ turns out to be numerically
same as that for $\dot\phi(0) <0$.

Such a variation of $N$ with initial anisotropy is tied to the behavior of the mean Hubble rate.
As discussed in the classical theory, an enhancement in the mean Hubble rate, results in
an increment in the amount of inflation if the inflaton is initially rolling down, and a decrement if it is
rolling up. In LQC, since the mean Hubble rate first increases, attains a maximum and then
decreases with increasing initial shear scalar (\fref{hublqc}), the amount of inflation varies in the
same way for $\dot\phi(0)<0$ (\fref{efoldlqc1}). That is, it first increases, attains a maximum and then falls down. The behavior of $N$ for $\dot\phi(0)>0$ is opposite to that of $\dot\phi(0)<0$.

\begin{table}[tbh!]
 	{\bf Number of e-foldings}\\[3pt]
	\begin{tabular}{|c|c|c|c|c|c|c|c|c|}
	\hline
	\multicolumn{1}{|c|}{$\dot\phi(0)(\mpl^2)$} &	\multicolumn{1}{|c|}{$\sigma^2(0)(\lp^{-2})$} & \multicolumn{6}{|c|}{$\phi(0)\, (\mpl)$}\\
	\cline{3-8}
	 &  & 1.0 & 2.0 & 3.0 & 3.14 & 4.0 & 5.0 \\
 	\hline
	&	 0.002 & 6.477 & 23.548 & 56.959 & 62.404 & 101.016 & 155.575 \\
	&	 0.277 & 6.486 & 23.555 & 56.971 & 62.426 & 101.03 & 155.592 \\
	& 	0.967 & 6.486 & 23.559 & 56.976 & 62.428 & 101.037 & 155.601 \\
	-0.002&	 1.441 & 6.487 & 23.561 & 56.979 & 62.429 & 101.041 & 155.606 \\
	&	3.184 & 6.486 & 23.562 & 56.981 & 62.428 & 101.044 & 155.61 \\
	&	 5.983 & 6.485 & 23.563 & 56.983 & 62.425 & 101.046 & 155.613 \\
	&	 6.831 & 6.484 & 23.564 & 56.984 & 62.424 & 101.048 & 155.615 \\
	\hline
	&	 0.002 & 6.654 & 23.887 & 57.447 & 62.833 & 101.659 & 156.365 \\
	&	0.277 & 6.645 & 23.880 & 57.436 & 62.812 & 101.645 & 156.347 \\
	&        0.967 & 6.644 & 23.877 & 57.431 & 62.809 & 101.638 & 156.339 \\
	0.002&         1.441 & 6.644 & 23.874 & 57.427 & 62.809 & 101.633 & 156.333 \\
	& 3.184 & 6.645 & 23.873 & 57.425 & 62.811 & 101.63 & 156.33 \\
        & 5.983 & 6.646 & 23.872 & 57.424 & 62.812 & 101.628 & 156.327 \\
 	& 6.831 & 6.647 & 23.871 & 57.423 & 62.813 & 101.627 & 156.325 \\
        \hline
 	& 0.049 & 5.765 & 23.548 & 56.959 & 60.278 & 101.016 & 155.575 \\
	&  0.287 & 5.787 & 23.555 & 56.970 & 60.352 & 101.03 & 155.592 \\
	&  0.976 & 5.795 & 23.559 & 56.975 & 60.377 & 101.037 & 155.601 \\
	-0.02 &  1.450 & 5.795 & 23.561 & 56.978 & 60.381 & 101.041 & 155.606 \\
	&  3.194 & 5.789 & 23.562 & 56.981 & 60.361 & 101.044 & 155.61 \\
	&  5.993 & 5.779 & 23.563 & 56.982 & 60.349 & 101.046 & 155.613 \\
	&  6.841 & 5.776 & 23.564 & 56.983 & 60.333 & 101.048 & 155.615 \\
        \hline
	&  0.049 & 7.409 & 27.221 & 59.486 & 65.001 & 105.226 & 161.673 \\
	&  0.287 & 7.385 & 27.233 & 59.502 & 64.924 & 105.248 & 161.700 \\
	&  0.976 & 7.376 & 27.217 & 59.479 & 64.898 & 105.218 & 161.663 \\
	0.02 &  1.450 & 7.376 & 27.203 & 59.457 & 64.893 & 105.188 & 161.626 \\
	&  3.194 & 7.382 & 27.191 & 59.439 & 64.915 & 105.164 & 161.596 \\
	& 5.993 & 7.394  & 27.182 & 59.425 & 64.924 & 105.145 & 161.572 \\
        & 6.841 & 7.397  & 27.175 & 59.415 & 64.943 & 105.131 & 161.554 \\
	 \hline
	  -0.905 & {\bf 0} & 12.595 & 2.087 & 1.648 & 3.103 & 17.997 & 47.994 \\
	  \hline
	 0.905 & {\bf 0} & 66.835 & 111.65 & 168.524 & 177.454 & 237.553 & 318.805 \\
	 \hline
	\end{tabular}
	\caption{This table summarizes the variation of number of e-foldings with varying shear scalar at the bounce, for various initial $\dot\phi$ at the bounce.}
	\label{efolddatavaryphi}
\end{table}

So far we have discussed the qualitative behavior of the amount of inflation with
varying initial anisotropy at the bounce, let us now turn to a more quantitative analysis.
An interesting feature of Bianchi-I spacetime lies in the comparison with the isotropic
spacetime. It is important to note that in the isotropic spacetime, if $\phi(0) = 3.14 \mpl$ at the bounce then the value of $\dot\phi$ at the
bounce is fixed to  $|\dot\phi(0)|\approx0.905\,\mpl^2$ as the bounce always occurs when
energy density saturates its upper maximum $\rho=\rho_{\rm max}=0.41\rho_{\rm Pl}$.
Whereas, in the Bianchi-I  spacetime, since the energy density may not be saturated at the bounce,
$|\dot\phi(0)|<0.905\mpl^2$ is allowed.  In this way,
there is a freedom in the choice of initial inflaton velocity at the bounce in Bianchi-I spacetime,
as compared to the isotropic spacetime.
Table-\ref{efolddatavaryphi} shows the number of e-foldings for various initial values
of the inflaton field with $\phi(0)=(1.0,\,2.0,\,3.0,\,3.14,\,4.0,\,5.0)\mpl$.
For each of $\phi(0)$, as we just discussed, there are several possible initial $\dot\phi$
in the Bianchi-I spacetime, unlike in the isotropic spacetime where the inflaton velocity at the
bounce is fixed for a given $\phi(0)$. This table shows the number of e-foldings for different initial velocities  $\dot\phi
(0)=(-0.002,\,0.002,\,-0.02,\,0.02)\mpl^2$ with varying initial shear scalar for each of the
$\phi(0)$ given at the bounce. The last two rows of Table - \ref{efolddatavaryphi} with
$|\dot\phi(0)|=0.905\,\mpl^2$ correspond to the isotropic spacetime, characterized by
$\sigma^2(0)=0$.
It is evident from the analysis of Table-\ref{efolddatavaryphi} that, if the inflaton is initially rolling down, then, the Bianchi-I
spacetime can produce significantly more number of e-foldings than in the isotropic
spacetime. For example, if the initial value of inflaton field is $\phi(0)=3.14\mpl$ then
the number of e-foldings in Bianchi-I spacetime is of the order of $60$ while for the same
initial value of inflaton the isotropic spacetime produces e-foldings as small as $3.103$, if the
inflaton is initially rolling down. In the same way, the maximum number of e-foldings for
$\phi(0) = 3.0\mpl$ in Bianchi-I spacetime is of the order of $50$ compared to the isotropic
spacetime where $N\approx 1.648$. In order to generate $N\approx60$, in the isotropic
spacetime for an inflaton initially rolling down, the initial value of the inflaton should be
$\phi(0)\approx5.50\mpl$ whereas, the same amount of e-foldings can be generated in the
Bianchi-I spacetime for a lower initial value of inflaton $\phi(0)<5.50\mpl$
(see Table-\ref{efolddatavaryphi}). In this way, presence of anisotropy widens up the window
of the values of the initial inflaton field, at the bounce, which produces a given amount of
inflation.

\subsubsection{Phase portrait and attractor behavior}

\begin{figure}[tbh!]
\includegraphics[width=0.47\textwidth]{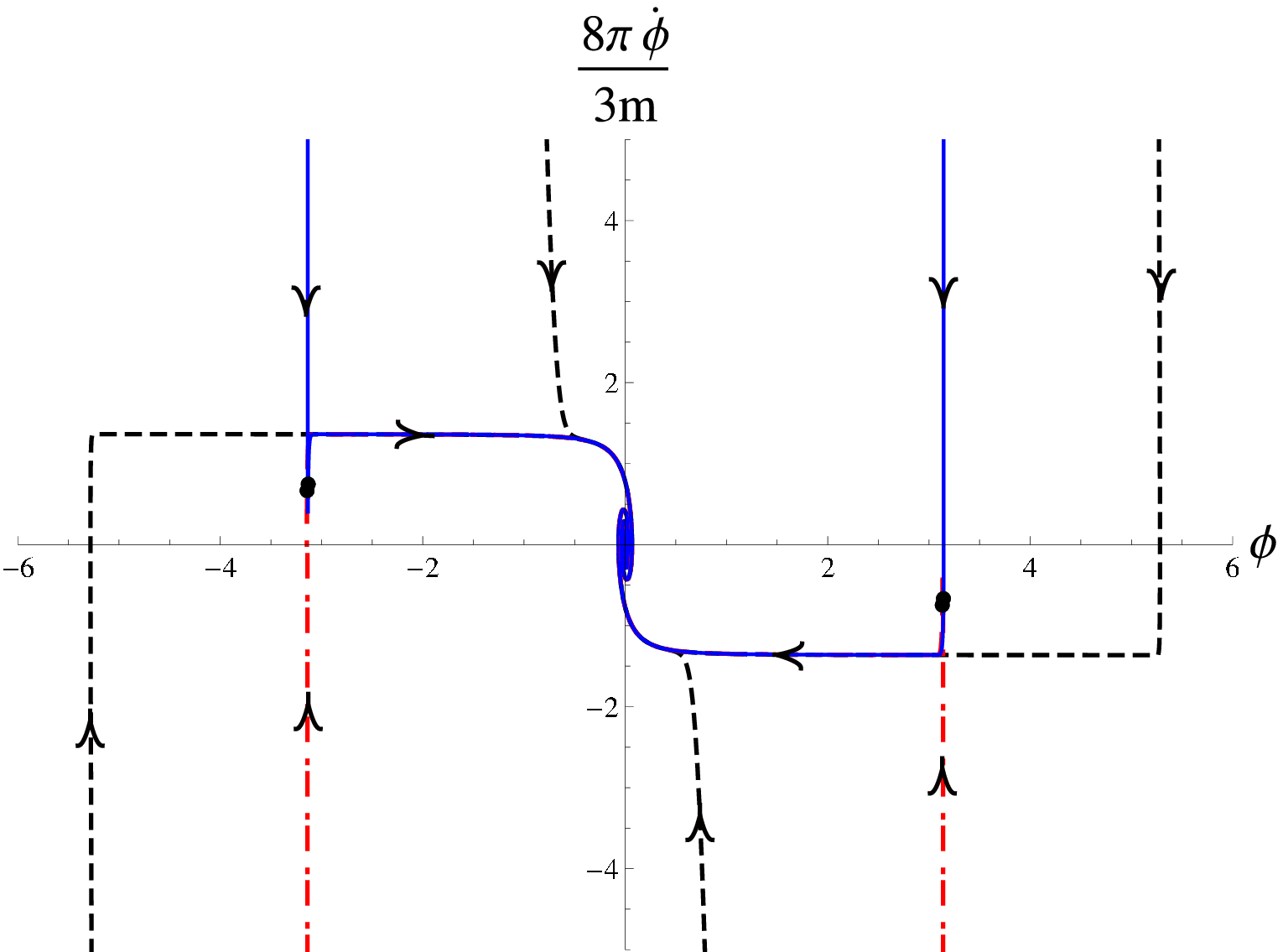}
\hskip0.5cm
\includegraphics[width=0.47\textwidth]{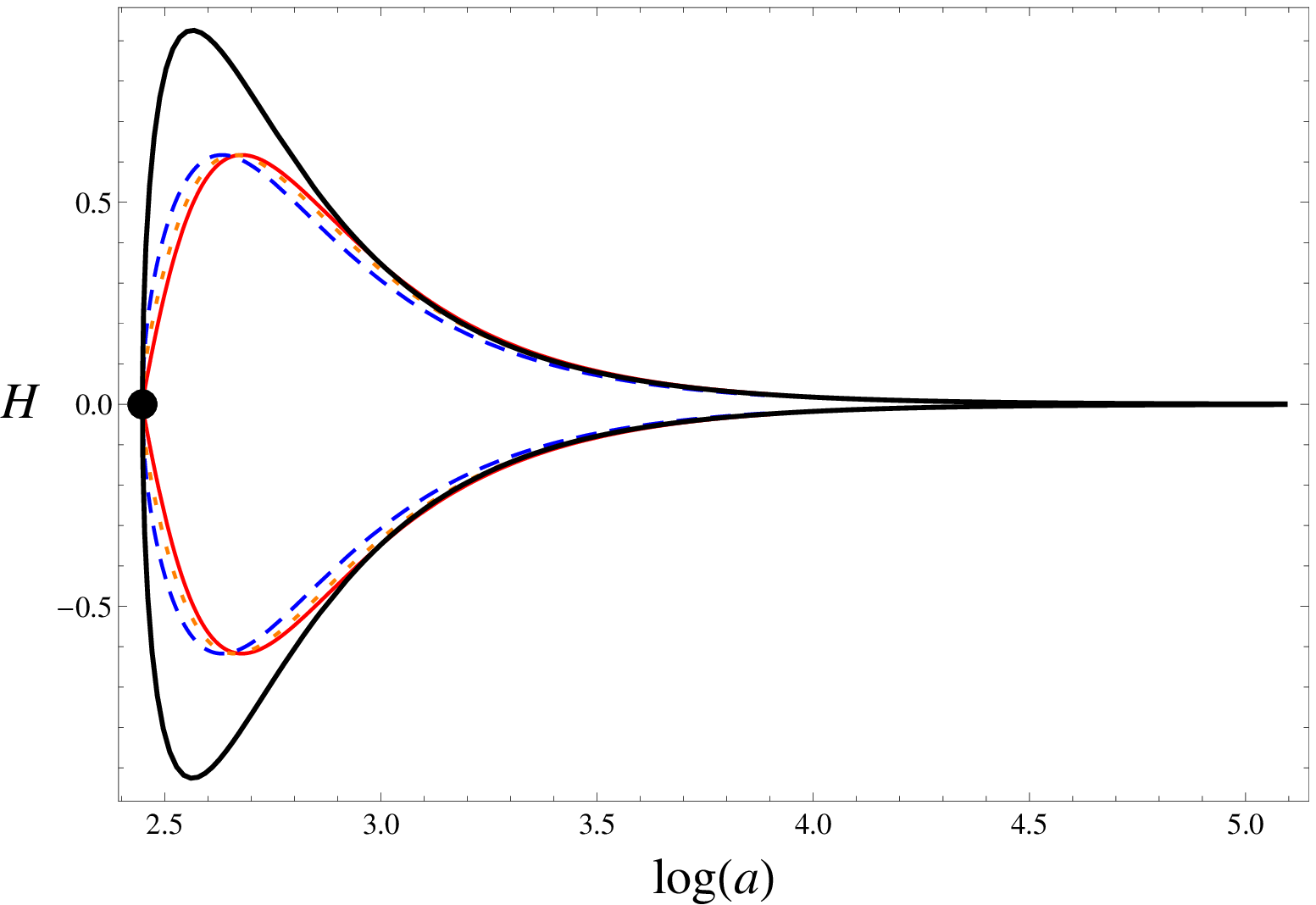}
\caption{The left figure shows the phase-space trajectory of the scalar field and the right plot is the plot of the trajectory in the gravitational phase-space ($\log(a)\,,H$).}
\label{lqcphase}
\end{figure}

Let us now explore the properties of the dynamical trajectories in LQC.
We will discuss the trajectories of both the matter and the gravitational sector i.e.
$(\phi, \dot\phi)$ for the inflaton and $(\log(a), H)$ for the gravitational part respectively.
A phase portrait of Bianchi-I spacetime in LQC is shown in \fref{lqcphase}.
The left plot corresponds to the inflaton in which the dashed thick lines show the isotropic
trajectories with the initial conditions
$\phi(0)=\pm3.14\,\mpl\, {\rm and}\, \dot\phi(0)\approx \pm 0.905\, \mpl^2 $ at the bounce.
The solid (blue) curve and the dotted-dashed (red) curves correspond to the LQC trajectories
with the initial conditions at the bounce as
$\phi(0)=\pm3.14\,\mpl\, {\rm and}\, \dot\phi(0) = \pm 0.002\,\mpl^2 $.
In this phase portrait, the first and third quadrant correspond to an inflaton which is
initially rolling up, while the second and fourth quadrant represent the trajectories
when the inflaton is rolling down the potential.
The slow-roll is characterized by the horizontal line, with an almost constant $\dot\phi$,
to which all the trajectories meet in their future evolution.
It is evident that for a rolling down inflaton with the given initial conditions, the Bianchi-I
trajectory remains on the slow-roll curve for a longer period of evolution than the isotropic
trajectory.
In this way, the presence of anisotropic shear at the bounce (even a small value)
leads the dynamical trajectory to meet the slow-roll curve sufficiently early which,
in contrast to the isotropic case, results in higher number of e-foldings if the inflaton is initially
rolling down the potential.

 Analysis of the phase-space trajectories of gravitational variables $\log(a)\, {\rm \it vs.}\, H$ gives information about the isotropization of the spacetime.\ The right plot in \fref{lqcphase} shows the comparison of isotropic trajectories with Bianchi-I trajectories for various values of initial shear at the bounce.
The thick (black) curve shows the isotropic trajectory, the dashed (blue) curve corresponds to Bianchi-I trajectory with initial shear scalar $\sigma^2(0)=5\times10^{-4}\, \lp^{-2}$, the  dashed (orange) curve with initial shear $\sigma^2(0)=5.7\times10^{-2}\, \lp^{-2}$ and the  solid (red) line with $\sigma^2(0)=6.81\, \lp^{-2}$ at the bounce, while the initial conditions on the inflaton are being fixed at $\dot\phi(0)=-0.002\,\mpl^2\,{\rm and}\, \phi(0)=3.14\, m_{\rm Pl}$. The black dot denotes the bounce point of the mean scale factor.
It is clear from the plot that, close to the bounce, the evolution of Bianchi-I spacetime
trajectories are different from that of the isotropic spacetime. Whereas, away from the bounce
when the curvature is much smaller, then all the trajectories tend to meet the isotropic
trajectory, leading the spacetime to isotropize.

\bfig[tbh!]
\includegraphics[width=0.67\textwidth]{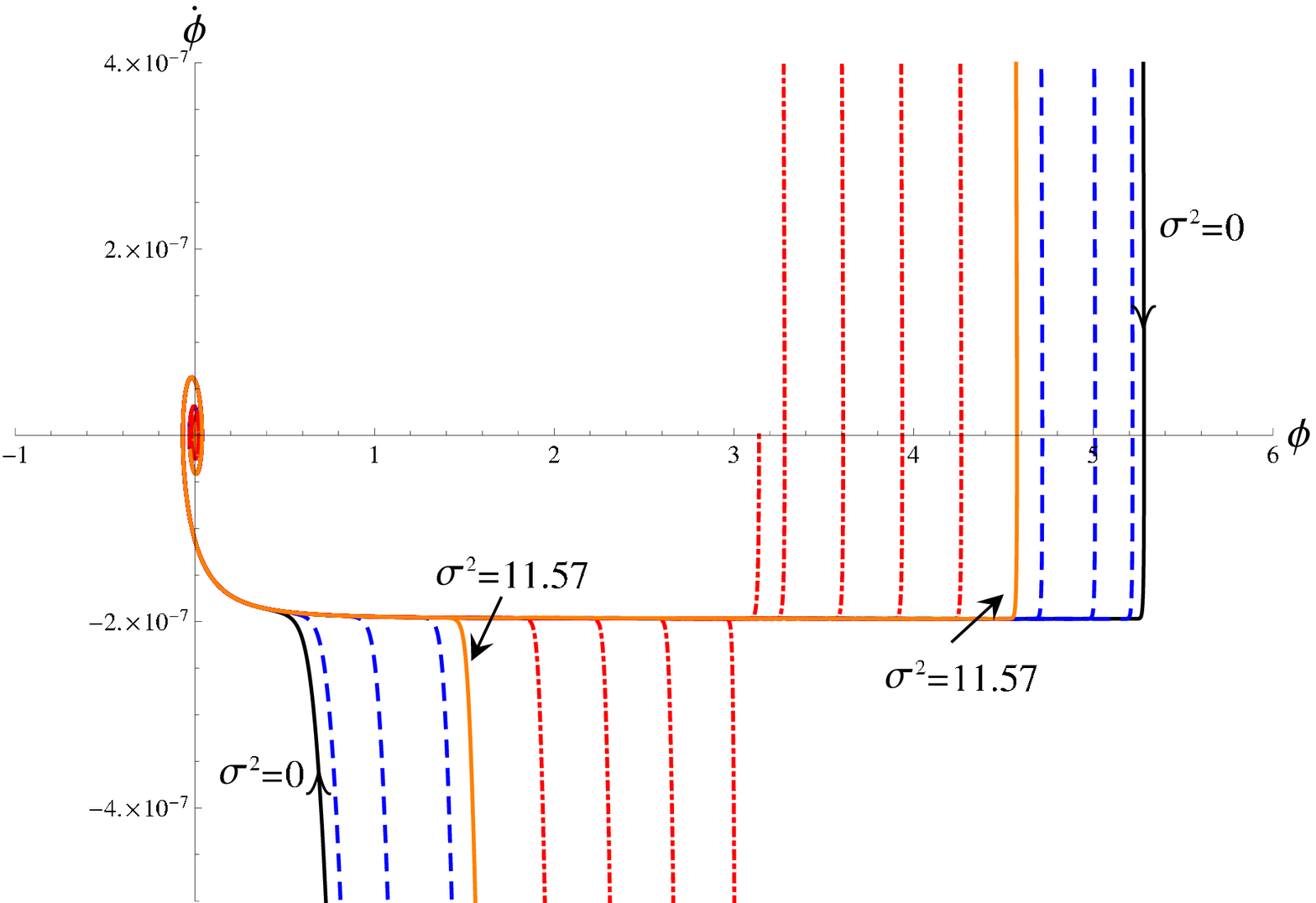}
\caption{This figure shows the evolution of various trajectories starting with different
$\dot\phi (0)= (-0.905,\,  -0.856,\, -0.727,\, -0.584,\, -0.536,\, -0.407,\, -0.289,\, -0.171,\, -0.053) \mpl^2$
starting in the fourth quadrant and
$\dot\phi(0)=(0.0,\, 0.053,\, 0.171,\, 0.289,\, 0.407,\, 0.536,\, 0.584,\, 0.727,\, 0.856,\, 0.905)\mpl^2$
in the first quadrant (from left to right). The solid black curves marked with $\sigma^2=0$
correspond to the isotropic trajectories and the solid (orange) curves denoted by
$\sigma^2=11.57$ (in Planck units) correspond to trajectories with maximum shear scalar
at the bounce. For all the trajectories, the initial value of the inflaton is kept fixed at $\phi(0)=3.14\,\mpl$. All the values shown in this plot are in Planck units.}
\label{phase2dlqc}
\efig

\bfig[tbh!]
\includegraphics[width=0.7\textwidth]{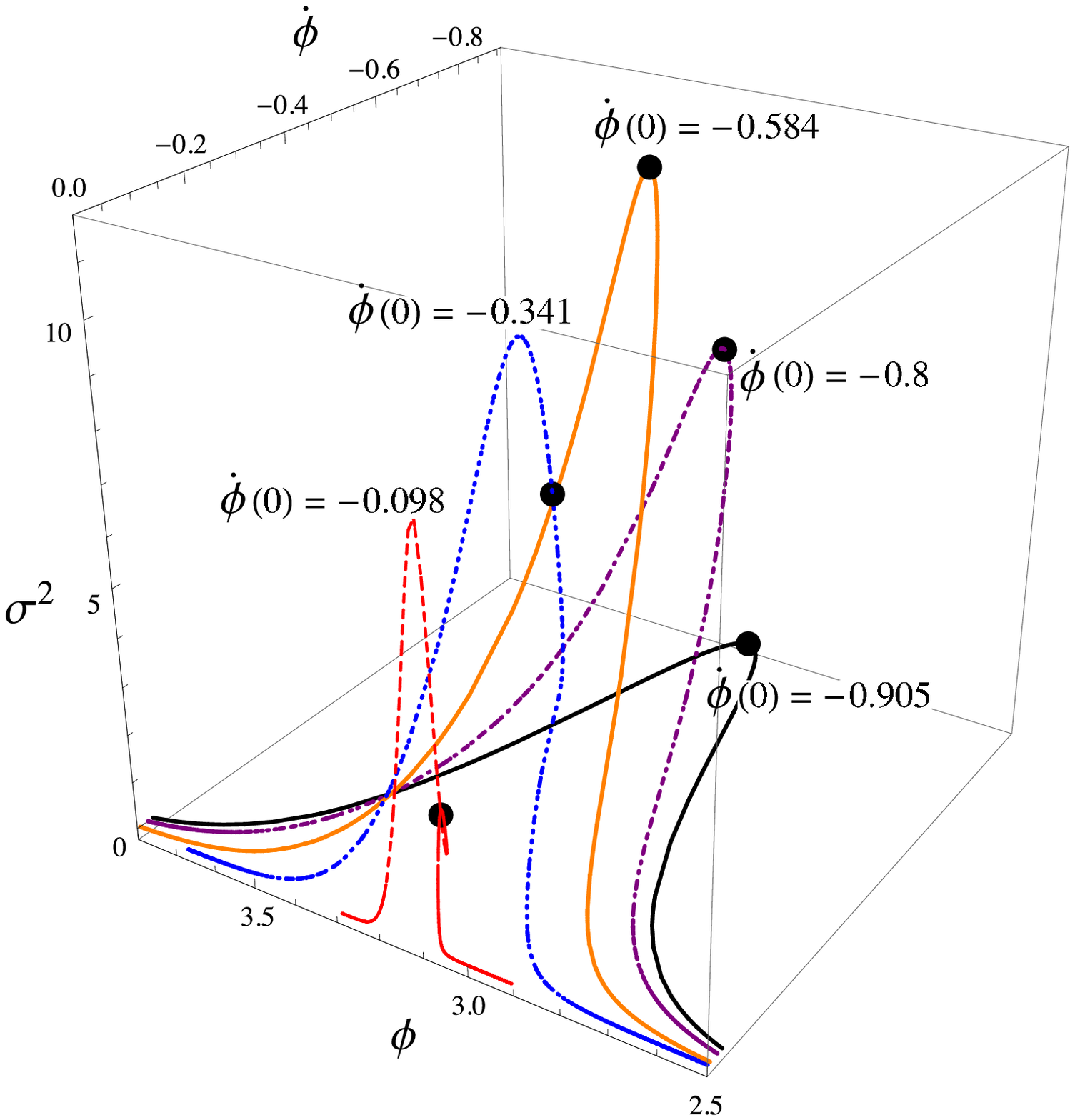}
\caption{This plot shows the 3D phase portrait of various trajectories starting with different $\dot\phi(0)<0$ in the near bounce region. The black dot marks the bounce point. The solid (orange) line corresponds to the maximum shear scalar at the bounce
$\sigma^2=\sigma^2_{\rm max}=11.57\,\lp^{-2}$. The thick black line in the $\sigma^2=0$ plane corresponds to the isotropic trajectory. Clearly, as the value of initial $\dot\phi$ tends to $-0.905\,\mpl^2$, the Bianchi-I trajectories approach the isotropic trajectory.}
\label{phase3dlqc}
\efig

We now discuss, in detail, a feature of the attractor behavior of Bianchi-I spacetime
in the effective description of LQC. \fref{phase2dlqc} shows the evolution of
trajectories starting with the same initial $\phi(0)=3.14\,\mpl$ but different initial
$\dot\phi$. For clarity, we only show the
first and the fourth quadrant of the phase portrait. The initial values of $\dot\phi$ for the
trajectories in the fourth quadrant are $\dot\phi(0)=(-0.905,\,  -0.856,\, -0.727,\, -0.584,\,
-0.536,\, -0.407,\, -0.289,\, -0.171,\, -0.053) \mpl^2$, and in the first quadrant,
$\dot\phi(0)=(0.0,\, 0.053,\, 0.171,\, 0.289,\, 0.407,\, 0.536,\, 0.584,\, 0.727,\, 0.856,\,
0.905)\mpl^2$. The left most trajectory, in the fourth quadrant, denoted by $\sigma^2=0$
corresponds to the isotropic spacetime with $\dot\phi(0)=-0.905\,\mpl^2$, and the right most
trajectory in the first quadrant marked with $\sigma^2=0$ represents the evolution of isotropic spacetime with
$\dot\phi(0)=0.905\,\mpl^2$. All the other trajectories correspond to Bianchi-I spacetime with
non-zero initial anisotropic shear scalar. Recall that in the effective description of Bianchi-I
spacetime the shear scalar has an upper bound, $\sigma^2_{\rm max}=11.57\,\lp^{-2}$.
Evolution of trajectories starting with initial $\sigma^2=\sigma^2_{\rm max}$ are shown by
solid (orange) curves denoted by ``$\sigma^2=11.57$''. All the other Bianchi-I trajectories
correspond to smaller values of initial shear scalar the bounce. The slow-roll curve is characterized by the horizontal line with an almost constant
$\dot\phi\approx1.97\times10^{-7}\mpl^2$, to which all the trajectories with various initial
$\dot\phi(0)$ and $\sigma^2(0)$ meet in their future evolution. Thus, like in the classical
theory, the slow-roll curve turns out to be an attractor for all the trajectories shown in the
phase portrait. This implies that irrespective of the initial anisotropic content of the spacetime,
all the Bianchi-I spacetime trajectories in the effective description of
LQC do meet the slow-roll curve in their future evolution, if the initial $\phi(0)$ is kept high
enough. Thus, we conclude that the presence of non-vanishing anisotropic
shear does not prevent the occurrence of slow-roll inflation with $\phi^2$ potential in Bianchi-I LQC.

In the effective description of LQC, the isotropic spacetime is characterized by
two conditions: the first condition requires the vanishing of the shear scalar $\sigma^2=0$ at
all times and the second condition requires the energy density to be $\rho=\rho_{\rm max}$ at
the bounce. This leads to an interesting feature of effective dynamical trajectories of Bianchi-I
spacetime regarding its isotropic limit. That is, in order to approach the isotropic limit of
Bianchi-I spacetime in effective dynamics of LQC, the energy density at the bounce must
tend to approach its maximum, i.e.\ $\rho\rightarrow\rho_{\rm max}$
($|\dot\phi(0)| \rightarrow 0.905\,\mpl^2$, if $\phi(0)=3.14\mpl$), along with the shear scalar vanishing, i.e.\
$\sigma^2(0)\rightarrow0$ at the bounce. This stands in contrast with the classical theory,
where $\sigma^2\rightarrow0$ is sufficient for Bianchi-I spacetime to approach the isotropic
limit, irrespective of the value of the energy density. As discussed earlier, in the classical theory, for a fixed initial
energy density, the isotropic limit can be approached by continuously
varying the initial $\sigma^2$ to zero as shown in \fref{2dphase}. Thus, there is an isotropic
trajectory corresponding to every initial energy density in the classical theory.
In contrast, in LQC, the energy density for an isotropic spacetime must be $\rho=\rho_{\rm max}$ at the bounce. In this way, for a given initial value of the inflaton field, there are
several classical isotropic trajectories, but there are only two isotropic LQC trajectories
(as shown in \fref{phase2dlqc}). These two trajectories correspond to different signs of initial $\dot\phi$, such that $\rho=\rho_{\rm max}$ at the bounce.
Therefore, for a given initial value of the inflaton field, there is a unique isotropic curve in each
quadrant of the $\phi-\dot\phi$ phase portrait of Bianchi-I spacetime in effective description of
LQC. The two isotropic curves corresponding to $\phi(0)=3.14\,\mpl$ are shown in
\fref{phase2dlqc}, one in first and the other in the fourth quadrant.
Starting from any given trajectory in \fref{phase2dlqc}, if one decreases $\dot\phi(0)$ to
$-0.905\,\mpl^2$, which is the value of $\dot\phi$ at the bounce in the isotropic
spacetime for an inflaton which is rolling down,
then the Bianchi-I trajectories approach towards the isotropic trajectory in the fourth
quadrant. While if one increases $\dot\phi(0)$ to $0.905\,\mpl^2$, the value of $\dot\phi$ at
the bounce in the isotropic spacetime for an inflaton which is rolling up, then the isotropic
trajectory in the first quadrant is approached. In this way, the isotropic spacetime  turns out to
be an ``isotropic attractor'' of Bianchi-I spacetime in the effective description of LQC. This
behavior is easy to understand from a 3D plot of the phase trajectories as shown in
\fref{phase3dlqc}, where we show the evolution of Bianchi-I trajectory in the vicinity of the
bounce. The solid black trajectory in the $\sigma^2=0$ plane corresponds to the isotropic
spacetime with $\dot\phi(0)=-0.905\,\mpl^2$ and all the other ones correspond to Bianchi-I
spacetime with different initial $\dot\phi$ and $\sigma^2$ at the bounce. It is clear to see from
this figure that as $\dot\phi(0)\rightarrow-0.905\,\mpl^2$, the trajectories tend to approach the
isotropic trajectory in the $\sigma^2=0$ plane.

It is important to note that the isotropic attractor behavior in LQC is obtained by varying
both the initial energy density at the bounce and the shear scalar, whereas in the classical theory the isotropic
spacetime can be approached for any fixed initial energy density. If one fixes the energy density, at
the bounce in LQC, to any other value than $\rho_{\rm max}$ by specifying $\phi(0)$ and
$\dot\phi(0)$ and tries to vary the shear scalar, then it turns out that the shear scalar can not
be varied to a value less than a non-zero minimum value which depends on the energy
density at the bounce. In this way, again, unless the energy density at the bounce tends to
$\rho_{\rm max}$, Bianchi-I trajectories do not approach the isotropic spacetime. Thus, in
LQC, isotropic attractor is not obtained for those Bianchi-I trajectories for which
$\rho\neq\rho_{\rm max}$ at the bounce.

To obtain a complete set of initial data, for the trajectories in \fref{phase2dlqc}, we give
$p_i(0),\,c_1(0),\,\phi(0)$ for varying $\dot\phi(0)$, and the values of $c_2(0),\,c_3(0)$ are
computed by solving the Hamiltonian constraint and the mean Hubble rate being,
$H=0$ at the bounce. In the numerical simulations, it turns out that for a fixed value of
$c_1(0)$, there is a fixed range of the values of $\dot\phi(0)$ for which the Hamiltonian constraint is
satisfied. That is, $\dot\phi(0)$ can not be continuously varied from $-0.905\,\mpl^2$ to
$0.905\,\mpl^2$ for a fixed value of $c_1(0)$, in the numerical simulations performed. In this
way, depending on various fixed values of $c_1(0)$, one can generate a continuous family of
trajectories by varying $\dot\phi(0)$ within the corresponding range, which depends on the value $c_1(0)$. In each
of these ranges, there will be a limiting curve with minimum shear scalar to which the other
trajectories in the range tend, as the shear scalar in the initial data decreases.
For the simulations shown in \fref{phase2dlqc}, one has to choose atleast two different
values of $c_1(0)$, one for $|\dot\phi(0)|<0.5839\,\mpl^2$ and another for
$|\dot\phi(0)|>0.5839\,\mpl^2$.

It is now clear from the discussion of the phase portraits of the inflationary Bianchi-I spacetime that:
\begin{itemize}
\item Trajectories starting with a wide variety of initial conditions meet the slow-roll trajectory in the future evolution. This indicates slow-roll is an attractor for Bianchi-I spacetime.

\item Isotropic spacetime behaves like an ``isotropic attractor'' for Bianchi-I spacetime
if the isotropic limit is taken by considering both $\sigma^2(0)\rightarrow0$ and
$\rho(0)\rightarrow\rho_{\rm max}$ at the bounce. For the trajectories, for which the initial energy density does not approach $\rho_{\rm max}$ at the bounce, the isotropic limit does not exist.

\end{itemize}

Hence, the two attractor behavior present in the classical Bianchi-I spacetime with $\phi^2$ potential, is also present
in the effective description of LQC,  with the above noted subtlety in the way isotropic limit of
Bianchi-I spacetime is taken. It is worth emphasizing that due to quantum geometric
effects in the pre-inflationary era, the physical trajectories of the effective description are very
different from those of the classical Bianchi-I spacetime. It is due to these differences that the
conditions to approach the isotropic limit are modified in the effective description of LQC as
compared to the classical theory.

\section{Discussion}
Though the inflationary paradigm has been successful in explaining the physics of the early universe, it is incomplete without understanding the way the initial singularity is resolved and the way inflation begins from generic initial conditions. In this work, we have analyzed these questions for the $\phi^2$ inflation in the setting of an anisotropic spacetime
described by the Bianchi-I model using the effective spacetime approach of LQC and made a comparative analysis with the classical theory.  The underlying quantum geometric
effects originating from LQG result in a non-singular evolution. The backward evolution in LQC is devoid of the
big bang singularity. The latter is replaced by a bounce of the mean scale factor when the spacetime curvature
becomes Planckian. The energy density and the shear scalar remain bounded throughout the evolution in LQC.
To our knowledge, this is the first such study of the inflationary scenario in Bianchi-I spacetime in a non-singular setting.

One of the key questions in the study of the evolution of the universe is its isotropization from a generic anisotropic universe in the pre-inflationary era. Previous studies of Bianchi-I inflation in the classical theory and brane world cosmology established that the presence of anisotropy does not prevent inflation, and it is actually conducive to the number of e-foldings obtained during the inflationary phase.
These studies for quadratic potential took the initial velocity of the inflaton to be negative (i.e.\ the inflaton initially rolling down the potential).
 It was found that the amount of inflation for such initial conditions always increased with increasing shear in the classical theory as well as in brane-world scenarios \cite{mss}.
We revisited  the analysis in the classical theory by taking a more generic initial data by including the case when inflaton is initially rolling up the potential. In this case, we found that in comparison to the corresponding isotropic evolution, the amount of inflation actually decreases with an increasing anisotropic shear.
This happens because of the greater anisotropy, the Hubble friction in the Klein-Gordon
equation increases which turns up the decay of kinetic energy of the inflaton. Due to the
faster decay of the kinetic energy, an inflaton which is initially rolling up, stops at a lower value
of the potential which decreases the amount of e-foldings in the subsequent slow-roll evolution.  
 It is worth emphasizing again that the presence of anisotropic shear does {\it not} rule out inflation.
We showed that for accelerated expansion to take place, the potential energy must win over the kinetic and the shear energy. Owing to this property, in the inflationary era, shear term is always less than the potential energy.
There may however be, a small period during the accelerated expansion when anisotropic shear contribution $(\sigma^2/16\pi G)$ while being less than potential term $(V(\phi))$, is still greater than the kinetic term $(\dot\phi^2/2)$. In this sense there may be a brief duration of shear dominance in the inflationary phase of Bianchi-I spacetime.
We also showed by explicit simulations that if one starts with highly anisotropic initial conditions such that two of the scale factors are contracting and third  one is expanding, in the inflationary phase all of the scale factors expand.\
This property can be attributed to the value of the anisotropy parameter $\epsilon_{\rm J}$ which always decreases in an expanding universe. Therefore, irrespective of large initial $\epsilon_{\rm J}$, the shear scalar decreases and due to the accelerated expansion there is a point, in the future evolution, where $\epsilon_{\rm J} < 2/\sqrt{3}$ which corresponds to all expanding directions of the Bianchi-I spacetime. Hence, isotropization is inevitable in an inflating classical Bianchi-I spacetime with a $\phi^2$ potential.

In the analysis of the loop quantum effective dynamics of the inflating Bianchi-I spacetime, the quantum gravitational corrections come into play in the near bounce phase in the pre-inflationary era.
Due to the presence of an upper bound on the anisotropic shear and the energy density, the set of initial conditions are quite different from those in the classical theory.
Also, to give initial conditions on the bounce point, along with satisfying the Hamiltonian constraint, one has to make sure that the mean Hubble rate is zero, a condition which was not present in the classical theory. Our analysis shows that the amount of inflation in LQC does not change monotonically with the initial anisotropic
shear at the bounce. It has a turning point at $\sigma^2_*$ whose value depends on the initial value of inflaton and its velocity (but is independent of the sign of the initial velocity). For an inflaton initially rolling down the potential, the number of e-foldings increase with anisotropy only when initial $\sigma^2$ is less than $\sigma^2_*$. On the other hand, if the inflaton is initially rolling up, then the number of e-foldings decrease with anisotropy only when initial $\sigma^2$ is less than $\sigma^2_*$. As in the classical theory, isotropization is inevitable in the inflationary LQC Bianchi-I spacetime starting from arbitrary anisotropic initial conditions.

Analysis of the phase space trajectories demonstrate that there is a double attractor behavior in effective dynamics of LQC for Bianchi-I spacetime as in the classical theory, however one has to be careful in taking the isotropic limit at the bounce. In the classical theory, phase space trajectories approach the slow-roll attractor in the future evolution. This holds true also in LQC. There is also a non slow-roll isotropic attractor which is reached by taking the limit $\sigma^2 \rightarrow 0$ for any value of energy density in the classical theory. In LQC, since energy density is saturated to its universal maximum value $\rho_{\rm max}$ at the bounce, the second  attractor is reached only for those trajectories starting at the bounce which satisfy  $\sigma^2 \rightarrow 0$ along with $\rho \rightarrow \rho_{\rm max}$.

We conclude our discussion by pointing out that  in the Bianchi-I spacetime sufficient number of e-foldings can be
obtained by starting at a significantly lower value of the inflaton field at the bounce than in the isotropic LQC when the inflaton is initially rolling down the potential. Further there is  more freedom in specifying the initial data at the bounce in the anisotropic model.
This can be of significance in the study of observational signatures of quantum gravitational effects in the pre-inflationary stage of LQC. As an example, it has been recently suggested that for the isotropic LQC, a narrow window depending on the value of the inflaton field at the bounce exists for which there can be potentially distinct signatures of LQC in the cosmic microwave background \cite{aan1,aan3}. Our analysis suggests that such a window potentially widens in the presence of anisotropies if the inflaton is initially rolling down in the pre-inflationary epoch.
Owing to widening the window of the value of the inflaton field in Bianchi-I spacetime for the same number of e-foldings in the isotropic case, there could be potentially additional corrections which will again be of quantum gravitational in origin. A complete quantum treatment of the field perturbations in Bianchi-I anisotropic background is expected to bring out additional possible observational signatures of loop quantum effects in the primordial power spectrum.

\acknowledgements
We thank Abhay Ashtekar, Jacobo Diaz-Polo, Jorge Pullin  and Edward Wilson-Ewing for useful discussions and suggestions. This work is supported by NSF grant PHYS1068743 and a grant by the John Templeton Foundation. The opinions expressed in this publication are those of the authors and do not necessarily reflect the views of the John Templeton Foundation. BG's research is partially supported by the Coates Scholar Research Award of Louisiana State University.

\end{document}